\documentclass[aps,prd,reprint,onecolumn,11pt,superscriptaddress,nofootinbib,floatfix]{revtex4-2}

\usepackage{graphicx}
\usepackage{multirow}
\usepackage{subcaption}
\usepackage{float}
\usepackage{bm}
\usepackage{comment}
\usepackage{color}
\usepackage{amsmath,amssymb}
\usepackage[T1]{fontenc}
\usepackage{lmodern}
\usepackage{graphicx,booktabs}
\usepackage[
    colorlinks=true,
    urlcolor=blue,
    citecolor=black,
    linkcolor=black,
    linktoc=none
]{hyperref}
\newcommand{\llangle}{\langle\mkern-4mu\langle}
\newcommand{\rrangle}{\rangle\mkern-4mu\rangle}
\begin{document}

\title{Massive Scalar Quasinormal Modes and Quasibound States of
Charged Black Holes with Logarithmic corrections}

\author{Prosenjit Paul}
\affiliation{AMC Engineering College, Bengaluru, India}

\author{Rickmoy Samanta}
\affiliation{Indian Institute of Technology Kharagpur,
Kharagpur, India}
\begin{abstract}
We study how long-range deformations in a black-hole metric affect
the oscillation, trapping, and decay of a neutral, minimally coupled
massive scalar field. We choose a logarithmic deformation of
Reissner-Nordstr\"om (RN) sourced by nonlinear electrodynamics,
arising as an exact solution of Einstein gravity. The same geometry
also arises in a phenomenological perfect-fluid dark-matter model.
We provide several exact completions of this geometry with finite Arnowitt-Deser-Misner (ADM) mass. Focusing on the original logarithmic geometry,
we analyze both quasinormal modes (QNMs), which describe relaxation,
and quasibound states (QBSs), which describe trapped modes with
small horizon absorption. We first show that a small scalar mass moves the potential barrier outward, raises it, and reduces its curvature. First-order WKB then predicts weaker damping at every fixed finite multipole and overtone. We then compare several approaches, namely sixth-order WKB, global matching, a generalized Leaver method (adapted to the logarithmic geometry) and time-domain evolution. All methods yield consistent QNM estimates within
their ranges of validity. Global matching and Leaver also agree closely for QBSs. Further analytic estimates with a reference trapped state show that a local Pad\'e
approximation gives a binding-energy error of about $0.04\%$,
while leading Gamow tunnelling underestimates damping by about
$25\%$.  Across the sampled QBSs, increasing the scalar
mass or nonlinear coupling strengthens binding but shortens the
lifetime through greater horizon absorption. Finally, using an appropriate horizon regularization and a common complex radial contour, we demonstrate orthogonality of  an independently computed QNM-QBS pair using a suitable relativistic bilinear framework.
\end{abstract}

\maketitle
\tableofcontents

\section{Introduction}
\label{sec:introduction}

Black-hole quasinormal modes (QNMs) describe how perturbations
oscillate and decay~\cite{Vishveshwara:1970zz,Kokkotas:1999bd,
Berti:2009kk,Konoplya:2011qq}. Observations of compact-binary
mergers~\cite{LIGOScientific:2016aoc,LIGOScientific:2016sjg,
LIGOScientific:2020stg,LIGOScientific:2017vwq,LIGOScientific:2021qlt} and open interferometer data~\cite{KAGRA:2023pio} motivate
understanding how this response depends on spacetime
geometry and matter content. Studies of modified gravity~\cite{Tattersall:2019pvx,Sirera:2023pbs,LuiChung2026QNM} and
matter environments~\cite{Liu:2021xfb,Das2023ess} have explored this dependence.
Related work examines detection prospects~\cite{Spieksma:2024voy} and the
connection between ringdown and horizon structure~\cite{Cardoso:2016rao}.
In this work we ask how long-range deformations in the metric affect both
radiative modes as well as trapped states outside a black hole.

We study this question using an exact solution of Einstein
gravity coupled to nonlinear electrodynamics (NED)~\cite{Mazharimousavi2023}.
The model is inspired by a logarithmic confinement potential
and has the metric function
\[
 f(r)=1-\frac{2M+A\ln(r/r_0)}{r}+\frac{q^2}{r^2},
 \qquad A=\frac{4q^{3/2}\zeta}{3}.
\]
The coupling $\zeta$ controls the logarithmic correction, and
$\zeta=0$ recovers Reissner-Nordstr\"om (RN). This allows us to
vary the deformation and compare the spectra with their RN limits. We use a neutral, minimally coupled massive test scalar, so the
electric field affects it only through the geometry. A nonzero
scalar mass $\mu$ introduces the length scale $\mu^{-1}$ and
makes the potential approach $\mu^2$ at large radius. An outer
well can then appear in addition to the inner barrier. QNMs probe
scattering and relaxation, while quasibound states (QBSs)
describe waves trapped in the well and gradually absorbed by
the horizon. Our metric also coincides, after a parameter redefinition, with
the positive-density Maxwell limit of phenomenological
perfect-fluid dark-matter (PFDM)
geometries~\cite{Das2023ess,Ma2024PFDM,Becar2026PFDM},
connecting our neutral-scalar analysis with the broader study
of massive scalar propagation in PFDM
backgrounds~\cite{Becar2024PFDM}.\\
The logarithmic term changes both the scattering potential and
the wave solutions at large distances. We must therefore retain
it in the outer boundary conditions, even for small coupling.
There is also a distinction between $M$ and the usual mass
measured at infinity. Although $f\to1$, the spherical mass
function grows logarithmically for $A>0$. We therefore treat
$M$ as a parameter at fixed reference length $r_0$, rather
than a finite standard Arnowitt-Deser-Misner (ADM) mass  \footnote{  We also construct several exact
electric solutions that reproduce the inner logarithmic geometry and approach a
finite ADM mass and a finite electrostatic potential  at
large radius (Sec.~\ref{subsec:finite_mass_completion}).}.\\
Our choice is part of a broader family of NED models. These
include Born-Infeld and Euler-Heisenberg theories~\cite{Born1934gh,Heisenberg1936nmg},
the framework of Pleba\'nski~\cite{Plebanski:1970zz}, charged solutions~\cite{Cataldo1999wr,Cai2004eh,Panahiyan2018gzv,Kruglov2017mpj},
and regular-black-hole constructions~\cite{Ayon-Beato:2000mjt,
Burinskii:2002pz,Fan:2016hvf,Bronnikov:2017sgg}.
Examples include logarithmic~\cite{Soleng1995kn,Hendi2013dwa,
Hendi2014mna,Kruglov2019ybs,Hendi:2012zz}, generalized and
double-logarithmic~\cite{Kruglov2014iqa,Gullu2020ant}, arcsin~\cite{Kruglov2014iwa,Kruglov2016ezw,Kruglov2019okd}, and other
electromagnetic models~\cite{Kruglov2016ymq,Kruglov2017xmb,Kar:2024zbo}, as well as extensions to
Gauss-Bonnet gravity~\cite{kruglov20224d,
Kruglov2021btd,Kruglov2021qzd}. We will use the terminology ``logarithmic-NED geometry'' to refer specifically to the logarithmic metric correction studied here. This is distinct from adding a logarithmic term to the Lagrangian. In addition, the neutral, minimally coupled scalar in our study obeys wave equation involving the
spacetime metric, whereas electromagnetic waves in NED can be
governed by an effective optical metric ~\cite{KumarWalia:2024yxn}.

Previous studies of this background have considered light
deflection~\cite{Sucu:2024xck}, circular orbits and accretion~\cite{Rehman:2024jqg},
and massless scalar QNMs and greybody factors~\cite{Liang2026}.
Our study is also motivated by the links between ringdown and
transmission~\cite{Oshita:2023cjz,Konoplya:2024lir}, their sensitivity to potential
deformations~\cite{Rosato:2024arw,Oshita:2024fzf}, and related correspondences in
wormholes~\cite{Bolokhov:2024otn}. Adding a scalar mass lets us study binding
and horizon leakage alongside the standard radiative modes.

Massive-field resonances have been studied in Schwarzschild~\cite{OhashiSakagami2004,KonoplyaZhidenko2005},
higher-dimensional~\cite{Zhidenko:2006rs}, Kerr~\cite{Konoplya:2006br}, brane-world~\cite{Zinhailo:2024jzt},
and higher-dimensional rotating~\cite{Lu:2023par} backgrounds.
Studies of charged RN and Kerr-Newman systems~\cite{Konoplya2002RN,Sampaio2014Clouds,
KonoplyaZhidenko2013KN},
de Sitter decay~\cite{Correa:2024xki,
Konoplya:2024ptj}, and near-extremal RN QBSs~\cite{Hod2017QBS}
also highlight the role of field couplings and boundary
conditions. For a charged massive scalar in the Ay\'on-Beato-Garc\'ia geometry,
the absorption cross section can diverge to negative values as the wave momentum
approaches zero~\cite{dePaula2024Absorption}, and quasibound states can grow
through superradiance~\cite{Dolan2024ABG}.
In our problem, both mode families are ingoing
at the horizon. At infinity, QNMs are outgoing, with
$k=\sqrt{\omega^2-\mu^2}$, while QBSs decay spatially, with
$\kappa=\sqrt{\mu^2-\omega^2}$ and
$\operatorname{Re}\kappa>0$. We include the logarithmic
corrections in both outer solutions.

We use analytical approximations to understand the spectra
before checking them with numerical calculations. The leading
Wentzel-Kramers-Brillouin (WKB) approach~\cite{SchutzWill1985} estimates
QNM frequencies from the height and curvature of the potential
barrier. Higher-order corrections~\cite{IyerWill1987,Iyer:1986nq,
Konoplya:2003ii,Konoplya:2004ip,Konoplya2019} and Pad\'e
resummation~\cite{Matyjasek:2017psv,Matyjasek:2019eeu,
Matyjasek:2026yiu,Konoplya:2026rjh} extend this approach. Other tools include
Borel and exact-WKB methods~\cite{Hatsuda:2019eoj,Miyachi:2025ptm} and large-multipole
expansions~\cite{CardosoEtAl2009,DolanOttewill2009}. These techniques have been applied to
regular geometries~\cite{Wahlang:2017zvk,Bolokhov:2023ruj,Rahmatov:2026qow,
DuttaRoy:2022ytr,Malik:2024qsz,Bolokhov:2025fto}, modified gravity~\cite{Rincon:2018sgd,Zinhailo:2018ska,Konoplya:2019hml,
Zinhailo:2019rwd,Matyjasek:2020bzc,Gogoi:2023kjt,Paul2024EsGB}, and
quantum-corrected backgrounds~\cite{Momennia:2022tug,
Konoplya:2024lch,Malik:2024nhy,Dubinsky:2024aeu,Dubinsky:2024mwd}.
Time-domain calculations provide a different approach by
constructing the waveform directly. They have been used to study
stability~\cite{Konoplya:2007jv,
Konoplya:2013sba,Cuyubamba:2016cug,Skvortsova:2023zca}, trapping and echoes~\cite{Aneesh:2018hlp,DuttaRoy:2019hij,
Roy:2021jjg,Bronnikov:2021liv,Konoplya:2025uiq}, overtone
extraction~\cite{Dubinsky:2024gwo}, and massive-field tails~\cite{Qian:2022kaq}.\\
We first perform a small mass expansion and note how it affects the barrier potential.
For nonnegative coupling and a simple outer horizon,
the massless potential has one exterior maximum with
nonzero curvature. At order $\mu^2$, the peak moves outward,
rises, and becomes less sharply curved in the tortoise coordinate. The full first-order WKB formula then
predicts weaker damping at each fixed finite multipole and
overtone near $\mu=0$. This is a local WKB result. The actual
spectrum must also be checked with global numerical methods.

We therefore compare higher-order WKB with global matching,
a generalized Leaver construction~\cite{Leaver1985,Benda2025} adapted to the
logarithmic geometry, and time-domain evolution~\cite{GundlachPricePullin1994}
followed by damped-exponential frequency extraction (Prony approach)~\cite{Berti:2007dg};
related matrix-pencil methods are discussed in Ref.~\cite{MLHuaSarkar}.
For the sampled fundamental QNM, increasing the scalar mass
raises the oscillation frequency and lengthens the decay time.
The global methods agree closely, and the accepted time-domain
estimates support this behavior.

For QBSs, we independently calculate the complex frequencies
using global matching and the decaying-mode Leaver construction.
We compare them with weak-field binding, Bohr-Sommerfeld
 formula, and local WKB-Pad\'e estimates~\cite{Konoplya2026}.
We treat leakage through the barrier separately using Gamow
tunnelling, as in semiclassical trapped-mode analyses~\cite{Patrick:2021oqk}. Across the sampled fundamental
QBSs, increasing the scalar mass or NED coupling strengthens
binding but also increases horizon absorption, shortening
the lifetime.

Finally, we ask whether the radiative QNM and trapped  QBS
are orthogonal under a suitable inner product. We use
the relativistic bilinear framework~\cite{Green2023Bilinear,Cannizzaro2024,Cannizzaro2025}, regularize the
horizon contribution, and integrate along a common complex
radial contour. We find numerical support for orthogonality
of an independently computed QNM-QBS pair under this
regularization. \\
We also construct several exact
electric solutions that reproduce the inner logarithmic geometry and approach a
finite ADM mass and a finite electrostatic potential  at
large radius (Sec.~\ref{subsec:finite_mass_completion}). In this work, we  restrict to the original geometry and reserve a comparison across
the several completions for future work.

Throughout, the scalar is treated as a test field. We use
$G=c=\hbar=1$ and time dependence $e^{-i\omega t}$.
Section~\ref{sec:log_ned_geometry} presents the background geometry,
Secs.~\ref{sec:massive-qnm} and \ref{sec:qbs} treat QNMs and QBSs,
Sec.~\ref{sec:ortho} examines orthogonality, Sec.~\ref{subsec:finite_mass_completion} provides the completions with finite ADM mass and
Sec.~\ref{sec:conclusions} gives our conclusions. The appendices
provide additional derivations, boundary expansions, and
detailed parameter scans and frequency tables.

\section{Background geometry: an exact logarithmic deformation of Reissner-Nordstr\"om}
\label{sec:log_ned_geometry}

We now introduce the background geometry chosen for our study.
We choose the solution of Ref.~\cite{Mazharimousavi2023} because
it combines an exact solution of Einstein gravity with an
adjustable logarithmic correction and a familiar
Reissner-Nordstr\"om (RN) limit. We can therefore compare both
radiative and trapped modes with their RN counterparts. Since our
test scalar is neutral, it responds to the geometry without
coupling directly to the electric field. This helps us isolate the geometric effects.

\subsection{Field equations and exact solution}

We consider Einstein gravity minimally coupled to nonlinear
electrodynamics (NED), with action
\begin{equation}
S=\int d^4x\,\sqrt{-g}
\left[\frac{\mathcal R}{16\pi}
+\frac{\mathcal L(\mathcal F)}{4\pi}\right],
\label{eq:NED_action}
\end{equation}
where we use $G=c=1$ and metric signature $(-+++)$, with
$\mathcal F=F_{\mu\nu}F^{\mu\nu}/4$ and
$F_{\mu\nu}=\partial_\mu\mathcal A_\nu-\partial_\nu\mathcal A_\mu$.
The Einstein and electromagnetic field equations are
\begin{align}
G_\mu{}^\nu&=8\pi T_\mu{}^\nu,
\label{eq:NED_Einstein_equation}\\
T_\mu{}^\nu&=\frac{1}{4\pi}
\left(\mathcal L\delta_\mu^\nu
-\mathcal L_{\mathcal F}F_{\mu\lambda}F^{\nu\lambda}\right),
\label{eq:NED_stress_general}\\
\nabla_\mu\left(\mathcal L_{\mathcal F}F^{\mu\nu}\right)&=0,
\qquad \nabla_{[\mu}F_{\nu\rho]}=0,
\label{eq:NED_Maxwell_general}
\end{align}
where $\mathcal L_{\mathcal F}\equiv
\partial\mathcal L/\partial\mathcal F$. The NED model is
specified by
\begin{equation}
\begin{aligned}
\mathcal L(\mathcal F)&=\frac{\chi^3}{6}(3\chi+2\zeta),\\
\chi&=\frac{\sqrt{\zeta^2+4\sqrt{-2\mathcal F}}-\zeta}{2}.
\end{aligned}
\label{eq:NED_exact_Lagrangian}
\end{equation}
The parameter $\zeta$ controls the nonlinear correction.
We take $\zeta\geq0$, consistent with the energy conditions.
When $\zeta\to0$, we recover $\mathcal L\to-\mathcal F$ and
$\mathcal L_{\mathcal F}\to-1$, so the theory reduces to
Einstein-Maxwell gravity.

The exact static, spherically symmetric solution has the form
\begin{equation}
ds^2=-f(r)dt^2+\frac{dr^2}{f(r)}+r^2d\Omega_2^2,
\label{eq:NED_line_element}
\end{equation}
where $d\Omega_2^2=d\theta^2+\sin^2\theta\,d\phi^2$ and
\begin{align}
f(r)&=1-\frac{2M}{r}+\frac{q^2}{r^2}
-\frac{A}{r}\ln\!\left(\frac{r}{r_0}\right),
\label{eq:NED_metric_A}\\
A&\equiv\frac{4\zeta q^{3/2}}{3}.
\label{eq:NED_A_def}
\end{align}
Here $q>0$ is the electric charge parameter, $M$ is a mass
integration parameter, and $r_0>0$ is a reference length.
The quantity $A$ sets the strength of the logarithmic term;
setting $\zeta=0$  recovers RN.

The corresponding gauge potential is $\mathcal A=U(r)dt$,
with $F=E(r)dt\wedge dr$ and
\begin{align}
U(r)&=\frac{q}{r}-\zeta\sqrt q\,
\ln\!\left(\frac{r}{r_0}\right),
\label{eq:NED_electric_potential}\\
E(r)&=-U'(r)=\frac{q}{r^2}+\frac{\zeta\sqrt q}{r}.
\label{eq:NED_electric_field}
\end{align}
Thus the nonlinear correction adds a slower $1/r$ contribution
to the usual Coulomb field. For this solution,
$\chi=\sqrt q/r$ and
$\mathcal L_{\mathcal F}=-\sqrt q/(\sqrt q+\zeta r)$.
These expressions satisfy the integrated electromagnetic
equation $r^2\mathcal L_{\mathcal F}E=-q$.

\subsection{Matter source and large-radius behavior}

We next describe the matter supporting this geometry. Its
energy density $\rho$, radial pressure $p_r$, and tangential
pressure $p_\perp$ enter the stress tensor as
$T_\mu{}^\nu=\operatorname{diag}(-\rho,p_r,p_\perp,p_\perp)$,
with
\begin{align}
\rho=-p_r&=\frac{q^2}{8\pi r^4}
+\frac{\zeta q^{3/2}}{6\pi r^3},
\label{eq:NED_density}\\
p_\perp&=\frac{q^2}{8\pi r^4}
+\frac{\zeta q^{3/2}}{12\pi r^3}.
\label{eq:NED_tangential_pressure}
\end{align}
The radial and tangential pressures differ by
\begin{equation}
p_\perp-p_r=\frac{q^2+\zeta q^{3/2}r}{4\pi r^4}.
\label{eq:NED_pressure_anisotropy}
\end{equation}

For our spectral analysis, the key feature is the slow decay of
the metric correction at large radius. Although $f(r)\to1$,
the term $\ln(r/r_0)/r$ falls off more slowly than the usual
$1/r$ mass term. Writing
$f=1-2m(r)/r$  defines the spherical mass function
\begin{equation}
m(r)=M-\frac{q^2}{2r}
+\frac{A}{2}\ln\!\left(\frac{r}{r_0}\right).
\label{eq:NED_mass_function}
\end{equation}
For $A>0$, $m(r)$ grows logarithmically. The metric therefore
approaches Minkowski spacetime, but it does not have a finite
standard Arnowitt-Deser-Misner (ADM) mass. Throughout this
work, we interpret $M$ as an integration parameter at fixed
$r_0$. The slow falloff also requires us to retain the logarithmic
terms in the outer wave boundary conditions used below. The reference length fixes the convention for $M$. Changing
$r_0\to\widetilde r_0$ leaves the metric unchanged if we also
shift $M\to M+(A/2)\ln(\widetilde r_0/r_0)$.
The same change adds only a gauge constant to $U$.\\\\
Note that the same geometry also arises in the Maxwell limit of the
Euler-Heisenberg solution with phenomenological perfect-fluid
dark matter (PFDM)~\cite{Ma2024PFDM,Becar2026PFDM}.
Identifying its electric charge with $q$, we recover our metric
by setting
\[
\alpha=-A,\qquad
M_{\mathrm P}=M+\frac{A}{2}\ln\!\left(\frac{A}{r_0}\right),
\quad A>0,
\]
where the mass shift accounts for the different reference
lengths in the logarithm \footnote{With $A=-\alpha$ (or $A=-k$),
Ref.~\cite{Ma2024PFDM} uses $A>0$ in its optical examples
and both signs in AdS thermodynamics.
Ref.~\cite{Becar2026PFDM} focuses on $A<0$ for QNMs and
scattering, but includes both signs in neutrino energy
deposition. Ref.~\cite{Becar2024PFDM} studies massive scalar
perturbations of an uncharged $A<0$ background.}. This solution has positive PFDM energy
density, $\rho_{\mathrm{PFDM}}=A/(8\pi r^3)$, and the PFDM
source satisfies the standard energy conditions despite its
anisotropic pressures. Since our neutral, minimally coupled
scalar responds only to the metric, our QNM and QBS results
also apply to this geometry with the same boundary conditions.

\subsection{Curvature, horizons, and the extremal limit}

We now examine the central singularity and the horizons.
The Ricci and Kretschmann scalars are
\begin{align}
\mathcal R&=\frac{A}{r^3},
\label{eq:NED_Ricci_scalar}\\
K&=[f''(r)]^2+\frac{4[f'(r)]^2}{r^2}
+\frac{4[1-f(r)]^2}{r^4},
\label{eq:kretschmann}
\end{align}
where $K=R_{\alpha\beta\gamma\delta}R^{\alpha\beta\gamma\delta}$.
For $q>0$, $K\sim56q^4/r^8$ as $r\to0$, so the center
remains a curvature singularity.

We locate the Killing horizons by finding the positive roots
of $f(r_h)=0$:
\begin{equation}
r_h^2-2Mr_h+q^2-Ar_h\ln\!\left(\frac{r_h}{r_0}\right)=0.
\label{eq:NED_horizon_equation}
\end{equation}
At extremality, the inner and outer horizons merge.
The conditions $f(r_e)=f'(r_e)=0$ give
$r_e^2-Ar_e-q^2=0$, and hence
\begin{align}
r_e&=\frac{A+\sqrt{A^2+4q^2}}{2},
\label{eq:NED_extremal_radius}\\
M_e&=r_e-\frac{A}{2}
\left[\ln\!\left(\frac{r_e}{r_0}\right)+1\right].
\label{eq:NED_extremal_mass}
\end{align}
For fixed $(q,\zeta,r_0)$, $M_e$ separates three cases:
two distinct horizons when $M>M_e$, one degenerate horizon
when $M=M_e$, and no horizon when $M<M_e$.

To analyze the near horizon geometry, we set
$M=M_e$ and expand around $r_e$:
\begin{equation}
f(r)=\mathcal C_e\left(\frac{r-r_e}{r_e}\right)^2
+O\!\left[\left(\frac{r-r_e}{r_e}\right)^3\right],
\label{eq:NED_extremal_expansion}
\end{equation}
with
\begin{equation}
\mathcal C_e\equiv\frac{r_e^2f''(r_e)}{2}
=1-\frac{A}{2r_e}
=\frac{\sqrt{A^2+4q^2}}{2r_e}>0.
\label{eq:NED_Ce}
\end{equation}
Performing the coordinate changes
$r=r_e(1+\epsilon x)$ and
$t=r_e\hat t/(\mathcal C_e\epsilon)$, followed by
$\epsilon\to0$, the metric becomes
\begin{equation}
ds^2=L_2^2\left(-x^2d\hat t^{\,2}+\frac{dx^2}{x^2}\right)
+r_e^2d\Omega_2^2,
\label{eq:NED_extremal_throat}
\end{equation}
where
\begin{equation}
L_2^2=\frac{r_e^2}{\mathcal C_e}
=\frac{2r_e^3}{\sqrt{A^2+4q^2}}.
\label{eq:NED_AdS2_radius}
\end{equation}
In Eq.~\eqref{eq:NED_extremal_throat}, the first term describes
two-dimensional anti-de Sitter space, and the second describes the usual sphere. Thus the extremal
near-horizon geometry is locally
$\mathrm{AdS}_2(L_2)\times S^2(r_e)$.
The logarithmic deformation preserves this product structure
but changes the relative curvature radii. For $A>0$,
$L_2>r_e$. In the RN limit $\zeta\to0$, we recover
$M_e=r_e=L_2=q$ and the familiar Bertotti-Robinson geometry.

\section{Quasinormal modes of a massive scalar field}
\label{sec:massive-qnm}

We now study massive scalar QNMs
in the logarithmic-NED geometry. We use five approaches:
leading-order WKB, sixth-order WKB, global matching, a generalized
Leaver method, and time-domain evolution with Prony extraction.

We begin with WKB, which relates the frequency to the shape of
the potential near its maximum. The leading-order result helps
us understand the effect of scalar mass, while sixth-order WKB
includes more information about the barrier. We then solve the
boundary-value problem with global matching and Leaver's method,
retaining the logarithmic corrections at large radius in both.
Their agreement provides a check beyond the local WKB
approximation. Finally, we evolve the field in time and use
Prony fits to extract frequencies from the ringdown. We describe
the numerical checks alongside each method.

\subsection{Leading-order WKB}

We begin with the massive Klein-Gordon equation, written here
for the scalar field before separation:
\begin{equation}
    \frac{1}{\sqrt{-g}} \partial_{\mu} \left( \sqrt{-g} g^{\mu\nu} \partial_{\nu} \Phi \right) - \mu^2 \Phi = 0
\end{equation}
Here $\mu$ is the field mass and $g$ is the metric determinant.
We decompose
$\Phi=e^{-i\omega t}\Psi(r)Y_\ell(\theta,\phi)/r$,
where $\Psi(r)$ is the radial function and $\ell$ is the
angular multipole. The radial equation becomes
\begin{equation}\label{eq:schrodinger}
    \frac{d^2 \Psi}{dr_{*}^{2}} + \left( \omega^2 -  V_{\ell\mu}(r) \right) \Psi = 0,
\end{equation}
where $r_*$ is the tortoise coordinate, defined by
$dr_*/dr=1/f$. The effective potential is
\begin{equation}
    V_{\ell\mu}(r) = f(r) \left( \frac{\ell(\ell + 1)}{r^2} + \frac{f'(r)}{r} + \mu^2 \right),
\end{equation}

\begin{figure*}[t]
    \centering
    \includegraphics[width=0.97\textwidth]{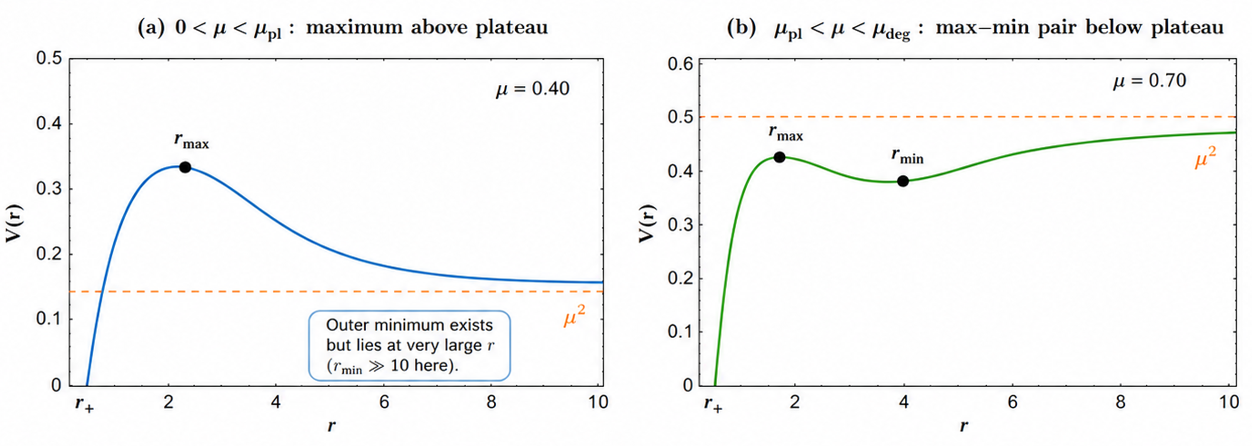}
    \caption{Schematic plot of massive-scalar potential for $M=r_0=1$, $q=0.5$,
$\zeta=0.05$, and $\ell=2$. Orange dashed lines mark $\mu^2$;
black dots mark extrema. (a) $\mu=0.40<\mu_{\rm pl}$: the
barrier is above the plateau, and the outer minimum is beyond
the plotted range. (b) $\mu_{\rm pl}<\mu=0.70<\mu_{\rm deg}$:
both extrema lie below the plateau. The two threshold masses
are defined in the text.}
    \label{Veff}
\end{figure*}

Figure~\ref{Veff} shows how this potential can support both
QNMs and quasibound states (QBSs). It vanishes at the horizon
and approaches $\mu^2$ at infinity. An outer well can also
appear beyond the inner barrier. QNMs describe scattering with
ingoing waves at the horizon and outgoing waves at infinity.
QBSs are trapped near the outer minimum and decay at infinity.
For weakly damped trapped states,
$V_{\min}<\omega_R^2<\min(V_{\max},\mu^2)$, where
$\omega_R=\operatorname{Re}\omega$. Their finite lifetime
comes from leakage through the inner barrier into the horizon.
We focus on QNMs here and study QBSs in Sec.~\ref{sec:qbs}.

Two scalar masses help describe the change in potential shape.
At $\mu=\mu_{\rm pl}$, the barrier reaches the outer plateau,
$V_{\max}=\mu_{\rm pl}^2$. At $\mu=\mu_{\rm deg}$, the maximum
and minimum merge, so
$\partial_rV_{\ell\mu}=\partial_r^2V_{\ell\mu}=0$
at the merging point.

\begin{figure*}[t]
    \centering
    \includegraphics[width=0.485\textwidth]{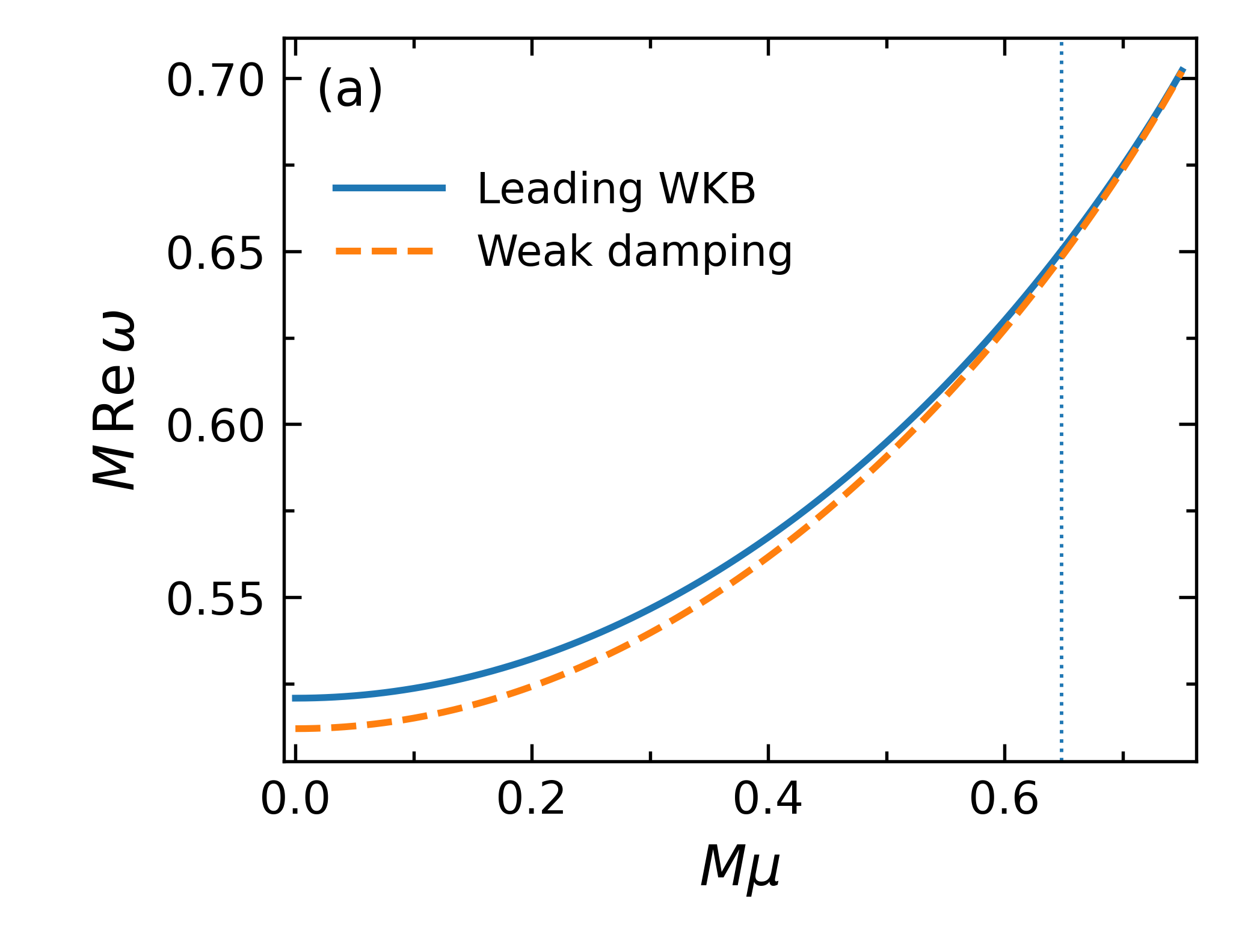}\hfill
    \includegraphics[width=0.485\textwidth]{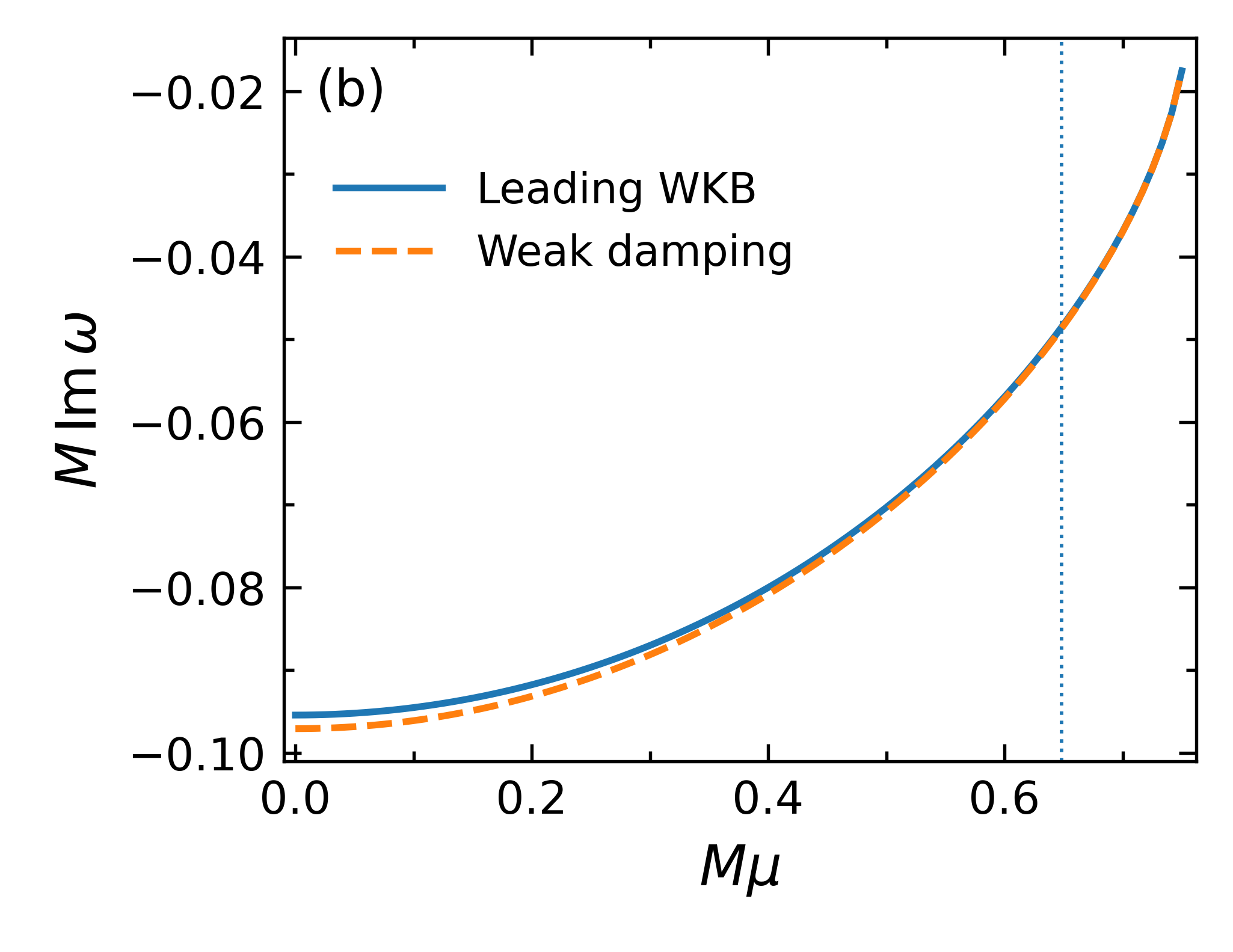}
    \caption{Leading-order WKB frequencies for $M=r_0=1$, $q=0.5$,
$\zeta=0.05$, $\ell=2$, and $n=0$: (a) real part and
(b) imaginary part. Solid curves use the full WKB square root;
dashed curves use the weak-damping approximation
\eqref{eq:uqnm_weak_damping}. The dotted line marks
$V_p=\mu^2$.}
    \label{fig:first_order_wkb}
\end{figure*}

At leading WKB order, the frequency is determined by
\begin{equation}
\omega^2 = V_p - i \alpha_n \sqrt{-2K_p},~~~~\alpha_n \equiv  \left(n+\frac{1}{2}\right)
\label{eq:full_wkb_master}
\end{equation}
Here $n=0,1,2,\ldots$ is the overtone number, $r_p$ is the
barrier maximum, $V_p=V(r_p)$ is its height, and
$K_p=(d^2V/dr_*^2)_{r_p}<0$ is its curvature in the tortoise
coordinate. Written in terms of the metric, this gives
\begin{align}
\omega_{\ell n}^2
={}&
f_p
\left[
\mu^2
+
\frac{\ell(\ell+1)}{r_p^2}
+
\frac{f'_p}{r_p}
\right]
-
i\left(n+\frac{1}{2}\right)
\sqrt{
-2f_p^2
\left[
f''_pG_p
+
2f'_pG'_p
+
f_pG''_p
\right]
},
\label{eq:full_wkb_master_explicit}
\end{align}
In this explicit expression, we set the reference-length
 $r_0=1$, so
\begin{equation}
f(r)
=
1
-
\frac{2M}{r}
+
\frac{q^2}{r^2}
-
\frac{4q^{3/2}\zeta}{3r}\ln r.
\label{eq:full_wkb_metric_function}
\end{equation}
and 
\begin{equation}
G(r)
\equiv
\mu^2+\frac{\Lambda}{r^2}+\frac{f'(r)}{r},
\label{eq:full_wkb_G}
\end{equation}
where we define $\Lambda=\ell(\ell+1)$. A subscript $p$ means evaluation at $r_p$, and primes denote
derivatives with respect to $r$. We find the peak by solving
$V'(r_p)=0$, or explicitly,
\begin{align}
f'_p
\left[
\mu^2
+
\frac{\Lambda}{r_p^2}
+
\frac{f'_p}{r_p}
\right] +
f_p
\left[
-\frac{2\ell(\ell+1)}{r_p^3}
+
\frac{f''_p}{r_p}
-
\frac{f'_p}{r_p^2}
\right]=0
\label{eq:full_wkb_peak_explicit}
\end{align}
For each scalar mass, we solve this equation numerically and
evaluate the WKB frequency at the resulting barrier maximum.
To separate the oscillation frequency from the damping rate,
we write
\begin{equation}
 Y_p=\alpha_n\sqrt{-2K_p}>0,
 \qquad \mathcal R_p=\sqrt{V_p^2+Y_p^2},
\end{equation}
and take the square root of Eq.~\eqref{eq:full_wkb_master}  with positive real part and negative imaginary part:
\begin{equation}
 \omega_R^{\rm WKB}=\sqrt{\frac{\mathcal R_p+V_p}{2}},
 \qquad
 \omega_I^{\rm WKB}=-\sqrt{\frac{\mathcal R_p-V_p}{2}}.
 \label{eq:uqnm_WKB_parts}
\end{equation}
If $Y_p/V_p\ll1$, the damping is weak and these expressions
simplify to
\begin{equation}
 \omega_R^{\rm wd}\simeq\sqrt{V_p},\qquad
 \omega_I^{\rm wd}\simeq
 -\alpha_n\sqrt{-\frac{K_p}{2V_p}}.
 \label{eq:uqnm_weak_damping}
\end{equation}
Equivalently, $\omega_R\simeq\sqrt{V_p}$ and
$\omega_I\simeq-Y_p/(2\sqrt{V_p})$. The peak height then sets the main oscillation frequency,
while damping depends on both the height and curvature.
Figure~\ref{fig:first_order_wkb} compares this additional
approximation with the full first-order WKB square root.

\subsection{Small-mass expansion}
\label{subsec:uqnm_mass_response}

We now explain the mass trend in Fig.~\ref{fig:first_order_wkb}
by a small mass expansion. We use $s=\mu^2$ and
keep the background, $\ell$, and $n$ fixed. We assume $M>0$,
$r_0>0$, $q\geq0$, $A=(4/3)q^{3/2}\zeta\geq0$, and a
nonextremal outer horizon. We separate the mass contribution
to the potential as
\begin{equation}
 V(r,\mu^2)=\mathcal V(r)+s f(r),\qquad
 \mathcal V=f\left[\frac{\ell(\ell+1)}{r^2}+\frac{f'}r\right],
 \label{eq:uqnm_mass_decomposition}
\end{equation}
Let $r_b$ be the exterior maximum in the massless case. Appendix~\ref{app:uqnm_damping_sign}
shows that this maximum is unique and has nonzero curvature
under our assumptions. A subscript $b$ denotes evaluation at $r_b$. The mass deformed peak, its height, and its curvature are
\begin{align}
 r_p&=r_b-s\frac{f'_b}{\mathcal V''_b}+O(s^2),\nonumber\\
 V_p&=v_b+s f_b+O(s^2),\qquad
 K_p=k_b+\mu^2 k_s+O(s^2),
 \label{eq:uqnm_mass_peak_summary}
\end{align}
Here $v_b=\mathcal V_b>0$, $k_b=f_b^2\mathcal V''_b<0$,
$K_p=(d^2V/dr_*^2)_{r_p}$, and $k_s=(dK_p/ds)_0$.
The appendix proves $f'_b>0$, $0<f_b<1$, and $k_s>0$.
These signs give a simple picture: a small scalar mass moves
the peak outward, raises it, and makes it flatter in the
tortoise coordinate. Lagos et al.~\cite{Lagos2020Decay} related the shape of
the barrier to the way the damping rates of different multipoles change
order as the scalar mass increases.

We next translate these changes into a frequency shift. We use
the full first-order WKB square root,
\begin{equation}
 \omega^{\rm WKB}(\mu^2)=\sqrt{V_p-i\alpha_n\sqrt{-2K_p}},
 \qquad \alpha_n=n+\tfrac12.
 \label{eq:uqnm_mass_wkb_definition}
\end{equation}
with $\omega^{\rm WKB}=\omega_R+i\omega_I$ and
$\omega_b^{\rm WKB}=\omega_{R,b}+i\omega_{I,b}$, choosing
$\omega_{R,b}>0$ and $\omega_{I,b}<0$. Differentiating the
squared frequency gives
\begin{align}
 \omega^{\rm WKB}(\mu^2)
 &=\omega_b^{\rm WKB}+\mu^2(C_{R,b}+iC_{I,b})+O(\mu^4),\nonumber\\
 C_{R,b}+iC_{I,b}
 &=\frac{f_b+i\alpha_n k_s/\sqrt{-2k_b}}{2\omega_b^{\rm WKB}}.
 \label{eq:uqnm_small_mass_complex}
\end{align}
Separating the real and imaginary parts gives
\begin{align}
 C_{R,b}&=\frac{f_b ~\omega_{R,b}}{2|\omega_b^{\rm WKB}|^2}
 \left(1-\frac{k_s\omega_{I,b}^2}{(-k_b)f_b}\right),
 \label{eq:mass_physical_real_factor}\\
 C_{I,b}&=\frac{-f_b\omega_{I,b}}{2|\omega_b^{\rm WKB}|^2}
 \left(1+\frac{k_s\omega_{R,b}^2}{(-k_b)f_b}\right)>0.
 \label{eq:massapp_response_parts}
\end{align}
Because $k_s>0$ and $\omega_{I,b}<0$, we have $C_{I,b}>0$:
the imaginary part becomes less negative, so damping weakens.
This holds at order $\mu^2$ for every fixed finite $(\ell,n)$
within the first-order WKB expression. However, the sign of the real
shift depends on the bracket in
Eq.~\eqref{eq:mass_physical_real_factor}.

\begin{figure*}[t]
 \centering
 \includegraphics[width=0.485\textwidth]{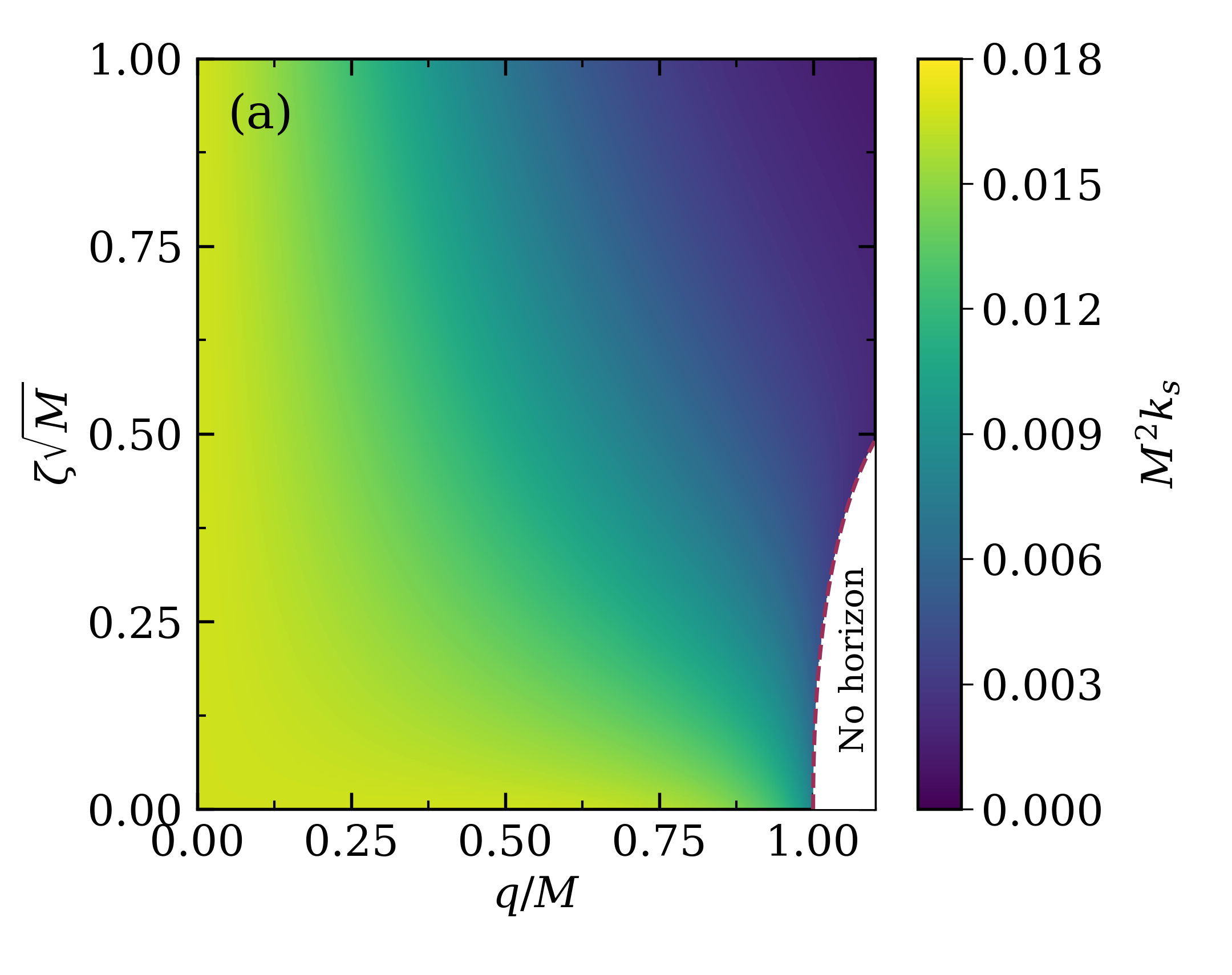}\hfill
 \includegraphics[width=0.485\textwidth]{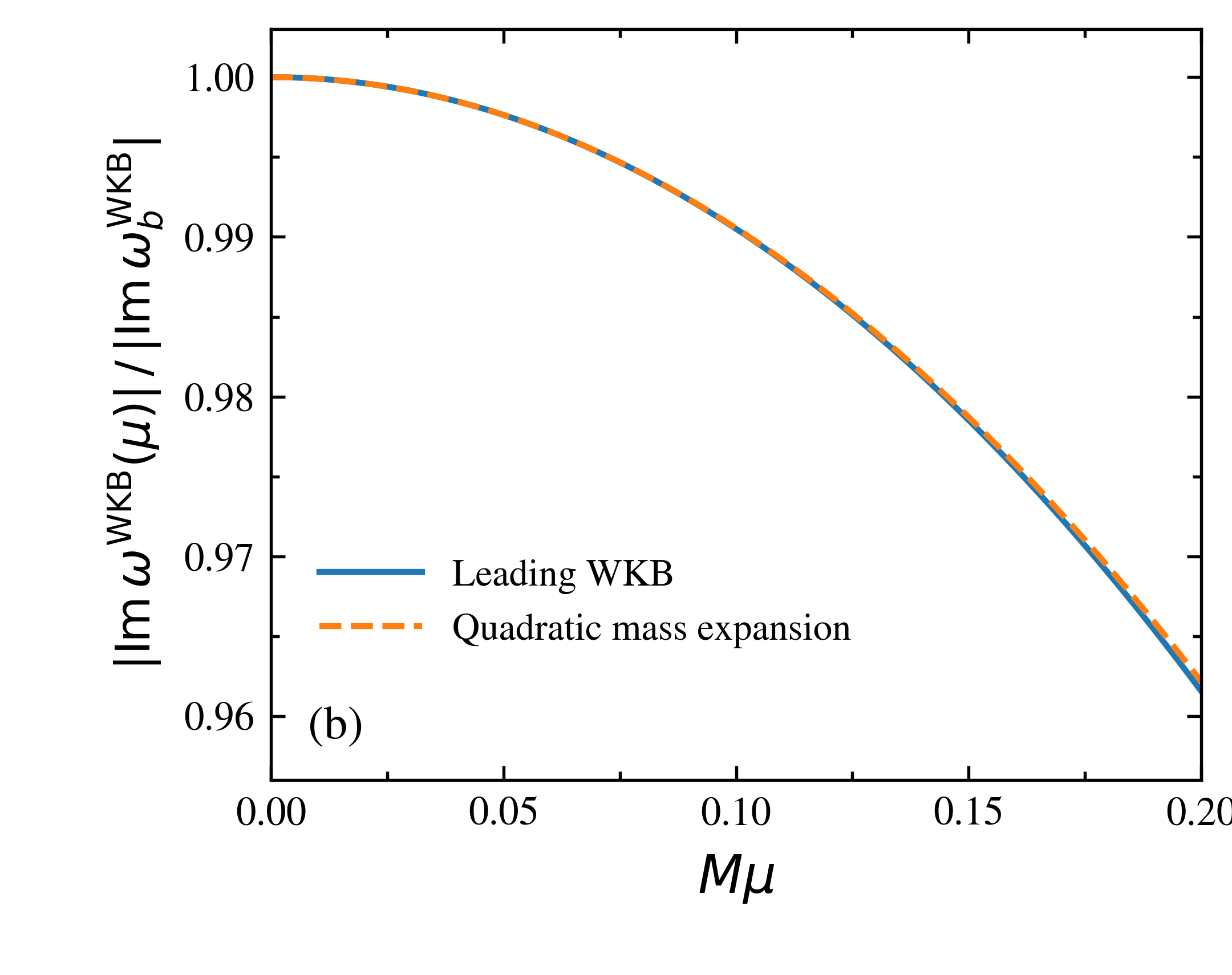}
 \caption{Small-mass decrease in WKB damping.
(a) $M^2k_s$ for $\ell=2$ and $r_0/M=1$. The colored region
has $k_s>0$; the dashed boundary is extremal, and the white
region has no horizon. (b) Damping relative to its massless
value, from first-order WKB (solid) and its quadratic mass
expansion (dashed), for $M=r_0=1$, $q=0.5$, $\zeta=0.05$,
$\ell=2$, and $n=0$.}
 \label{fig:uqnm_mass_sign}
\end{figure*}

For $M=r_0=1$, $q=0.5$, $\zeta=0.05$, $\ell=2$, and $n=0$,
we set $x=M\mu$ and obtain
\begin{align}
 M\omega^{\rm WKB}
 ={}&0.52075915-0.09544262\,i\nonumber\\
 &+(0.28203129+0.09028656\,i)x^2+O(x^4).
 \label{eq:uqnm_small_mass_numerical_wkb}
\end{align}
Thus, for this example, the oscillation frequency increases
and damping decreases. Figure~\ref{fig:uqnm_mass_sign}(b) plots the damping ratio
$|\operatorname{Im}\omega^{\rm WKB}(\mu)|/
|\operatorname{Im}\omega_b^{\rm WKB}|$ from the quadratic
expansion and the unexpanded WKB result. Panel (a) illustrates
$k_s>0$ in a parameter scan;
the proof for every finite $\ell\geq0$ is given in
Appendix~\ref{app:uqnm_damping_sign}.
The physical picture is even simpler when damping is weak,
$\eta_p=\alpha_n\sqrt{-2K_p}/V_p\ll1$. In this limit,
\begin{equation}
 \omega_R^{\rm wd}=\sqrt{V_p},\qquad
 \omega_I^{\rm wd}=-\alpha_n\sqrt{-\frac{K_p}{2V_p}}.
 \label{eq:uqnm_mass_weak_damping_definitions}
\end{equation}
and the small-mass expansion becomes
\begin{align}
 \omega_R^{\rm wd}
 &=\sqrt{v_b}+\frac{\mu^2f_b}{2\sqrt{v_b}}+O(\mu^4),
 \label{eq:uqnm_small_mass_real}\\
 \omega_I^{\rm wd}
 &=-\alpha_n\sqrt{-\frac{k_b}{2v_b}}
 \left[1+\frac{\mu^2}{2}B_s\right]+O(\mu^4),
 \label{eq:uqnm_small_mass_imag}\\
 B_s&=\frac{k_s}{k_b}-\frac{f_b}{v_b}<0.
 \label{eq:uqnm_small_mass_damping_coefficient}
\end{align}
Since $k_s>0$, $k_b<0$, and $f_b/v_b>0$, we have $B_s<0$.
Both frequency corrections are then positive: the real part
rises and the imaginary part becomes less negative. The
amplitude therefore decays more slowly. These are local
small-mass WKB results; the global calculations below test
the spectrum beyond this approximation.

\subsection{Higher-order WKB analysis}
\label{subsec:higher_order_wkb}
\begin{figure*}[t]
  \centering
  \includegraphics[width=0.485\textwidth]{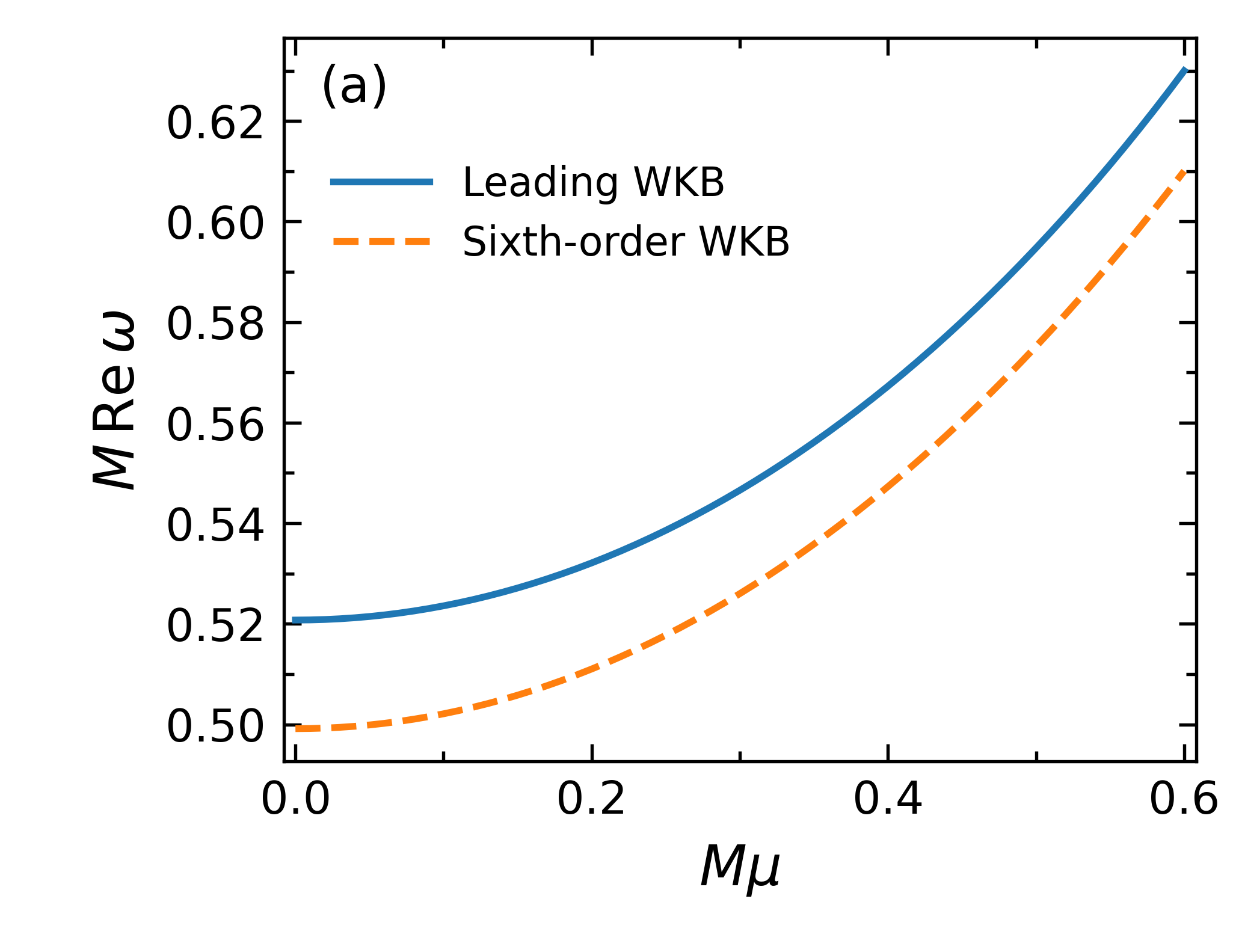}\hfill
  \includegraphics[width=0.485\textwidth]{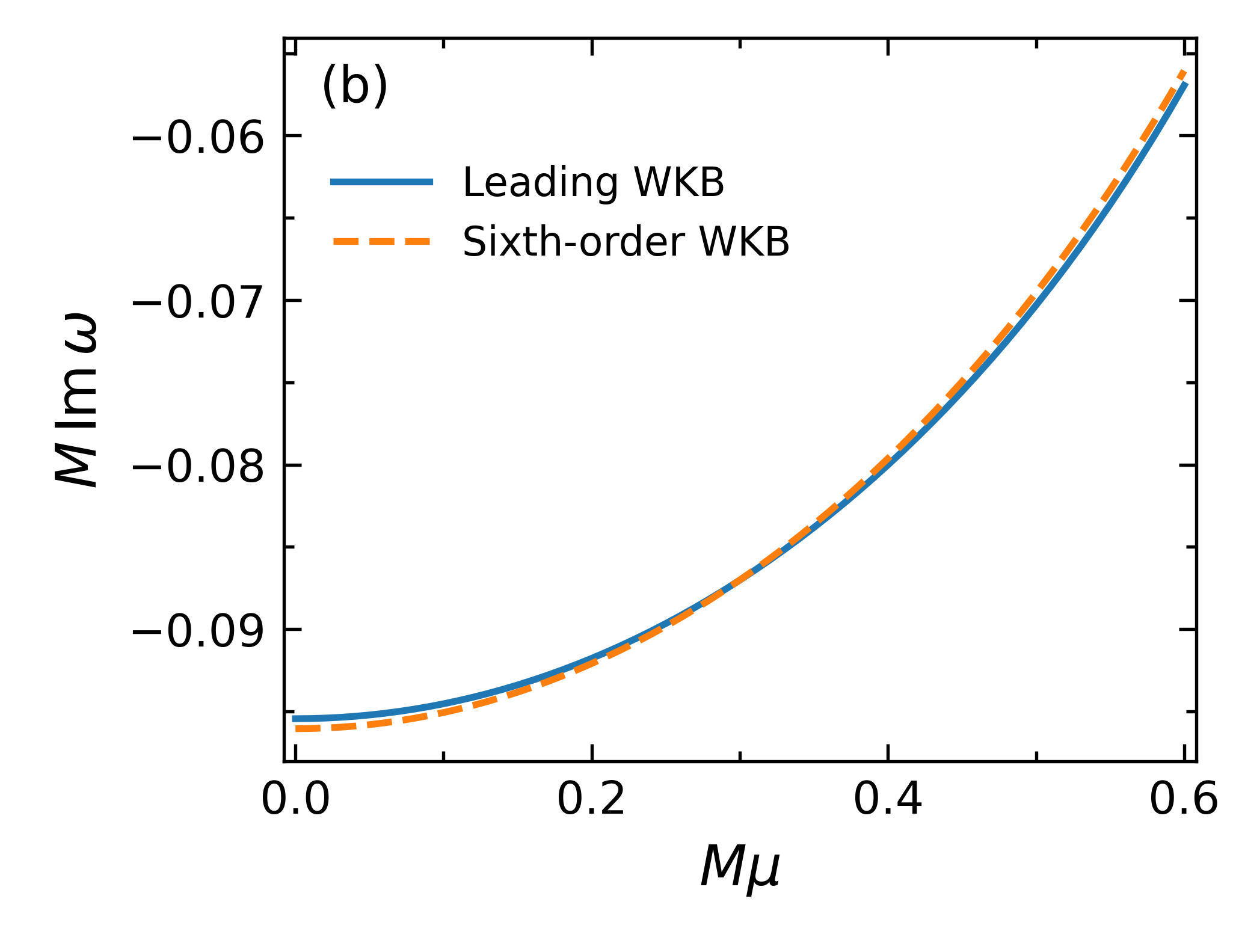}
  \caption{First-order (solid) and sixth-order (dashed) WKB
frequencies for $M=r_0=1$, $q=0.5$, $\zeta=0.05$, $\ell=2$,
and $n=0$: (a) real part and (b) imaginary part. Both use
the same potential maximum. The scan ends at $M\mu=0.60$,
below the barrier-plateau crossing $M\mu_{\rm pl}\simeq0.648503$.}
  \label{fig:wkb_first_sixth_log_ned}
\end{figure*}

We now include higher derivatives of the barrier to see how
much they change the leading-order estimate. We define the
derivatives with respect to the tortoise coordinate by
\begin{equation}
 V_j
 \equiv
 \left.
 \left(
 f\frac{d}{dr}
 \right)^j
 V(r)
 \right|_{r=r_p}
 =
 \left.
 \frac{d^jV}{dr_*^j}
 \right|_{r=r_p},
 \label{eq:wkb_tortoise_derivatives}
\end{equation}
at the same nondegenerate exterior maximum $r_p$. In particular,
\begin{equation}
 V_0=V_p,
 \qquad
 V_1=0,
 \qquad
 V_2=K_p<0.
\end{equation}
The WKB frequency at order $N$ can be written as
\begin{equation}
 \omega_N^2
 =
 V_0+\sum_{j=1}^{N}{\cal C}_j,
 \label{eq:wkb_hierarchy}
\end{equation}
where the first correction is
\begin{equation}
 {\cal C}_1
 =
 -i\alpha_n\sqrt{-2V_2},
 \qquad
 \alpha_n=n+\frac12,
\end{equation}
Thus $N=1$ gives Eq.~\eqref{eq:full_wkb_master}.
The terms ${\cal C}_2,\ldots,{\cal C}_6$ depend on higher
derivatives at the peak; sixth order requires derivatives
through $V_{12}$. At each mass, we locate the maximum of the
full potential and use the same peak for both calculations.
We restrict the comparison to
\begin{equation}
 0\leq M\mu\leq0.60,
 \label{eq:wkb6_mass_range}
\end{equation}
so that the barrier remains above the outer plateau. For the
chosen parameters, the crossing
\begin{equation}
 V_p=\mu^2
\end{equation}
occurs at
\begin{equation}
 M\mu_{\rm pl}\simeq0.648503,
\end{equation}
which lies beyond our scan. Throughout the scan, we check
\begin{equation}
 r_p>r_+,
 \qquad
 V_p>\mu^2,
 \qquad
 V_2<0,
 \qquad
 \operatorname{Re}(\omega_6^2)>\mu^2.
 \label{eq:wkb6_checks}
\end{equation}
These checks keep the comparison away from the barrier-plateau
crossing shown in Fig.~\ref{Veff}. Table~\ref{tab:wkb1_wkb6}
and Fig.~\ref{fig:wkb_first_sixth_log_ned} show the results.
Sixth-order WKB lowers the real frequency throughout this
interval, while the change in the imaginary part is smaller.
\begin{table}[t]
\centering
\caption{First- and sixth-order WKB values of $M\omega$ for
$M=r_0=1$, $q=0.5$, $\zeta=0.05$, $\ell=2$, and $n=0$.}
\label{tab:wkb1_wkb6}
\begin{ruledtabular}
\begin{tabular}{c cc}
$M\mu$
&
$M\omega_{1}$
&
$M\omega_{6}$
\\
\hline
$0.00$
&
$0.52075915-0.09544262\,i$
&
$0.49916022-0.09604850\,i$
\\
$0.20$
&
$0.53212222-0.09177429\,i$
&
$0.51102519-0.09209014\,i$
\\
$0.40$
&
$0.56725439-0.08002216\,i$
&
$0.54728640-0.07962929\,i$
\\
$0.50$
&
$0.59475237-0.07031654\,i$
&
$0.57517576-0.06955625\,i$
\\
$0.60$
&
$0.62997840-0.05695410\,i$
&
$0.61004787-0.05606591\,i$
\end{tabular}
\end{ruledtabular}
\end{table}
To measure these changes, we use the sixth-order values as
the reference and define
\begin{align}
 \delta_R(\mu)
 &=
 \frac{
 \left|
 \operatorname{Re}\omega_1-
 \operatorname{Re}\omega_6
 \right|
 }{
 \left|
 \operatorname{Re}\omega_6
 \right|
 },
 \\
 \delta_I(\mu)
 &=
 \frac{
 \left|
 \operatorname{Im}\omega_1-
 \operatorname{Im}\omega_6
 \right|
 }{
 \left|
 \operatorname{Im}\omega_6
 \right|
 }.
 \label{eq:wkb_relative_differences}
\end{align}
Their largest values over Eq.~\eqref{eq:wkb6_mass_range} are
\begin{equation}
 \max\delta_R\simeq4.33\%,
 \qquad
 \max\delta_I\simeq1.58\%.
 \label{eq:wkb1_wkb6_maxdiff}
\end{equation}
For this parameter set, the higher-order terms therefore
affect the oscillation frequency more than the damping rate.
We also compare fifth and sixth order and find
\begin{equation}
\max_{\,0\leq M\mu\leq0.60}
 \frac{|\omega_6-\omega_5|}
      {|\omega_6|}
 \simeq9.24\times10^{-5}.
 \label{eq:wkb5_wkb6_change}
\end{equation}
The small change shows that these last two WKB estimates are
close over the chosen range. It is a check of successive
approximations. We
next compare WKB with methods that impose both boundary
conditions directly.

\subsection{Global and time-domain determination of quasinormal modes}
\label{sec:logned-qnm}

We now calculate the same modes using global matching,
Leaver's method, and time-domain evolution with Prony
extraction. For these comparisons, we use the metric
$ds^2=-f\,dt^2+f^{-1}dr^2+r^2d\Omega^2$, with
\begin{equation}
 f(r)=1-\frac{2M+A\ln(r/r_0)}{r}+\frac{q^2}{r^2},
 \qquad A=\frac{4}{3}q^{3/2}\zeta .
 \label{eq:logned-lapse}
\end{equation}
Writing the neutral scalar as $\Phi=\Psi(t,r)Y_{\ell m}/r$,
we obtain
\begin{equation}
 \begin{split}
 &\left(\partial_t^2-\partial_{r_*}^2+V_\ell\right)\Psi=0,
 \qquad \frac{dr_*}{dr}=\frac{1}{f},\\
 &V_\ell=f\left[\mu^2+\frac{\ell(\ell+1)}{r^2}+\frac{f'}{r}\right].
 \end{split}
 \label{eq:logned-wave}
\end{equation}
Primes denote derivatives with respect to $r$. We use
$\Psi=e^{-i\omega t}\psi(r)$, so damped modes have
$\operatorname{Im}\omega<0$. We set $M=r_0=1$, $q=0.5$,
$\zeta=0.05$, and $\ell=2$, and follow the fundamental
mode $n=0$ over $0\leq M\mu\leq0.60$.

\subsubsection{Global matching method}
\label{subsec:logned-global}

The idea is to construct a solution from each boundary and
find the frequencies at which they join smoothly. Near the
outer horizon $r_+$, we use the ingoing expansion
\begin{equation}
 \psi_H=\left(\frac{\rho}{M}\right)^{-i\omega/f'(r_+)}
 \sum_{j=0}^{N_H}a_j\rho^j,
 \qquad \rho=r-r_+,
 \label{eq:logned-horizon}
\end{equation}
The radial equation determines the coefficients $a_j$, while
the overall normalization is arbitrary. We keep terms through
$N_H=8$ and use the series to start the outward integration.

At infinity, we select the outgoing value of
$k=+\sqrt{\omega^2-\mu^2}$ by continuing from $k>0$ for real
$\omega>\mu$. We include the long-range corrections through
the logarithmic derivative
\begin{equation}
 \begin{split}
 Y_\infty(r)&\equiv\frac{\psi_\infty'}{\psi_\infty}
 =ik+\sum_{j=1}^{N_\infty}\frac{p_j(L)}{r^j},\quad
 L=\ln(r/r_0),\\
 p_1(L)&=\frac{i(2\omega^2-\mu^2)}{2k}(2M+AL).
 \end{split}
 \label{eq:logned-far}
\end{equation}
We find the polynomials $p_j$ in turn by substituting into
$Y'+Y^2+(f'/f)Y+(\omega^2-V_\ell)/f^2=0$.
This keeps both the mass-dependent and logarithmic phase
corrections in the outer boundary condition. We use
$N_\infty=6$ and compare with $N_\infty=4,8$ to check
sensitivity to the truncation. Appendix~\ref{app_series}
gives the first few terms of both boundary expansions.

We integrate the horizon solution outward to a real matching
radius $r_m$. We integrate the outer solution inward along
$r=r_m+s e^{i\theta}$, choosing
$\operatorname{Im}(ke^{i\theta})>0$ so that the outgoing
wave decays along the ray toward infinity.
Appendix~\ref{app_contour} explains this contour choice.
The two solutions match when their logarithmic derivatives
agree:
\begin{equation}
 D(\omega)=
 \left.\frac{f\psi_H'}{\psi_H}\right|_{r_m}
 -\left.\frac{f\psi_\infty'}{\psi_\infty}\right|_{r_m}=0.
 \label{eq:logned-match}
\end{equation}
This is equivalent to a vanishing Wronskian when the matching
amplitudes are nonzero. We start with the massless mode and
follow the root in steps of $\Delta(M\mu)=0.01$. At
$M\mu=0,0.2,0.4,0.5,0.6$, we check sensitivity to the matching
radius, horizon cutoff, outer endpoint, ray angle, and both
series orders.

\subsubsection{Generalized Leaver method}
\label{subsec:logned-leaver}

We also calculate the frequencies with Leaver's continued-fraction
method~\cite{Leaver1985}. Here ``generalized'' means that we
adapt the boundary factors and coefficient recurrence to the
logarithmic geometry. The continued-fraction construction
itself is unchanged. Related details of the approach are
discussed in Ref.~\cite{Benda2025}.

For a simple outer horizon, we use $z=1-r_+/r$, so the horizon
is at $z=0$ and infinity is at $z=1$. With $a=A/r_+$ and
$b=q^2/r_+^2$, the metric becomes
\begin{equation}
 \begin{split}
 f(z)&=(1-b)z+bz^2+a(1-z)\ln(1-z)=zF(z),\\
 F_0&=1-b-a,\qquad F_1=b+\frac a2,\qquad
 F_j=\frac{a}{j(j+1)}\quad(j\geq2),
 \end{split}
 \label{eq:logned-leaver-lapse}
\end{equation}
where $F(z)=\sum_{j\geq0}F_jz^j$. We use the same outgoing
 $k$ as in global matching and factor out the
boundary behavior:
\begin{equation}
 \begin{split}
 \psi(z)&=z^\nu
 \exp\!\left[ikr+i\chi\left(2ML+\frac A2L^2\right)\right]
 \sum_{j=0}^{\infty}a_jz^j,\\
 \nu&=-\frac{i\omega}{f'(r_+)},\qquad
 \chi=\frac{2\omega^2-\mu^2}{2k},\qquad L=\ln(r/r_0).
 \end{split}
 \label{eq:logned-leaver-ansatz}
\end{equation}
The factor $z^\nu$ imposes the ingoing horizon condition.
The exponential contains the outgoing behavior, including
the $L^2$ phase obtained by integrating the leading
large-radius correction in Eq.~\eqref{eq:logned-far}  (see also Appendix~\ref{app_series}).
The remaining series supplies the radial dependence between
the boundaries.

\begin{figure*}[t]
 \centering
 \includegraphics[width=\textwidth]{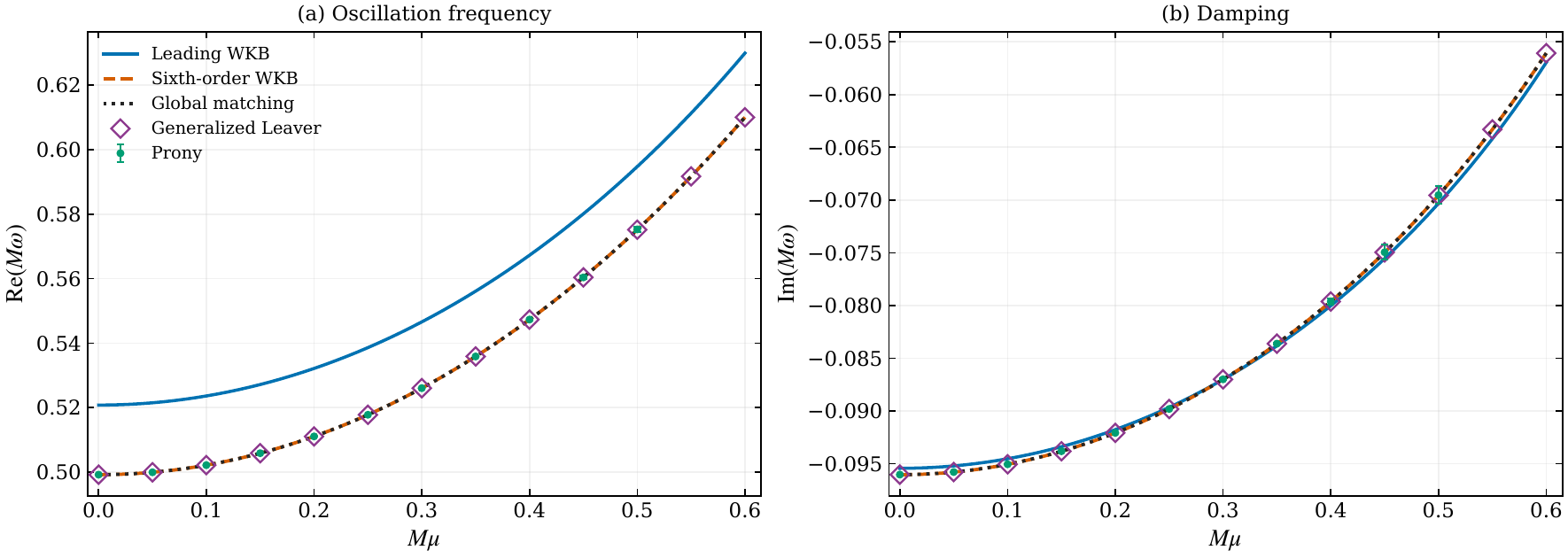}
 \caption{Fundamental QNM frequency and damping versus scalar mass
 for $M=r_0=1$, $q=0.5$, $\zeta=0.05$, and $\ell=2$.
 Leaver diamonds overlap the global-matching curve. Green circles
 show the eleven accepted Prony estimates; their bars indicate
 sensitivity to the numerical checks.}
 \label{fig:logned-comparison}
\end{figure*}

Substitution into the radial equation gives
$\sum_{j=0}^{n+1}H_{nj}(\omega)a_j=0$, which couples many
series coefficients. Gaussian elimination reduces this
relation to three terms:
\begin{equation}
 \alpha_na_{n+1}+\beta_na_n+\gamma_na_{n-1}=0,
 \qquad a_{-1}=0,
 \label{eq:logned-leaver-recurrence}
\end{equation}
Here $n$ labels the series coefficients, and
$\alpha_n(\omega)$, $\beta_n(\omega)$, and $\gamma_n(\omega)$
are the coefficients after elimination. We introduce the
ratio $R_n=a_{n+1}/a_n$, which obeys
\begin{equation}
 R_{n-1}
 =-\frac{\gamma_n}{\beta_n+\alpha_nR_n},
 \qquad n=1,\ldots,N.
 \label{eq:logned-ratio-recurrence}
\end{equation}
At a finite series depth $N$, we choose a terminal ratio
$R_N=\tau_N$ and work backward. Combining the resulting
$R_0$ with $\beta_0+\alpha_0R_0=0$ gives
\begin{equation}
 \mathcal C_N(\omega)=\beta_0-
 \cfrac{\alpha_0\gamma_1}{\beta_1-
 \cfrac{\alpha_1\gamma_2}{\beta_2-\cdots-
 \cfrac{\alpha_{N-1}\gamma_N}{\beta_N+\alpha_N\tau_N}}}=0,
 \qquad \tau_N=\frac{a_{N+1}}{a_N}.
 \label{eq:logned-leaver-cf}
\end{equation}
We use $N=800$ and $\tau_N=0$, and also test $\tau_N=1$
and consistency with the RN recurrence relations. Across all
61 sampled masses, the largest difference from global matching
is $|M(\omega_{\rm L}-\omega_g)|<3.6\times10^{-13}$.
Figure~\ref{fig:logned-comparison} compares these frequencies
with the other methods.\\
Let's note that both global matching and Leaver's method can find the
fundamental mode and its overtones. At fixed background,
$\mu$, and $\ell$, the frequencies $\omega_{\ell n}$ are
distinct roots of the matching or continued-fraction equation.
Here $n=0$ denotes the fundamental mode and $n=1,2,\ldots$
its overtones. In the main text, we follow the $\ell=2$,
$n=0$ mode continuously from the massless limit.
Appendix~\ref{app:systematic-scans} gives the first two
overtones and the wider parameter scans.

\subsubsection{Time-domain evolution and Prony extraction}
\label{subsec:logned-prony}
We now study the ringdown directly by evolving the scalar wave
equation~\eqref{eq:logned-wave}. We use a second-order scheme in
the null coordinates $u=t-r_*$ and $v=t+r_*$, with initial data
$\Psi(0,v)=g(v)-g(-v)$ and $\Psi(u,0)=0$, where
$g(v)=\exp[-(v-10M)^2/(18M^2)]$. At each scalar mass, we place
$r_*=0$ at the potential maximum $r_p$. We record the waveform at
$r_*^{\rm obs}/M=-5,0,10,20$ and use two grid spacings,
$h/M=0.1$ and $0.05$, to check numerical sensitivity.

The idea behind Prony analysis is to describe a portion of the
waveform as a sum of damped oscillations. For samples separated
by $\Delta t$, we write
\begin{equation}
 \Psi(t_0+j\Delta t)\simeq\sum_{a=1}^{P}C_a z_a^j,
 \qquad \omega_a=\frac{i}{\Delta t}\log z_a .
 \label{eq:logned-prony}
\end{equation}
Here $C_a$ sets the amplitude and phase of each contribution,
while $z_a$ determines its oscillation frequency and damping.
We use the matrix-pencil version of Prony analysis, with singular
value decomposition to stabilize the fit. Note that the fit size $P$ counts
exponential terms; it is not an overtone number.

In the main text, we focus on the fundamental mode. We retain
frequencies that remain stable when we change the fitting interval,
$P$, the observer position, and the grid spacing.\footnote{We set the acceptance criterion as follows: a stable
group must recur in at least ten fitting windows, at three values
of $P$, and at five starting times. The median absolute deviation in
each component of $M\omega$ must be below $4\times10^{-4}$, with
such groups present at two or more observers on both grids. We
quote the fine-grid median and combine the observer and grid shifts
with twice the within-group median absolute deviation to obtain the
sensitivity bars.}
We also check sensitivity to the initial pulse width. Requiring
the sensitivities in both parts of $M\omega$ to be below $10^{-3}$
leaves eleven accepted estimates over $0\leq M\mu\leq0.50$. No WKB or global frequency
is supplied to the fit.

For the extensive scans in Appendix~\ref{app:systematic-scans},
we extend this procedure to search for the first two overtones.
Since these modes decay more rapidly, we also fit earlier,
shorter portions of the waveform and allow more exponential terms.
We look for additional frequencies that recur across fits and
check their stability against changes in the grid, observer, and
initial pulse. Only after extraction do we compare the candidates
with the global frequencies to assign their mode labels. Thus the
frequencies come from the waveform, while their identification
uses the frequency-domain spectrum. The appendix gives the wider
parameter scans and overtone tables, including flagged or unresolved
estimates; its stricter pulse and stability checks can flag an
estimate accepted in the main-text comparison. Separate selections can also shift the median, as at $M\mu=0.5$.

\begin{table*}[t]
 \caption{Selected frequencies $M\omega$ from
 Fig.~\ref{fig:logned-comparison}. Global matching and Leaver agree
 at all displayed digits. Prony values are rounded to six decimals
 for comparison; the printed digits do not imply this precision.
 A dash means that no Prony estimate passed the acceptance checks.}
 \label{tab:logned-frequencies}
 \begingroup\footnotesize
 \setlength{\tabcolsep}{2pt}
 \begin{ruledtabular}
 \begin{tabular}{cccccc}
 $M\mu$ & Leading WKB & Sixth-order WKB & Global QNM & Prony & Generalized Leaver \\
0.00 & $0.520759-0.095443i$ & $0.499160-0.096049i$ & $0.499162-0.096042i$ & $0.499165-0.096040i$ & $0.499162-0.096042i$ \\
0.10 & $0.523585-0.094536i$ & $0.502117-0.095067i$ & $0.502118-0.095060i$ & $0.502122-0.095058i$ & $0.502118-0.095060i$ \\
0.20 & $0.532122-0.091774i$ & $0.511025-0.092090i$ & $0.511027-0.092084i$ & $0.511031-0.092081i$ & $0.511027-0.092084i$ \\
0.30 & $0.546564-0.087021i$ & $0.526010-0.087013i$ & $0.526011-0.087007i$ & $0.526017-0.087003i$ & $0.526011-0.087007i$ \\
0.40 & $0.567254-0.080022i$ & $0.547286-0.079629i$ & $0.547288-0.079623i$ & $0.547292-0.079621i$ & $0.547288-0.079623i$ \\
0.50 & $0.594752-0.070317i$ & $0.575176-0.069556i$ & $0.575178-0.069550i$ & $0.575278-0.069534i$ & $0.575178-0.069550i$ \\
0.60 & $0.629978-0.056954i$ & $0.610048-0.056066i$ & $0.610063-0.056078i$ & \textemdash & $0.610063-0.056078i$ \\
 \end{tabular}
 \end{ruledtabular}
 \endgroup
\end{table*}

\begin{figure*}[t]
 \centering
 \includegraphics[width=\textwidth]{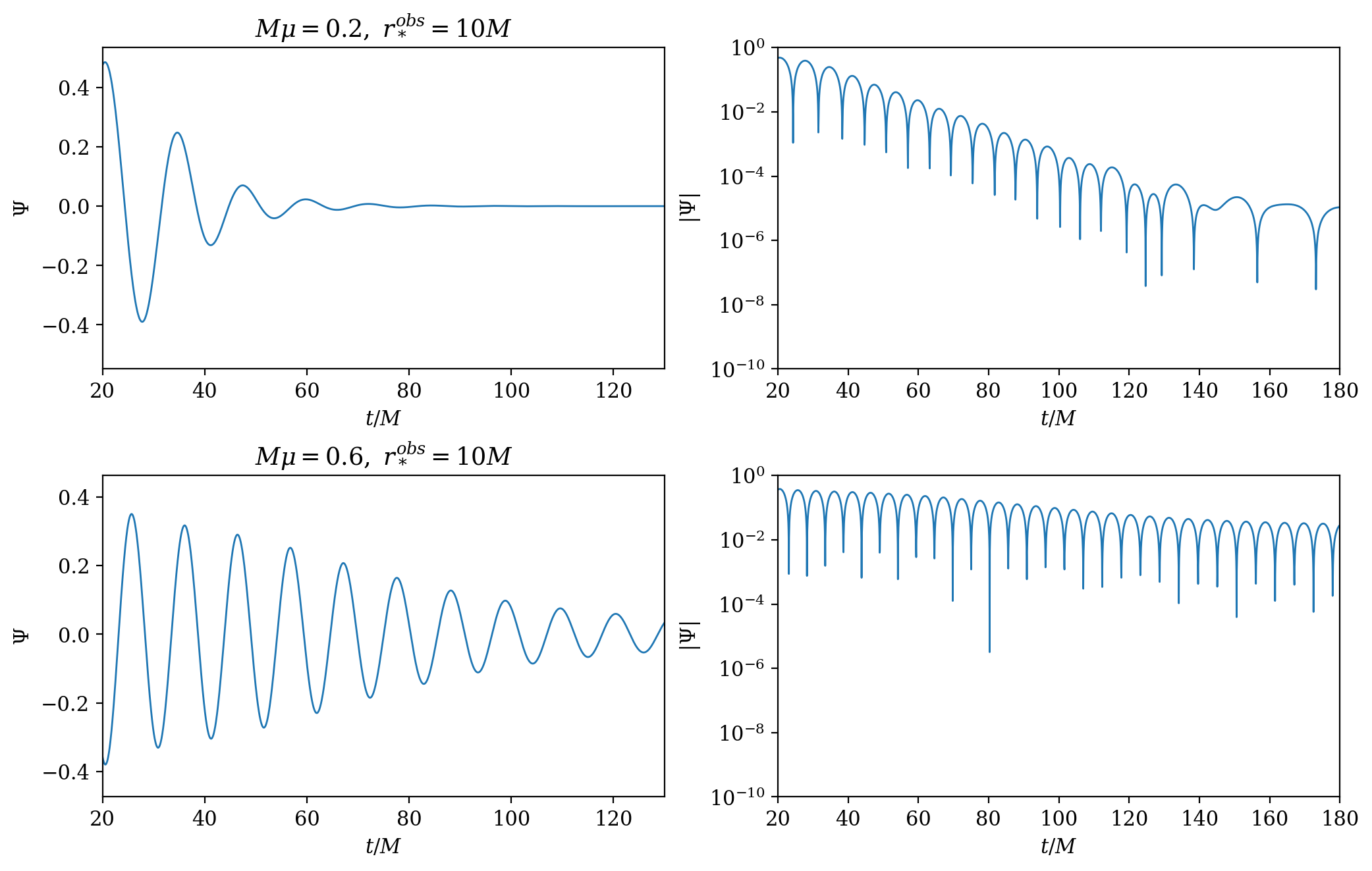}
 \caption{Waveforms for $M\mu=0.2$ (top) and $0.6$ (bottom),
 using the same initial pulse, $r_*^{\rm obs}=10M$, and $h=0.05M$.
 Other parameters are as in Fig.~\ref{fig:logned-comparison}.
 Left panels show $\Psi$; right panels show $|\Psi|$ on a logarithmic
 scale. The sharp dips mark zero crossings of the real waveform.}
 \label{fig:logned-waveforms}
\end{figure*}

\subsection{Comparison of approaches and mass dependence}
\label{subsec:logned-results}

We now compare the five approaches in
Fig.~\ref{fig:logned-comparison} and Table~\ref{tab:logned-frequencies}.
The frequency-domain calculations cover $0\leq M\mu\leq0.60$
in steps of $0.01$, with accepted Prony estimates available through
$M\mu=0.50$. Throughout this interval, the potential maximum
lies above the asymptotic level $\mu^2$; the two become equal at
$M\mu_{\rm pl}\simeq0.648503$.

As the scalar mass increases, the tracked fundamental mode
oscillates faster and decays more slowly. Global matching and
generalized Leaver agree at all displayed digits: the real part
of $M\omega$ rises from $0.499162$ to $0.610063$, while the
damping magnitude falls from $0.096042$ to $0.056078$.
The amplitude decay time $\tau=1/|\operatorname{Im}\omega|$
therefore grows from $10.41M$ to $17.83M$.

\begin{figure*}[t]
 \centering
 \includegraphics[width=\textwidth]{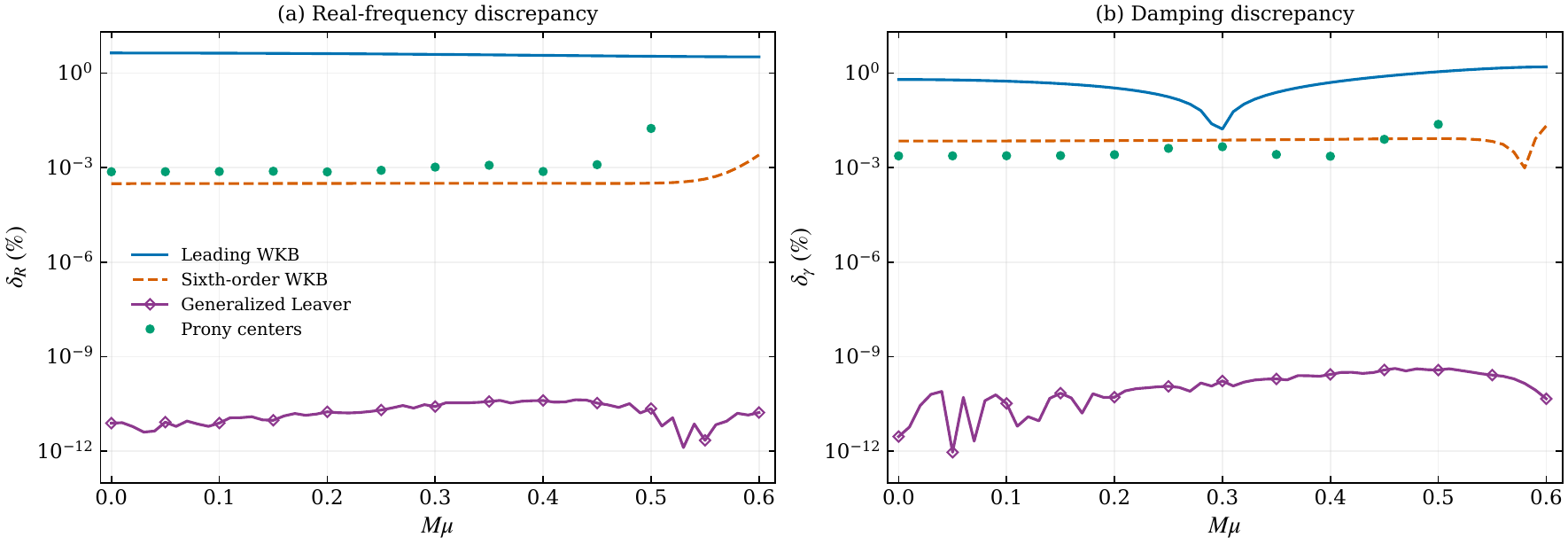}
 \caption{Percentage differences from global matching,
 defined in Eq.~\eqref{eq:logned-discrepancy}, for the real frequency
 (left) and damping (right). Green circles show differences of the
 accepted Prony estimates, not their sensitivity bars.}
 \label{fig:logned-errors}
\end{figure*}

To compare the methods more closely, we use global matching as
the reference and measure the percentage difference in each
frequency component:
\begin{equation}
 \delta_X^{(j)}=100\frac{|X_j-X_g|}{|X_g|},
 \qquad j\in\{1,6,\mathrm L,\mathrm P\},
 \label{eq:logned-discrepancy}
\end{equation}
where $X=\operatorname{Re}\omega$ or $\gamma=-\operatorname{Im}\omega$.
The subscript $g$ denotes global matching, and $1,6,\mathrm L,\mathrm P$
denote leading WKB, sixth-order WKB, generalized Leaver, and Prony.
Figure~\ref{fig:logned-errors} and Table~\ref{tab:logned-errors}
show that sixth-order WKB substantially improves the leading
estimate. The largest difference in the real frequency falls
from $4.33\%$ to $0.00252\%$, and that in the damping falls from
$1.56\%$ to $0.0211\%$. Leaver and global matching agree much
more closely. All accepted Prony estimates agree
with global matching within their sensitivity bars, which widen
toward the upper end of the accepted mass range.

\begin{table}[t]
 \caption{Largest percentage differences from global matching.
 WKB and Leaver use 61 masses over $0\leq M\mu\leq0.60$;
 Prony uses eleven accepted estimates over $0\leq M\mu\leq0.50$.}
 \label{tab:logned-errors}
 \begin{ruledtabular}
 \begin{tabular}{lcc}
 Method & $\max\delta_R$ (\%) & $\max\delta_\gamma$ (\%) \\
 Leading WKB & 4.326731 & 1.562713 \\
 Sixth-order WKB & 0.002525 & 0.021144 \\
 Generalized Leaver & $4.3\times10^{-11}$ & $4.2\times10^{-10}$ \\
 Prony & 0.017475 & 0.023568 \\
 \end{tabular}
 \end{ruledtabular}
\end{table}

We next examine the waveforms in Fig.~\ref{fig:logned-waveforms}.
At $M\mu=0.2$, the intermediate-time signal shows a clear damped
oscillation, consistent with
$M\omega_0\simeq0.511027-0.092084i$. At $M\mu=0.6$, the
oscillations are faster and persist longer. Global matching and
Leaver give $M\omega_0\simeq0.610063-0.056078i$ at this mass. The amplitude decay time increases from $10.86M$ at
$M\mu=0.2$ to $17.83M$ at $M\mu=0.6$.

These decay times describe one fundamental mode, whereas the waveform can
contain several contributions. In particular, the full signal
at $M\mu=0.6$ decays more slowly than this fundamental mode
alone would predict. A massive field also has a dispersive
response, in which different frequencies propagate differently;
the asymptotic potential $V_\ell\to\mu^2$ sets the threshold
$\omega=\mu$~\cite{KonoplyaZhidenko2013KN}. At $M\mu=0.6$,
this threshold is close to the real part of the tracked QNM
frequency. The later changes in the envelope and oscillation
period at $M\mu=0.2$ likewise indicate that other contributions
become important. In Euler-Heisenberg electrodynamics,
L\"utf\"uo\u{g}lu~\cite{Lutfuoglu2025EH} found that a massive-field tail can
mask long-lived QNMs in the waveform even when the Leaver method resolves their frequencies.

At both masses, grid refinement is consistent with second-order
convergence: halving the spacing reduces the leading numerical
error by about a factor of four. A converged waveform does not,
however, guarantee a reliable separation of its modes. We therefore
show the $M\mu=0.6$ waveform but mark its Prony entry with a dash,
because no estimate passes our selection criterion.

\section{ QBS spectrum of a massive scalar field }
\label{sec:qbs}

We now turn to quasibound states (QBSs). These waves are trapped
in an outer potential well, but gradually lose amplitude as part
of the wave passes through the inner barrier and enters the horizon.
We ask two questions: how strongly is the wave bound, and how long
does it survive? Writing $\omega=\omega_R-i\Gamma$, with
$\Gamma>0$, we measure binding by $\mu-\omega_R$ and the amplitude
lifetime by $\Gamma^{-1}$.

We first calculate the complex frequencies using global matching
and the generalized Leaver method. We then compare these results
with simpler estimates of binding and decay. Finally, we examine
how the states change with scalar mass, NED coupling, charge, and
radial level. Additional results are tabulated in
Appendix~\ref{app:systematic-scans}.

\subsection{Global matching method}

We use the background and radial equation of
Eqs.~\eqref{eq:logned-lapse} and \eqref{eq:schrodinger}, with
the potential and tortoise coordinate in Eq.~\eqref{eq:logned-wave}.
The matching procedure follows Sec.~\ref{subsec:logned-global}.
The main change from a QNM is the outer boundary condition:
a QBS decays as $e^{-\kappa r}$ at large radius, while remaining
ingoing at the simple outer horizon,
\begin{equation}
 \Psi\sim e^{-i\omega r_*}\quad(r\to r_+),\qquad
 \kappa=\sqrt{\mu^2-\omega^2},\quad\Re\kappa>0.
 \label{eq:qbs-prd-bc}
\end{equation}
For the modes considered here, $0<\omega_R<\mu$ and
$\omega_I<0$. We also retain the logarithmic correction to the
decaying solution. With $L=\ln(r/r_0)$ and $\kappa\ne0$, it gives
\begin{equation}
 \frac{\Psi'}\Psi=-\kappa+
 \frac{(2\omega^2-\mu^2)(2M+AL)}{2\kappa r}
 +O\!\left(\frac{1+L^2}{r^2}\right).
 \label{eq:qbs-prd-tail}
\end{equation}

We integrate the horizon solution outward and the decaying
solution inward, then compare them at $r_m$. We adjust the complex
frequency until their logarithmic derivatives agree, or equivalently
until $D(\omega)=0$, where
\begin{equation}
 D(\omega)=
 \left.\frac{f\Psi_H'}{\Psi_H}\right|_{r_m}
 -\left.\frac{f\Psi_\infty'}{\Psi_\infty}\right|_{r_m},
 \label{eq:qbs-prd-match}
\end{equation}
Here $\Psi_H$ and $\Psi_\infty$ satisfy the horizon and outer
conditions, respectively. We evolve $(\Psi,f\Psi')$ and choose
$r_m$ where the amplitudes are nonzero. We normally use boundary
expansions through order eight, and vary their order, the endpoints,
and $r_m$ to check convergence.

\subsection{Generalized Leaver determination of QBSs}
\label{subsec:qbs-leaver}

We now calculate the same QBS frequencies using the Leaver method
of Sec.~\ref{subsec:logned-leaver}. To make the wave decay at
infinity, we choose
\begin{equation}
 k_{\rm B}=i\kappa,\qquad
 \kappa=\sqrt{\mu^2-\omega^2},\qquad \Re\kappa>0,
 \label{eq:qbs-leaver-sheet}
\end{equation}
so that the leading radial factor becomes
$e^{ik_{\rm B}r}=e^{-\kappa r}$. Dolan used this decaying boundary condition
in a continued-fraction calculation of massive scalar bound states on
Kerr~\cite{Dolan2007Bound}.
We use the full complex $\omega$ in $\kappa$ and keep
$\Re\kappa>0$ throughout the frequency search.
For a weakly damped positive-frequency QBS, this gives
$\Re k_{\rm B}<0$. With $k=i\kappa$, the radial ansatz becomes
\begin{equation}
 \begin{split}
 \Psi(z)&=z^\nu
  \exp\!\left[-\kappa r+
  \eta\left(2ML+\frac A2L^2\right)\right]y(z),\\
 \eta&=\frac{2\omega^2-\mu^2}{2\kappa},\qquad
 y(z)=\sum_{j=0}^{\infty}a_jz^j,\qquad a_0=1,
 \end{split}
 \label{eq:qbs-leaver-ansatz}
\end{equation}
We use the same $z=1-r_+/r$, $L=\ln(r/r_0)$, and
$\nu=-i\omega/f'(r_+)$ as before. The prefactors describe the
horizon and outer behavior, leaving $y(z)$ to be found as a series.
Integrating Eq.~\eqref{eq:qbs-prd-tail} gives the displayed outer
exponent, including its logarithmic term. At $A=0$, this factor
reduces to $e^{-\kappa r}(r/r_0)^{M(2\omega^2-\mu^2)/\kappa}$.
Note that at $A\ne0$, $y$ may still contain logarithmic terms near $z=1$.

We reuse the metric coefficients in
Eq.~\eqref{eq:logned-leaver-lapse}, the recurrence, and its
Gaussian reduction, evaluated with Eq.~\eqref{eq:qbs-leaver-sheet}.
We then solve Eq.~\eqref{eq:logned-leaver-cf} for a complex root.
This determines the oscillation frequency and damping rate together,
without a separate tunnelling estimate.

We check all 67 distinct QBS states tabulated in Appendix~\ref{app:systematic-scans} by increasing the series depth,
changing its terminal condition, and comparing with independent
global method discussed before.\footnote{We begin with $N=300,500,800$ and,
where needed, extend to $N=1200,1600,2200$. Six particularly weakly damped states require extra care because
their imaginary frequencies are very small. For five low-mass
excited states, we increase the series depth to $N=4400$; for
the $\ell=5$ state, we use $N=1600$. We also compare terminal
ratios $\tau_N=0$ and $1$. We set the acceptance criterion as follows: a retained
damping rate must be positive, change by less than $10^{-3}$
relatively under each final convergence check, and agree with global method to better than $10^{-3}$ relatively. The global QBS
calculation uses the real outer axis, with separate endpoint and
matching-radius checks.}
All displayed entries pass these tests, usually by a much wider
margin than required.
\subsection{Analytic binding and semiclassical approaches}

We now introduce the simpler estimates, one method at a time.
The first three estimate only the real frequency. We then combine
well quantization with an estimate of leakage through the barrier,
and finally consider an expansion about the RN pole.
Table~\ref{tab:qbs-prd-benchmark} compares all these approaches
with global matching and Leaver.

\textit{Coulomb approximation.}
We treat the scalar as a hydrogen-like state in the
weak gravitational potential $-M\mu/r$. With
$N=n_r+\ell+1$ and $n_r\geq0$, this gives
$\omega_R^{\mathrm C}=\mu-M^2\mu^3/(2N^2)$.
This estimate neglects both the logarithmic deformation and
horizon absorption.

\textit{First-order logarithmic correction.}
We next include the potential $-A\mu\ln(r/r_0)/(2r)$ as a small
perturbation of the Coulomb state. Taking its expectation value gives
the estimate labelled ``NR + log'' in Table~\ref{tab:qbs-prd-benchmark}:
\begin{align}
 C_{N\ell}&=\psi(N+\ell+1)+\ln\frac{N}{2M\mu^2r_0},\nonumber\\
 \omega_R^{\mathrm{NR+log}}
 &=\mu-\frac{M^2\mu^3}{2N^2}
 -\frac{AM\mu^3}{2N^2}C_{N\ell}.
 \label{eq:qbs-prd-nr}
\end{align}
Here $\psi$ is the digamma function. We require weak-field
localization and $|AC_{N\ell}/M|\ll1$. This improves the binding
estimate but still does not describe absorption. Pengpan~\cite{Pengpan2026PFDM}
obtained a similar first-order correction to the hydrogenic levels from
the logarithmic PFDM potential of a rotating dilaton-axion black hole.

\textit{Variational approximation.}
We also allow the radial size of a Coulomb trial wavefunction to
adjust through a parameter $B>0$, keeping $M$ fixed. For
$n_r=0$ and $N=\ell+1$, minimizing the approximate energy gives
\begin{align}
 B&=M+\frac A2\left[C_{N\ell}-\ln(B/M)-1\right],\nonumber\\
 \omega_R^{\mathrm{var}}
 &=\mu-\frac{\mu^3}{2N^2}(B^2+AB).
 \label{eq:qbs-prd-var}
\end{align}
We solve the first equation for $B$ and insert it into the second.
Since this auxiliary bound-state problem leaves out the absorbing
horizon, the result gives neither a damping rate nor a rigorous
bound on the real black-hole resonance frequency.

\begin{figure}[!t]
 \centering
 \includegraphics[width=0.6\textwidth]{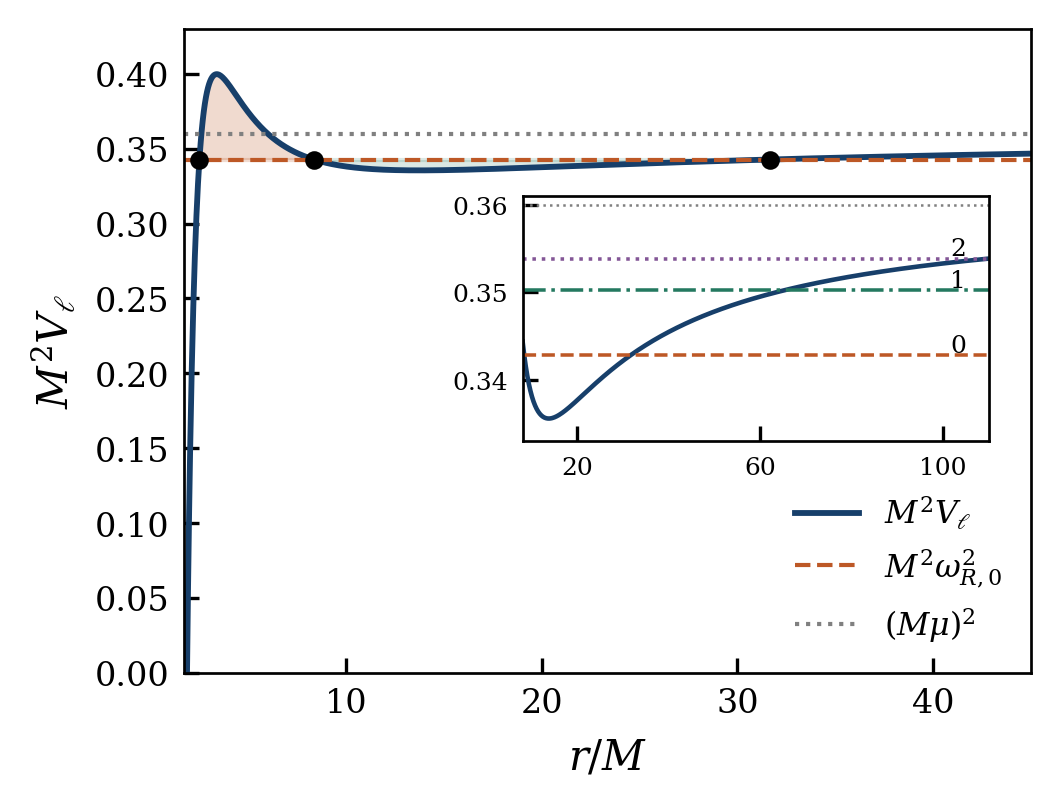}
 \caption{Trapping potential at the reference parameters of
Eq.~\eqref{eq:qbs-prd-reference}. The dashed line marks the fundamental
level $\omega_{R,0}^2$, and the dotted line marks $\mu^2$.
Shading shows the inner barrier and outer well; dots mark the turning
points. The inset shows the first three radial levels.}
 \label{fig:qbs-prd-trapping}
\end{figure}

Before introducing the semiclassical estimates, we specify the
trapping region. We have the  exact complex expressions:
\begin{equation}
\omega^2=\omega_R^2-\Gamma^2-2i\omega_R\Gamma,
\qquad
\operatorname{Re}(\omega^2)=\omega_R^2-\Gamma^2.
\label{eq:qbs_real_spectral_level}
\end{equation}
For weak damping, $\Gamma/\omega_R\ll1$, we neglect damping in
locating the real turning points and use $E\simeq\omega_R^2$.
The difference from $\Re(\omega^2)$ is only $\Gamma^2$.
The three turning points satisfy $V_\ell(r_i)=E$, with
$r_+<r_1<r_2<r_3$ and
\begin{equation}
 V_{\min}<E<\min(V_{\max},\mu^2).
 \label{eq:qbs-prd-trapping}
\end{equation}
The wave oscillates between $r_2$ and $r_3$. The inner barrier
between $r_1$ and $r_2$ separates this well from the horizon,
while the outer tail decays toward infinity
(Fig.~\ref{fig:qbs-prd-trapping}). These real turning points
describe the weak-damping approximation; the exact condition
uses $V_\ell(r)=\omega^2$.

\textit{Bohr-Sommerfeld quantization formula and Gamow tunnelling.}
We treat the well as a cavity with a small leakage through its
inner barrier. First, we determine the real energy by quantizing
the action across the well ($n_r$ denotes the radial level):
\begin{align}
 \mathcal I(E)&=\int_{r_2}^{r_3}\frac{\sqrt{E-V_\ell}}f\,dr,
 &\mathcal I(E_{n_r}^{\mathrm{BS}})&=\pi(n_r+\tfrac12),
 \label{eq:qbs-prd-bs}\\
 \mathcal S(E)&=\int_{r_1}^{r_2}\frac{\sqrt{V_\ell-E}}f\,dr,
 &\mathcal I_E&=\frac12\int_{r_2}^{r_3}\frac{dr}{f\sqrt{E-V_\ell}}.
 \label{eq:qbs-prd-actions}
\end{align}
Here $\mathcal I_E=\partial\mathcal I/\partial E$ and
$\mathcal S$ measures suppression through the barrier. The
factors $1/f$ convert the integrals to tortoise radius.
We use the full potential in these integrals. We then estimate horizon absorption using the leading Gamow
formula, valid for small transmission $T=e^{-2\mathcal S}\ll1$:
\begin{equation}
 \omega\simeq\sqrt E-i\frac{e^{-2\mathcal S(E)}}
 {8\sqrt E\,\mathcal I_E(E)},\qquad
 \Gamma\equiv-\omega_I>0.
 \label{eq:qbs-prd-gamow}
\end{equation}
For ``BS + Gamow'', we evaluate this expression at
$E=E_{n_r}^{\mathrm{BS}}$. The amplitude lifetime is thus
$\Gamma^{-1}$.

\textit{Local WKB with $[2/2]$ Pad\'e resummation.}
We can also estimate the energy by expanding the potential near
the well minimum. Defining
$V_k=[(f\partial_r)^kV_\ell]_{r_{\min}}$, with $V_2>0$, we write
\begin{align}
 E(h)&=V_0+\sum_{j\geq1}c_jh^j,\nonumber\\
 c_1&=(n_r+\tfrac12)\sqrt{2V_2},\nonumber\\
 c_2&=\frac{(2n_r^2+2n_r+1)V_4}{16V_2}-\frac{(30n_r^2+30n_r+11)V_3^2}{144V_2^2}.
 \label{eq:qbs-prd-local}
\end{align}
The parameter $h$ keeps track of expansion order and is set to
one at the end. Following Refs.~\cite{Konoplya2019,Konoplya2026},
we replace the series by a Pad\'e approximant: a ratio of two
polynomials that reproduces the same expansion. For $P_{p/q}(h)$,
the degrees are at most $p$ and $q$, and the series is matched
through $h^{p+q}$. Thus $[2/2]$ uses terms through $h^4$ and gives
$E^{[2/2]}=P_{2/2}(1)$ and
$\omega_R^{[2/2]}=\sqrt{E^{[2/2]}}$.
We insert this energy into Eq.~\eqref{eq:qbs-prd-gamow} to
estimate damping. We use this same ``$[2/2]$ + Gamow'' prescription
throughout the parameter plots and comparisons below.

\textit{Higher Pad\'e orders.}
We also test $[8/8]$, which matches the local series through
$h^{16}$ and gives $E^{[8/8]}=P_{8/8}(1)$. We retain it and
neighboring approximants as checks and in the extended tables.
Only real energies obeying Eq.~\eqref{eq:qbs-prd-trapping} are
used for tunnelling. In every case, damping is still obtained
from the same leading Gamow formula.

\textit{First-order expansion about the RN limit.}
Finally, we start from a global RN mode and ask how a small
logarithmic deformation shifts its complex frequency. For an
isolated simple root, at fixed $M$, $q$, $\mu$, $\ell$, $n_r$,
and $r_0$, we use
\begin{equation*}
 \omega(A)=\omega_{\mathrm{RN}}
 +A\left.\frac{d\omega}{dA}\right|_{A=0}+O(A^2),
 \qquad \omega_{\mathrm{RN}}\equiv\omega(0).
\end{equation*}
The ``RN expansion'' entry  in Table~\ref{tab:qbs-prd-benchmark} retains terms upto first order in
$A$. We obtain both the RN frequency and its complex derivative
from global matching. Differentiating $D(\omega,A)=0$ gives
$d\omega/dA=-(\partial_A D)/(\partial_\omega D)$, evaluated at
$(\omega_{\mathrm{RN}},0)$. This method thus predicts binding
and damping together, without using the Gamow formula.\\
The same derivative follows from the relativistic bilinear
framework of Ref.~\cite{Cannizzaro2024}. We set
$x=r-r_+(A)$ and $F(x,A)=f(r_+(A)+x,A)$, and write the
radial equation as $\mathcal L\Psi=\omega^2W\Psi$, where
$\mathcal L=-\partial_x(F\partial_x)+V_\ell/F$ and $W=F^{-1}$.
We differentiate with respect to $A$ at fixed $x$, multiply
by $\Psi$ without complex conjugation, and integrate using
the regularized bilinear prescription. Integration by parts
transfers $\mathcal L$ from $\partial_A\Psi$ to $\Psi$;
the original mode equation then cancels the terms containing
$\partial_A\Psi$, giving
\begin{equation}
 \left.\frac{d\omega}{dA}\right|_{A=0}
 =
 \left.
 \frac{\displaystyle
 \operatorname{Reg}\int_0^\infty
 \Psi(\partial_A\mathcal L-\omega^2\partial_AW)\Psi\,dx}
 {\displaystyle
 2\omega\,\operatorname{Reg}\int_0^\infty W\Psi^2\,dx}
 \right|_{A=0}.
\end{equation}
Here $\operatorname{Reg}$ denotes analytic regularization
at the horizon; together with QBS decay at infinity, this
removes the boundary terms. Both the operator and its
spectral weight are varied, with the horizon displacement
included through the fixed-$x$ derivatives. Guo et al.~\cite{Guo2026Environment}
used relativistic perturbation theory to calculate how a galactic halo or
an accretion disk shifts both the oscillation frequency and the decay rate
of a quasibound state.

\subsection{Reference state and comparison of approaches}

We now compare the methods at
\begin{equation}
 M=r_0=1,\quad q=0.5,\quad\zeta=0.01,\quad
 \ell=2,\quad\mu=0.6.
 \label{eq:qbs-prd-reference}
\end{equation}
We follow the fundamental state, $n_r=0$, from the weak-binding
regime at $M\mu=0.3$ to $M\mu=0.6$, increasing the mass in
small steps while holding the other parameters fixed. This is
the lowest radial level at fixed $\ell$.\footnote{The starting
value $M\mu=0.3$ is a numerical choice. A nonzero mass is needed
for these outer-well QBSs, whereas barrier QNMs also exist at
$\mu=0$.}
Global matching gives
\begin{equation}
\begin{aligned}
 M\omega_0 &\simeq 0.5855151163197978-3.589904833649897\times10^{-5}i,
\end{aligned}
 \label{eq:qbs-prd-mode}
\end{equation}
and the independent Leaver calculation gives
\begin{equation}
\begin{aligned}
 M\omega_{\rm L} &\simeq 0.5855151163197972
-3.5899048338048783\times10^{-5}i.
\end{aligned}
 \label{eq:qbs-leaver-reference}
\end{equation}
These values differ by approximately $1.66\times10^{-15}$
in $M\omega$, while their damping rates differ relatively
by approximately $4.32\times10^{-11}$. We also compare the methods
across the parameter scans in Appendix~\ref{app:systematic-scans}.
Figure~\ref{fig:qbs-prd-trapping} shows where this state lies
in the potential. The horizon is at $r_+/M=1.86920176$, the
barrier peak at $r_{\max}/M=3.38763141$, and the well minimum
at $r_{\min}/M=13.72522844$. The real level satisfies
\[
M^2V_{\min}=0.33557006
< M^2\omega_{R,0}^2
< M^2\mu^2=0.36
< M^2V_{\max}=0.39984057.
\]
Using $E=\omega_{R,0}^2$, we find
\[
(r_1,r_2,r_3)/M=(2.504387,\,8.350488,\,31.659771),
\]
with $\mathcal S=1.662596$ and
$\mathcal I_E/M^2=285.87220$ for the tunnelling calculation. We compare binding and damping separately, using global matching
as the reference:
\begin{equation}
 \epsilon_B=\frac{|\omega_R^{\mathrm{app}}-\omega_R^{\mathrm{glob}}|}
 {\mu-\omega_R^{\mathrm{glob}}},\qquad
 \epsilon_\Gamma=\frac{|\Gamma^{\mathrm{app}}-\Gamma^{\mathrm{glob}}|}
 {\Gamma^{\mathrm{glob}}}.
 \label{eq:qbs-prd-errors}
\end{equation}
Note that the binding error is measured relative to the small binding
frequency $\mu-\omega_R$, rather than the full frequency.
All percentage errors below are calculated before rounding. Table~\ref{tab:qbs-prd-benchmark} shows how the binding estimate
improves. NR + log underestimates binding by $15.93\%$.
Bohr-Sommerfeld reduces this error to $1.107\%$, and $[2/2]$
reduces it further to $0.03809\%$. The latter corresponds to
only $0.000942\%$ of the full real frequency. Damping is less
accurate: BS and Pad\'e energies combined with leading Gamow
decay underestimate it by about $25\%$. Accurate binding alone
therefore does not guarantee an accurate lifetime.

\begin{table*}[!t]
\caption{QBS estimates at the reference parameters
of Eq.~\eqref{eq:qbs-prd-reference}. The last two columns give
percentage binding and damping errors relative to global matching.
BS and Pad\'e estimates use leading Gamow decay. A dash indicates
that the method does not provide a damping rate.}
\label{tab:qbs-prd-benchmark}
\centering\small
\setlength{\tabcolsep}{4pt}
\begin{tabular}{lrrrr}
\toprule
Method & $M\omega_R$ & $10^5M\Gamma$ & $100\epsilon_B$ & $100\epsilon_\Gamma$\\
\midrule
Coulomb & 0.5880000000 & --- & 17.16 & ---\\
NR + log & 0.5878227575 & --- & 15.93 & ---\\
Variational & 0.5878224549 & --- & 15.93 & ---\\
BS + Gamow & 0.5853548212 & 2.67585 & 1.107 & 25.46\\
$[2/2]$ + Gamow & 0.5855206342 & 2.68621 & 0.03809 & 25.17\\
$[8/8]$ + Gamow & 0.5855346846 & 2.68705 & 0.1351 & 25.15\\
RN expansion & 0.5855160488 & 3.58205 & 0.006437 & 0.2189\\
Leaver & 0.5855151163 & 3.58990 & $3.832\times10^{-12}$ & $4.317\times10^{-9}$\\
Global & 0.5855151163 & 3.58990 & 0 & 0\\
\bottomrule
\end{tabular}
\end{table*}

\begin{figure}[!t]
 \centering
 \includegraphics[width=0.6\textwidth]{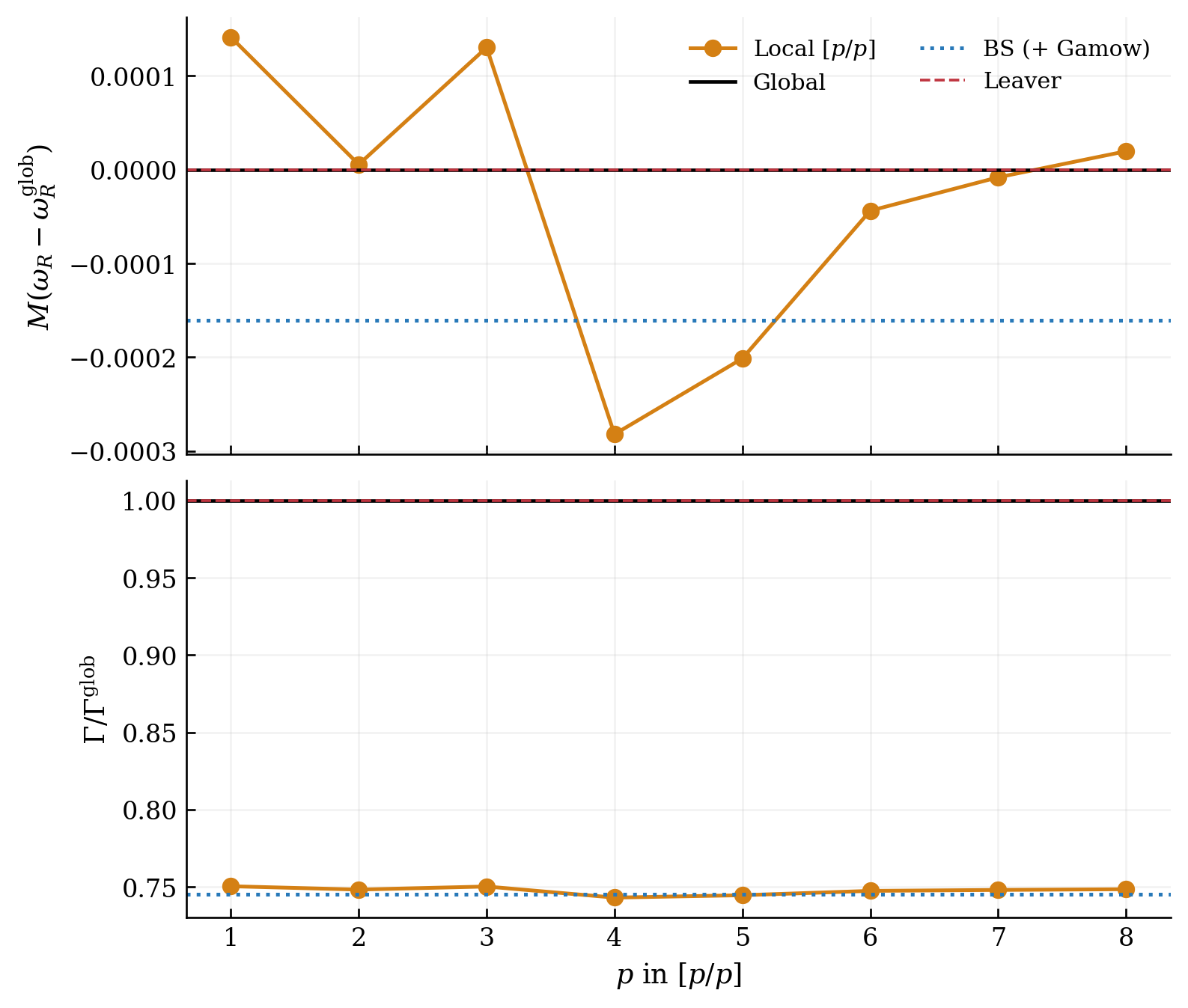}
 \caption{Effect of Pad\'e order at the reference parameters
of Eq.~\eqref{eq:qbs-prd-reference}. Top: real-frequency difference
$M(\omega_R-\omega_R^{\rm glob})$. Bottom: damping rate divided by
the global result. BS and Pad\'e estimates use leading Gamow decay.
The global and Leaver reference lines overlap. Higher Pad\'e order
does not consistently improve agreement.}
 \label{fig:qbs-prd-order}
\end{figure}

We next check whether a higher Pad\'e order improves the result
(Fig.~\ref{fig:qbs-prd-order}). The answer is not monotonic:
$[2/2]$ is more accurate than $[8/8]$ for this state. At total
order sixteen, the neighboring $[7/9]$ and $[9/7]$ approximants
give $M\omega_R=0.5855556208$ and $0.5855030853$, respectively.
The damping estimates remain close to one another because they
all use the same leading tunnelling formula.

\subsection{Parameter dependence of the QBS }

\begin{figure*}[!t]
 \centering
 \includegraphics[width=\textwidth]{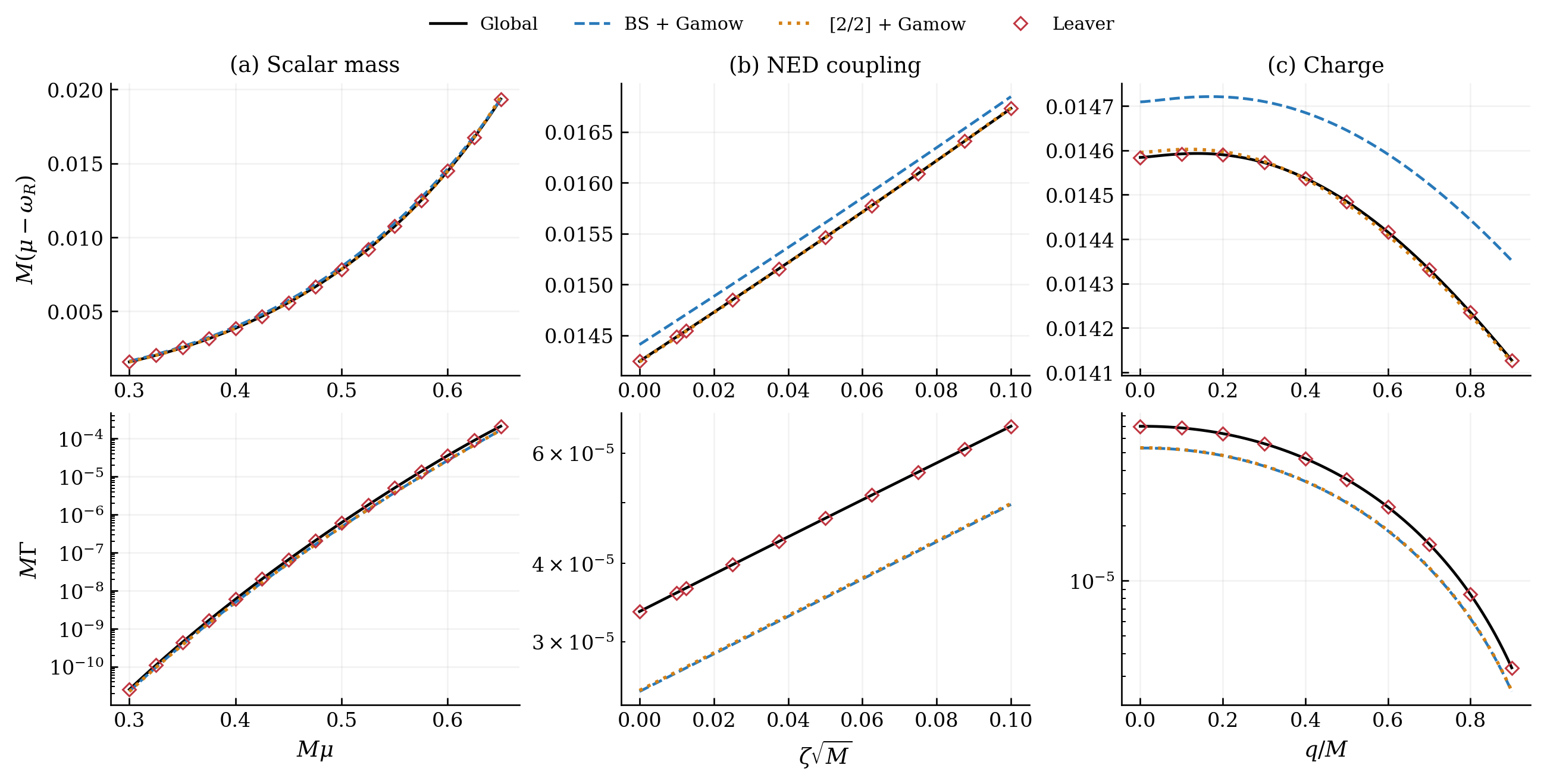}
 \caption{Binding $M(\mu-\omega_R)$ (top) and damping
$M\Gamma$ (bottom) of the fundamental QBS as we vary scalar mass,
NED coupling, and charge. The other parameters retain their values
in Eq.~\eqref{eq:qbs-prd-reference}. In the charge scan, $\zeta$
is fixed, so the NED amplitude $A$ also changes.}
 \label{fig:qbs-prd-parameters}
\end{figure*}

We now vary the scalar mass, NED coupling, and charge, one at a
time, keeping the other parameters at their reference values.
Figure~\ref{fig:qbs-prd-parameters} shows the resulting binding
and damping of the fundamental state. Global matching and Leaver
closely agree, while the approximations let us assess how well
binding and absorption are captured separately.

\begin{figure*}[!t]
 \centering
 \includegraphics[width=\textwidth]{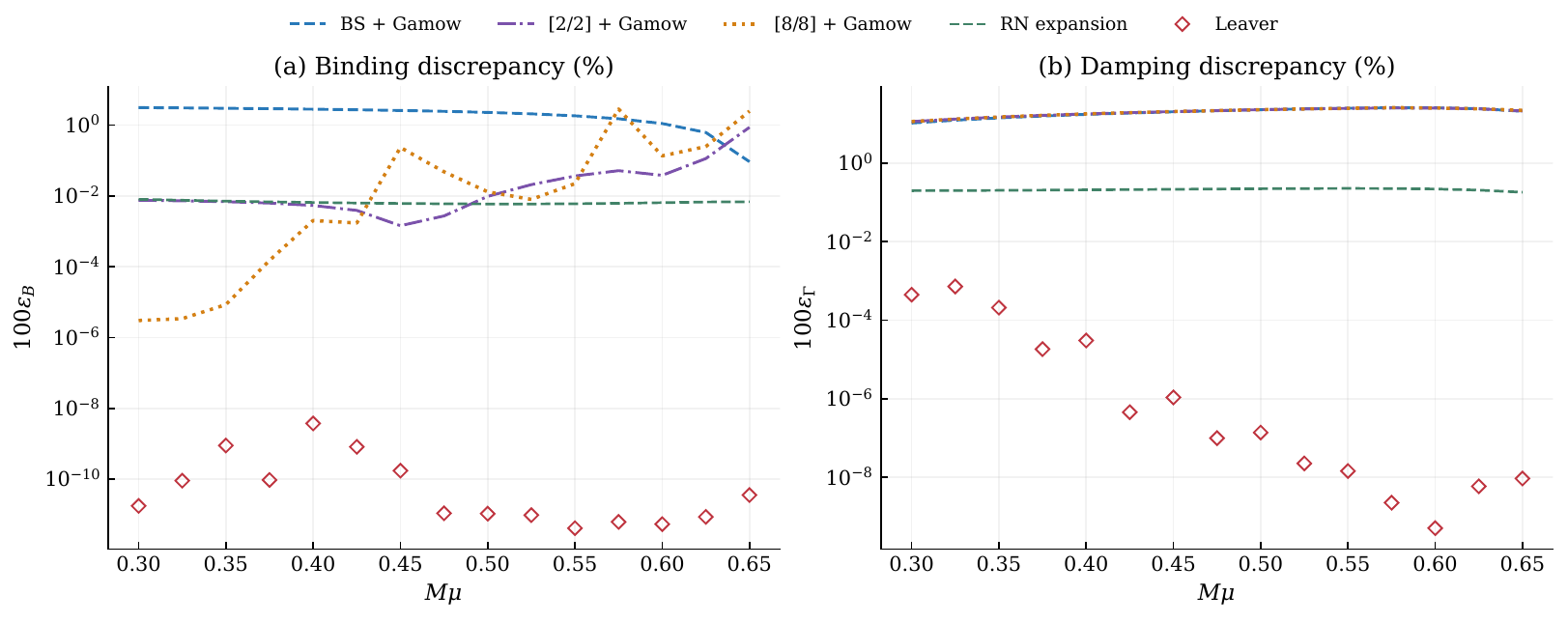}
\caption{Percentage errors in (a) binding and
(b) damping along the fundamental QBS mass scan, relative to global
matching. Errors are defined in Eq.~\eqref{eq:qbs-prd-errors};
both vertical axes are logarithmic. Parameters other than mass
are fixed at Eq.~\eqref{eq:qbs-prd-reference}.}
 \label{fig:qbs-method-errors}
\end{figure*}

As we increase the mass over $0.30\leq M\mu\leq0.65$, the state
becomes more strongly bound but also decays faster. The global
damping rate rises from $2.52\times10^{-11}/M$ to
$2.12\times10^{-4}/M$, while the $[2/2]$ barrier action falls
from $7.57$ to $0.89$. All tabulated approximations retain three
turning points.
We check the smallest damping rate using a larger outer domain
and higher-order boundary data. Across all 15 masses, Leaver and
global damping rates differ relatively by less than
$8\times10^{-6}$, or $8\times10^{-4}\%$.
Table~\ref{tab:qbs-prd-mass} lists selected values, and
Fig.~\ref{fig:qbs-method-errors} compares the errors of the
different approaches. The binding accuracy of $[2/2]$ does not
improve monotonically with mass.

\begin{table*}[!t]
\caption{Selected fundamental QBSs as the scalar mass
varies, with the other parameters fixed at
Eq.~\eqref{eq:qbs-prd-reference}. Global and Leaver real frequencies
agree to the digits shown. The last three columns give percentage
errors relative to global matching; $[2/2]$ uses leading Gamow decay.}
\label{tab:qbs-prd-mass}
\centering\small
\setlength{\tabcolsep}{3pt}
\begin{tabular}{rrrrrrr}
\toprule
$M\mu$ & $M\omega_R^{\rm glob,L}$ & $M\Gamma^{\rm glob}$ & $M\Gamma^{\rm L}$ & $100\epsilon_\Gamma^{\rm L}$ & $100\epsilon_B^{[2/2]}$ & $100\epsilon_\Gamma^{[2/2]}$\\
\midrule
0.30 & 0.298415228 & $2.523\times10^{-11}$ & $2.523\times10^{-11}$ & $4.475\times10^{-4}$ & 0.007539 & 11.38\\
0.40 & 0.396142615 & $5.959\times10^{-9}$ & $5.959\times10^{-9}$ & $3.057\times10^{-5}$ & 0.005358 & 17.79\\
0.45 & 0.444406366 & $6.587\times10^{-8}$ & $6.587\times10^{-8}$ & $1.139\times10^{-5}$ & 0.001438 & 20.44\\
0.50 & 0.492151041 & $6.190\times10^{-7}$ & $6.190\times10^{-7}$ & $2.582\times10^{-8}$ & 0.009678 & 22.8\\
0.60 & 0.585515116 & $3.590\times10^{-5}$ & $3.590\times10^{-5}$ & $4.317\times10^{-9}$ & 0.03809 & 25.17\\
0.65 & 0.630654178 & $2.117\times10^{-4}$ & $2.117\times10^{-4}$ & $9.449\times10^{-11}$ & 0.8644 & 21.16\\
\bottomrule
\end{tabular}
\end{table*}

We find a similar trade-off when increasing the NED coupling.
At fixed $M\mu=0.6$ and $q/M=0.5$, raising $\zeta\sqrt M$
from zero to $0.1$ strengthens binding and increases $M\Gamma$
from $3.35098\times10^{-5}$ to $6.62180\times10^{-5}$, shortening
the lifetime. Increasing charge has the opposite effect on
damping: as $q/M$ rises from zero to $0.9$ at fixed
$\zeta\sqrt M=0.01$, $\Gamma$ falls from
$6.99744\times10^{-5}/M$ to $3.34648\times10^{-6}/M$.
Note that changing $q$ also changes $A=4q^{3/2}\zeta/3$.
The endpoint $q=0$ is Schwarzschild, whereas $\zeta=0$ at
nonzero $q$ gives RN. On each coupling and charge scan, Leaver
and global damping rates agree relatively to better than
$3\times10^{-10}$, supporting these trends independently of
the Gamow estimate.

We also follow the first three radial levels from $M\mu=0.3$
to $0.6$ (Table~\ref{tab:qbs-prd-radial}). Higher levels lie
closer to the mass threshold and decay more slowly in this sample.
At $M\mu=0.6$, the Leaver/global complex-frequency differences
are below $3\times10^{-12}$ in $M\omega$, and the relative
damping differences are below $1.3\times10^{-7}$.
The $[2/2]$ binding errors increase from $0.03809\%$ to
$0.1141\%$ and $1.183\%$, while its Gamow damping errors are
$25.17\%$, $22.07\%$, and $20.08\%$, respectively.

\begin{table*}[!t]
\caption{First three radial levels at the reference
parameters of Eq.~\eqref{eq:qbs-prd-reference}. We compare global
matching, Leaver, and the $[2/2]$ real-frequency estimate. The last
column gives the percentage binding error of $[2/2]$.}
\label{tab:qbs-prd-radial}
\centering\small
\setlength{\tabcolsep}{3pt}
\begin{tabular}{rrrrrrrr}
\toprule
$n_r$ & $M\omega_R^{\rm glob}$ & $10^5M\Gamma^{\rm glob}$ & $M\omega_R^{\rm L}$ & $10^5M\Gamma^{\rm L}$ & $100\epsilon_\Gamma^{\rm L}$ & $M\omega_R^{[2/2]}$ & $100\epsilon_B^{[2/2]}$\\
\midrule
0 & 0.585515116 & 3.58990 & 0.585515116 & 3.58990 & $4.317\times10^{-9}$ & 0.585520634 & 0.03809\\
1 & 0.591833355 & 2.93457 & 0.591833355 & 2.93457 & $3.144\times10^{-8}$ & 0.591824037 & 0.1141\\
2 & 0.594841533 & 1.93442 & 0.594841533 & 1.93442 & $1.208\times10^{-5}$ & 0.594780489 & 1.183\\
\bottomrule
\end{tabular}
\end{table*}

We use $[2/2]$ consistently in these plots. Its energies remain
inside the trapping interval and give positive Gamow damping
rates throughout the sampled ranges. We also check that its
denominator stays away from zero.
\section{Global QNM-QBS matching and orthogonality}
\label{sec:ortho}

Having determined the QNM and QBS spectra, we now ask whether
these radiative and trapped modes are orthogonal under a common
definition of inner product. This is a useful step toward describing
both families within the same framework. The question is
nontrivial because a QNM is outgoing at infinity, whereas a QBS
decays there, and the usual radial overlap can diverge.
The logarithmically deformed RN geometry provides an interesting
spacetime for this test: it supports both families and modifies
their long-range behavior. We examine one independently computed
QNM-QBS pair.

\subsection{Selecting the mode pair}

We return to the reference parameters of
Eq.~\eqref{eq:qbs-prd-reference} and use the matching methods of
Secs.~\ref{subsec:logned-global} and \ref{sec:qbs}. Each mode
retains its own boundary conditions and logarithmic outer
corrections. The mismatch $D$ in
Eqs.~\eqref{eq:logned-match} and \eqref{eq:qbs-prd-match}
measures how well the horizon and outer solutions join at
$r_m$. Solving $D=0$ gives
\begin{align}
 M\omega_Q&\simeq0.6148438615-0.05782840305i,
 \label{eq:ortho_QNM_frequency}\\
 M\omega_B&\simeq0.5855151163-3.589904834\times10^{-5}i,
 \label{eq:ortho_QBS_frequency}
\end{align}
where $Q$ denotes the QNM and $B$ the fundamental QBS of
Eq.~\eqref{eq:qbs-prd-mode}.

\begin{figure*}[t]
  \centering
  \includegraphics[width=0.95\textwidth]{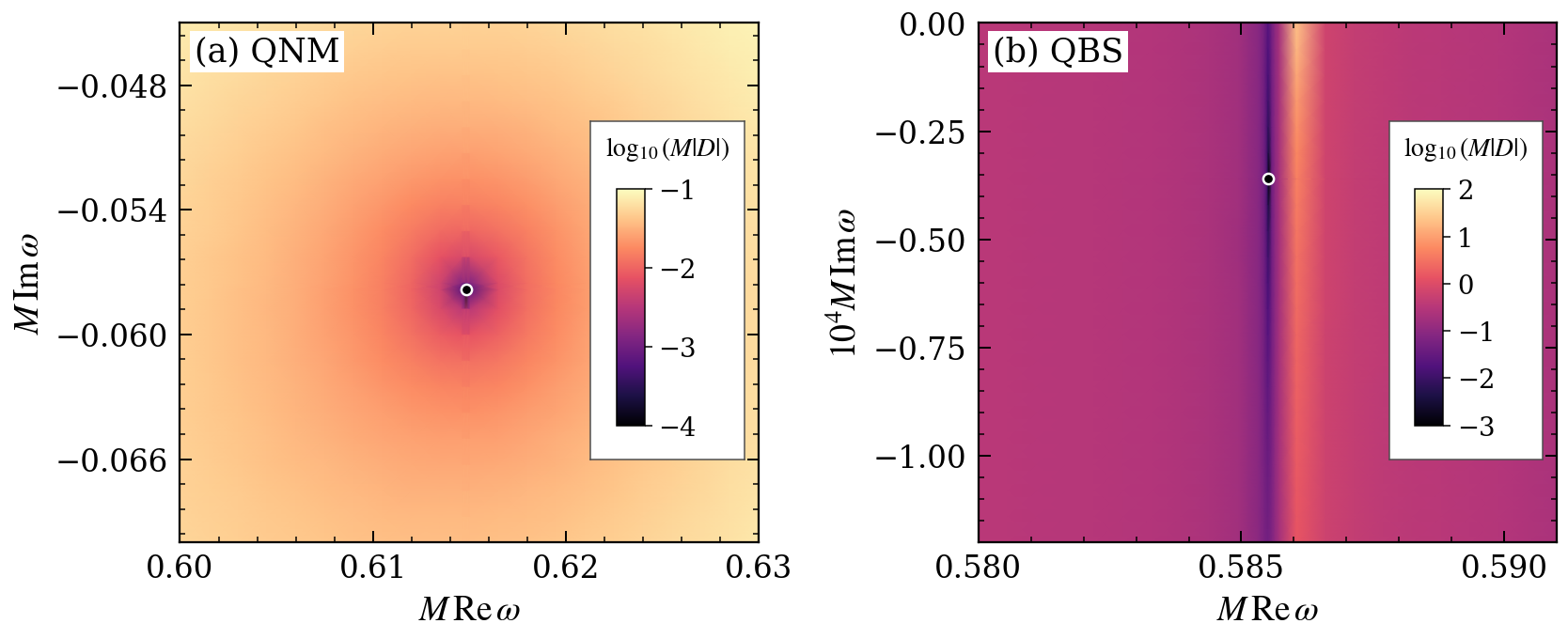}
  \caption{Matching the reference (a) QNM and (b) QBS.
  Color shows $\log_{10}(M|D|)$; darker regions indicate better
  matching. Circles mark the selected frequencies. The vertical
  scales differ between panels.}
  \label{fig:logned-spectral-maps}
\end{figure*}

\begin{figure*}[t]
  \centering
  \includegraphics[width=0.95\textwidth]{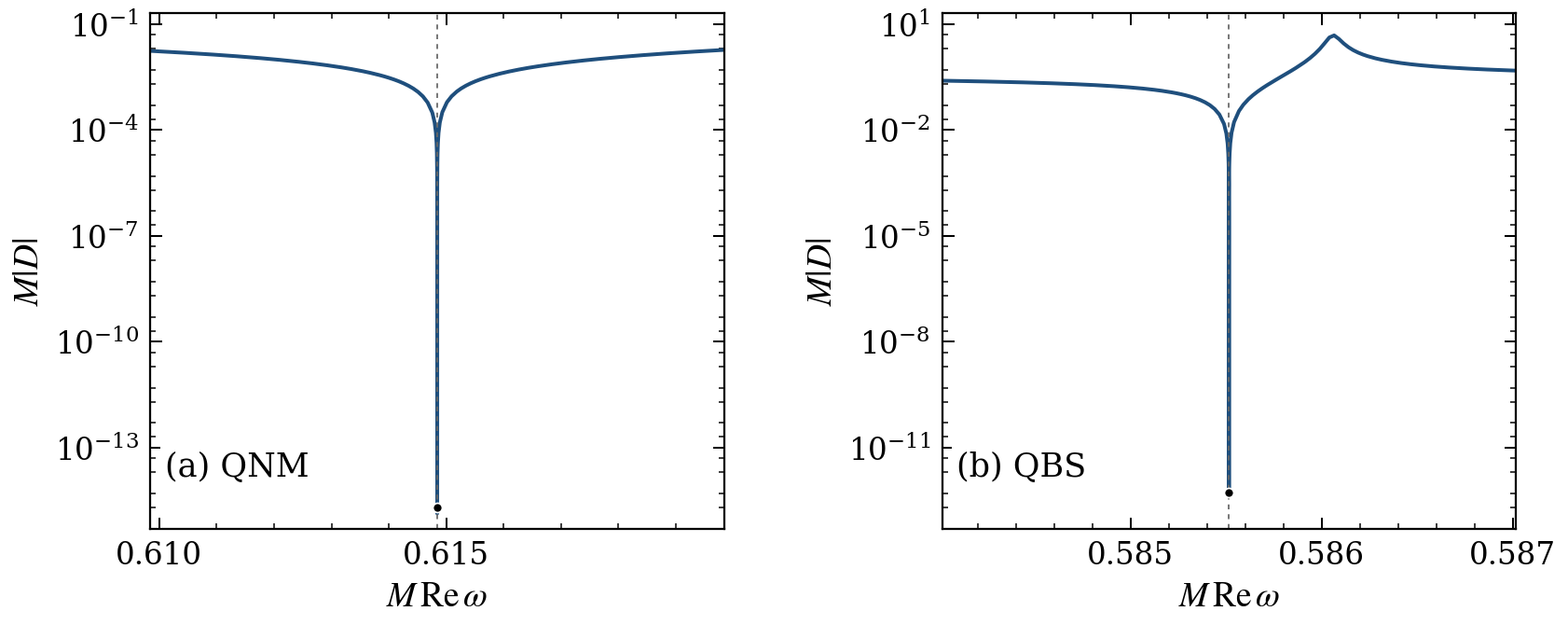}
  \caption{Mismatch $M|D|$ versus real frequency for the reference
  (a) QNM and (b) QBS, with the imaginary part fixed at its
  matched value.
  The minima and dashed lines identify the matched frequencies.}
  \label{fig:logned-spectral-cuts}
\end{figure*}

Figures~\ref{fig:logned-spectral-maps} and
\ref{fig:logned-spectral-cuts} show how the mismatch falls near
each frequency. Both frequencies remain stable when we vary
the matching radius and boundary.\footnote{We use
$r_m=5M$, $\epsilon/M=10^{-3}$, $R/M=240$, $\theta=60^\circ$,
and horizon and outer expansions through order eight. The
parameter variations give changes of at most $5\times10^{-14}$
in $M\omega_Q$ and a real-frequency spread below $2\times10^{-14}$ in
$M\Re\omega_B$. The maximum mismatches satisfy
$M|D_Q|<1.4\times10^{-14}$ and
$M|D_B|<1.4\times10^{-10}$.}

\subsection{Defining the bilinear}

We use the symmetric bilinear form of
Refs.~\cite{Cannizzaro2024,Cannizzaro2025}. For two modes
$\Phi_j=e^{-i\omega_jt}Y_{\ell m}\psi_j/r$ with the same
unit-normalized angular harmonic, it becomes
\begin{equation}
 \llangle\Phi_1,\Phi_2\rrangle
 =-i(\omega_1+\omega_2)J_{12},\qquad
 J_{12}=\operatorname{FP}\!\int_{\mathcal C}
 \frac{\psi_1\psi_2}{f}\,dr.
 \label{eq:ortho_bilinear}
\end{equation}
The contour $\mathcal C$ specifies the radial integration path,
and $\operatorname{FP}$ means that we remove the known divergent
horizon terms before taking the limit. We describe this procedure
below. The radial functions are multiplied without complex
conjugation, so this bilinear is not a positive-definite
norm.\footnote{The factor $\omega_1+\omega_2$ follows from
reflecting the first field by $(t,\phi)\mapsto(-t,-\phi)$
in the symplectic current.}

The reason to expect orthogonality follows from the radial
equation~\eqref{eq:schrodinger}. Combining the equations for
the two modes gives
\begin{equation}
 \begin{aligned}
 W_{12}&=\psi_2\partial_{r_*}\psi_1
             -\psi_1\partial_{r_*}\psi_2,\\
 \partial_{r_*}W_{12}&=(\omega_2^2-\omega_1^2)\psi_1\psi_2.
 \end{aligned}
 \label{eq:ortho_W_identity}
\end{equation}
When we integrate this identity along $\mathcal C$, the overlap
is determined by the endpoint values of $W_{12}$. If these
endpoint terms vanish under a common regularization, we obtain
\begin{equation}
 (\omega_1-\omega_2)\llangle\Phi_1,\Phi_2\rrangle=0.
 \label{eq:ortho_analytic}
\end{equation}
Distinct frequencies therefore have zero overlap under these
conditions. We now show how to define the common regularization
for our pair and check this cancellation numerically.

\subsection{Evaluating the regularized overlap}

We split the radial integral at $r_m$ into a horizon part and
an outer part. Near the horizon, the ingoing
expansion~\eqref{eq:logned-horizon}, with $x=(r-r_+)/M$, gives
\begin{equation}
 \frac{\psi_Q\psi_B}{f}
 =x^{\gamma-1}\sum_{n\geq0}h_nx^n,\qquad
 \gamma=-\frac{i(\omega_Q+\omega_B)}{f'(r_+)}.
 \label{eq:ortho_mixed_Frobenius}
\end{equation}
For our pair, $\Re\gamma\simeq-0.1168$, so the horizon
integral diverges as its lower endpoint approaches $r_+$.
We cancel the divergent terms explicitly and retain the finite
part:
\begin{equation}
 J_H=\lim_{\epsilon\to0^+}\left[
 M\sum_{n=0}^{N}\frac{h_n(\epsilon/M)^{\gamma+n}}{\gamma+n}
 +\int_{r_++\epsilon}^{r_m}\frac{\psi_Q\psi_B}{f}\,dr
 \right].
 \label{eq:ortho_JH}
\end{equation}
The sum removes the cutoff-dependent contribution, leaving a
well-defined limit.\footnote{This prescription is analytic
continuation from $\Re\gamma>0$. We take $\ln x$ real for
$x>0$ and choose $N$ so that $\Re\gamma+N+1>0$; none of the
denominators $\gamma+n$ vanishes for this pair.}

For the outer part, we continue both radial solutions along
the same complex ray,
\begin{equation}
 \begin{gathered}
 r=r_m+s e^{i\theta},\qquad s\geq0,\\
 \Im(k_Qe^{i\theta})>0,\qquad
 \Re(\kappa_Be^{i\theta})>0.
 \end{gathered}
 \label{eq:ortho_outer_ray}
\end{equation}
keeping the values of $k_Q$ and $\kappa_B$ used in the
matching calculations. The choice $\theta=60^\circ$ makes
both solutions decay, so the outer integral converges. We then
combine the two parts:
\begin{equation}
 J_\infty=\int_{r_m}^{\infty_\theta}
 \frac{\psi_Q\psi_B}{f}\,dr,\qquad
 J_{\rm reg}=J_H+J_\infty=J_{QB}.
 \label{eq:ortho_Jreg}
\end{equation}
The contour choice and its relation to the separately matched
frequencies are explained in Appendix~\ref{app_contour}.

We normalize each horizon and outer solution to unity at
$r_m$. 
With $r_m=5M$, $\epsilon/M=10^{-3}$, $s_{\max}/M=220$,
and $\theta=60^\circ$, we integrate the two parts separately
and find
\begin{align}
 J_H/M&\simeq2.442444786057+0.127821300081i,\nonumber\\
 J_\infty/M&\simeq-2.442444786049-0.127821300071i.
 \label{eq:ortho_overlap_parts}
\end{align}
The outer contribution is evaluated on the finite ray.
Although each part is nonzero, their sum is only
$|J_{\rm reg}|/M\simeq1.3\times10^{-11}$, giving
\begin{equation}
 \left|\llangle\Phi_Q,\Phi_B\rrangle\right|
 \simeq1.5\times10^{-11}.
 \label{eq:ortho_full_bilinear_numeric_def}
\end{equation}
This near cancellation supports orthogonality of the independently
matched QNM-QBS pair, despite their different conditions at
infinity. The conclusion applies to the stated bilinear form
and regularization. Note that the modes can still couple
under external perturbations, which is beyond the scope of this work.
For Kerr QNMs, Lestingi et al.~\cite{Lestingi2026Perturbation} showed that
the first-order correction to a mode contains a continuous-spectrum
contribution alongside the sum over discrete QNMs.

\section{Some finite-mass completions}
\label{subsec:finite_mass_completion}

We now provide a few completions of the geometry studied in this paper. Equation~\eqref{eq:NED_mass_function} shows why the mass grows without
bound and motivates such completions. This also raises an interesting physical question. How much of the scalar response comes
from the geometry near the black hole, and how much depends on extending
the logarithmic tail to arbitrarily large distances? A 
quasibound state can sample radii well beyond the inner potential
barrier. To make that question accessible, we show
several ways to make this extra density fall more rapidly beyond a radius
$R_c$. If $r_+\ll R_c$, the black hole still has an exterior region
$r_+<r\ll R_c$ in which the original logarithmic metric is a good
approximation. We can then change the distant geometry and ask which
features of the wave spectrum depend on extending the logarithm to
infinity. Below we present seven such completions using the electric sector.\\
\noindent\textit{The metrics and electric fields.}
We take $q>0$ and retain the Einstein-NED form of the
action~\eqref{eq:NED_action}, choosing a corresponding matter
Lagrangian $\mathcal L_i(\mathcal F)$ for each completion.
We denote the electric displacement by
$\mathcal D=-\mathcal L_{i,\mathcal F}E_i$, where
$\mathcal L_{i,\mathcal F}=d\mathcal L_i/d\mathcal F$.
Gauss's law then gives $\mathcal D(r)=q/r^2$, which can be used to explicitly compute the action using the electric field expressions in the table below\footnote{For the purely electric solutions, we have $\mathcal F=-E_i^2/2$, so the
chain rule gives
\begin{equation}
\frac{d\mathcal L_i}{dE_i}
=
\frac{\partial\mathcal L_i}{\partial\mathcal F}
\frac{d\mathcal F}{dE_i}
=
-\mathcal L_{i,\mathcal F}E_i
=
\mathcal D,
\end{equation}
where $\mathcal L_{i,\mathcal F}\equiv
\partial\mathcal L_i/\partial\mathcal F$, and the couplings
$\zeta$ and $\mathcal D_c$ are held fixed. This can be integrated to get the electromagnetic part of the action.}.
Each theory has fixed couplings $\zeta\geq0$ and
$\mathcal D_c>0$. The latter sets the displacement scale at
which the electric response changes. For a solution with charge
$q$, this occurs at
\[
R_c=\sqrt{\frac{q}{\mathcal D_c}},
\qquad
\mathcal D(R_c)=\mathcal D_c.
\]
 We present the solutions in a table. To read the table, start with
\[
 f_{\rm RN}(r)=1-\frac{2M_{\rm ADM}}{r}+\frac{q^2}{r^2}
\]
and add the middle column to $f_{\rm RN}(r)$ to obtain the exact metric
function in Eq.~\eqref{eq:NED_line_element}. The last column gives the
full electric field, including its Coulomb part.

\begin{center}
\begingroup\small\renewcommand{\arraystretch}{1.15}
\begin{tabular}{@{}l@{\quad}l@{\quad}l@{}}
\toprule
Model & Term added to $f_{\rm RN}(r)$ & Electric field $E_i(r)$\\
\midrule
C0 &
$\displaystyle \frac{A}{r}\!\left[\operatorname{arsinh}\frac{R_c}{r}
-\frac{R_c(\sqrt{r^2+R_c^2}+2r)}{3(\sqrt{r^2+R_c^2}+r)^2}\right]$ &
$\displaystyle \frac{q}{r^2}+\frac{\zeta\sqrt q\,R_c}{r\sqrt{r^2+R_c^2}}$\\[1.1ex]
C1 &
$\displaystyle \frac{A}{r}\ln\!\left(1+\frac{R_c}{r}\right)$ &
$\displaystyle \frac{q}{r^2}+\frac{\zeta\sqrt q\,R_c(3R_c+4r)}
  {3r(R_c+r)^2}$\\[1.1ex]
C2 &
$\displaystyle \frac{A}{r}\operatorname{arsinh}\frac{R_c}{r}$ &
$\displaystyle \frac{q}{r^2}+\frac{\zeta\sqrt q\,R_c(3R_c^2+4r^2)}
  {3r(R_c^2+r^2)^{3/2}}$\\[1.1ex]
C3 &
$\displaystyle \frac{A}{2r}\ln\!\left(1+\frac{R_c^2}{r^2}\right)$ &
$\displaystyle \frac{q}{r^2}+\frac{\zeta\sqrt q\,R_c^2(3R_c^2+5r^2)}
  {3r(R_c^2+r^2)^2}$\\[1.1ex]
C4 &
$\displaystyle \frac{A}{r}\!\left[\operatorname{arsinh}\frac{R_c}{r}
-\frac{R_c(8\sqrt{r^2+R_c^2}+7r)}{6(\sqrt{r^2+R_c^2}+r)^2}\right]$ &
$\displaystyle \frac{q}{r^2}+\frac{\zeta\sqrt q\,R_c}
  {r(\sqrt{r^2+R_c^2}+r)}$\\[1.1ex]
C5 &
$\displaystyle \frac{A}{2r}\!\left[\operatorname{artanh}\!\frac{R_c}{\sqrt[4]{r^4+R_c^4}}
+\arctan\!\frac{R_c}{\sqrt[4]{r^4+R_c^4}}\right]$ &
$\displaystyle \frac{q}{r^2}+\frac{\zeta\sqrt q\,R_c(3R_c^4+4r^4)}
  {3r(R_c^4+r^4)^{5/4}}$\\[1.1ex]
C6 &
$\displaystyle \frac{A}{r}\!\left[\ln\!\left(1+\frac{R_c}{r}\right)
+\frac{R_c}{r+R_c}\right]$ &
$\displaystyle \frac{q}{r^2}+\frac{\zeta\sqrt q\,R_c(3R_c^2+9R_cr+8r^2)}
  {3r(R_c+r)^3}$\\[0.7ex]
\bottomrule
\end{tabular}\endgroup
\end{center}
For each row, the metric gives the mass function
$m_i(r)=r[1-f_i(r)]/2$. The radial and angular Einstein equations,
together with Gauss's law, determine the electric field through
\[
 4\pi\rho_i=\frac{m_i'}{r^2},\qquad
 4\pi p_{\perp i}=-\frac{m_i''}{2r},\qquad
 E_i(r)=\frac{4\pi(\rho_i+p_{\perp i})}{\mathcal D(r)}
       =\frac{2m_i'-r m_i''}{2q},\qquad p_{ri}=-\rho_i .
\]
Primes on radial functions denote derivatives with respect to $r$. Substituting
the middle column gives exactly the field in the last column. For
example, the C2 density is
\[
 \rho_2(r)=\frac{q^2}{8\pi r^4}
    +\frac{A}{8\pi r^3}\frac{R_c}{\sqrt{r^2+R_c^2}} .
\]
The second term falls as $r^{-3}$ for $r\ll R_c$ and as $r^{-4}$ for
$r\gg R_c$. This is how C2 ends the logarithmic growth of the mass.
For every row we recover the matter Lagrangian on the electric branch from
$\mathcal L_i=\mathcal D E_i-4\pi\rho_i$ and
$\mathcal F=-E_i^2/2$. To express the action in terms of the field,
we use $r=\sqrt{q/\mathcal D}$ and $R_c=\sqrt{q/\mathcal D_c}$.
The charge $q$ cancels from the resulting constitutive law, and
$E_i=\partial_{\mathcal D}(4\pi\rho_i)$ at fixed couplings.
In all seven cases $E_i$ increases with $\mathcal D$, so this prescription
specifies a single-valued $\mathcal L_i(\mathcal F)$. Related electric
reconstructions are discussed in Refs.~\cite{MkrtchyanSvazas2022,Hale2023}.

The potential is also available directly from the metric. In the gauge
$U_i(\infty)=0$, integrating $E_i=-U_i'$ gives
\[
 U_i(r)=\frac{3[M_{\rm ADM}-m_i(r)]+r m_i'(r)}{2q}.
\]
This is the radial reconstruction of Fan and Wang~\cite{Fan:2016hvf}
in our sign convention and gauge. Comparing C0 and C2 shows the
location of the inverse hyperbolic sine in each solution,
\[
 \begin{aligned}
 U_0(r)&=\frac qr+\zeta\sqrt q\,\operatorname{arsinh}\frac{R_c}{r},\\
 f_2(r)&=1-\frac{2M_{\rm ADM}}r+\frac{q^2}{r^2}
            +\frac Ar\operatorname{arsinh}\frac{R_c}{r},\qquad
 U_2(r)=U_0(r)+\frac{\zeta\sqrt q\,R_c}{3\sqrt{r^2+R_c^2}} .
 \end{aligned}
\]
For C4 the matter law takes the simple form
$E_4=\mathcal D+\zeta(\sqrt{\mathcal D+\mathcal D_c}-\sqrt{\mathcal D_c})$.
We can therefore recover $\mathcal D$ from $E_4$ by solving a quadratic.

For a distant observer, we match each solution to the same inner mass parameter $M$ and
reference length $r_0$ in Eq.~\eqref{eq:NED_metric_A}. Once the common Coulomb and logarithmic terms are matched,
we fix $M_{\rm ADM}$ by requiring $r[f_i(r)-f(r)]\to0$ as $r\to0$.
At infinity, write $Z_i=\lim_{r\to\infty}r^2E_i(r)/q$. The same
coefficient appears in the electric potential and the metric,
\[
 f_i(r)=1-\frac{2M_{\rm ADM}}r+\frac{Z_iq^2}{r^2}+o(r^{-2}),\qquad
 U_i(r)=\frac{Z_i q}{r}+o(r^{-1}).
\]
The following two rows give the mass shift and the field coefficient
for $\zeta>0$. For $\zeta=0$ all seven reduce to RN with
$M_{\rm ADM}=M$ and $Z_i=1$.
\begin{center}
\begingroup\small\renewcommand{\arraystretch}{1.55}
\begin{tabular}{@{}l@{\quad}ccccccc@{}}
\toprule
 & C0 & C1 & C2 & C3 & C4 & C5 & C6\\
\midrule
$\displaystyle\frac{2(M_{\rm ADM}-M)}{A}-\ln\frac{R_c}{r_0}$ &
$\ln2-\tfrac13$ & $0$ & $\ln2$ & $0$ & $\ln2-\tfrac43$ &
$\tfrac\pi8+\tfrac34\ln2$ & $1$\\[0.5ex]
$\displaystyle\frac{\sqrt{\mathcal D_c}}{\zeta}(Z_i-1)$ &
$1$ & $4/3$ & $4/3$ & $0$ & $1/2$ & $4/3$ & $8/3$\\[0.4ex]
\bottomrule
\end{tabular}\endgroup
\end{center}
All seven have a finite ADM mass and a finite potential difference
between the horizon and infinity. C3 alone retains the bare Coulomb
coefficient $q^2$ at nonzero coupling. C2 and C5 have the same outer
field coefficient but different accuracy in the inner region. At
fixed $M,q,\zeta,R_c$ and $r\ll R_c$, the leading metric corrections
are $f_0-f\simeq3Ar/(4R_c^2)$, $f_2-f\simeq Ar/(4R_c^2)$, and
$f_5-f\simeq Ar^3/(16R_c^4)$. C5 thus preserves the inner logarithm
to the highest order among these seven choices.

\noindent\textit{Energy conditions, curvature, and stability.}
The mass derivatives above give $\rho_i>0$, $p_{ri}=-\rho_i$, and
$p_{\perp i}>0$ throughout $r>0$. The null, weak, and strong energy
conditions follow. The dominant condition also requires
$p_{\perp i}\leq\rho_i$. Six profiles satisfy it everywhere. For C3
the sign follows directly from the Ricci scalar,
\[
 \mathcal R_3(r)=16\pi(\rho_3-p_{\perp 3})
   =\frac{A R_c^2(R_c^2-r^2)}{r^3(r^2+R_c^2)^2}.
\]
For $A>0$ the tangential pressure equals the density at $R_c$ and
exceeds it for $r>R_c$. All seven charged centers are timelike and
singular. Equation~\eqref{eq:kretschmann} gives
$K_i\sim56q^4/r^8$ at the center and $K_i\sim48M_{\rm ADM}^2/r^6$
at infinity for a black hole. The horizons can be counted by solving
$f_i(r_h)=0$ for the required ADM mass. As a function of $r_h$ this
mass diverges at both ends and has second derivative
$8\pi r_h p_{\perp i}(r_h)>0$. Its minimum gives the extremal mass. There are two horizons
above this mass and one double horizon at equality. Below it the
charged singularity is exposed.

Outside a simple outer horizon we have $0<f_i<1$ and $f_i'>0$.
Indeed $m_i'>0$, and $m_i-rm_i'=r^2f_i'/2$ increases outward since
$m_i''<0$. The neutral scalar potential
$f_i[\ell(\ell+1)/r^2+f_i'/r+\mu^2]$ is therefore positive for
$\mu^2\geq0$. For coupled electromagnetic and gravitational
perturbations, the Moreno-Sarbach criterion~\cite{MorenoSarbach2003}
can be tested directly on the fields in the table. Since
$\mathcal D=q/r^2$, the logarithmic response of the field is
\[
 \frac{\partial\ln E_i}{\partial\ln\mathcal D}
 =-\frac{r E_i'(r)}{2E_i(r)} .
\]
With $\rho_i>0$ and $E_i/\mathcal D>0$, it suffices for the product of
this response with $f_i$ to lie in $(0,3]$. Differentiating the radial
fields gives a response in $(0,1]$ for C0, C1, C2, C4, C5, and C6,
and in $(0,3/2)$ for C3. The neutral scalar and the coupled Einstein-NED system therefore admit
no exponentially growing linear modes outside a nonextremal outer
horizon. This includes C3 even when its exterior contains a region
where the dominant energy condition fails.

Related solutions also arise in the magnetic
sector~\cite{Fan:2016hvf,GarciaSalcedo2024Rational,Kruglov2024Magnetic}.
To the best of our knowledge, C0, C2, and C4 appear to be
new in their realization as complete black-hole solutions with
the corresponding fixed-coupling electric NED actions.
We plan to investigate the QNM and QBS spectra of these
completions in future work.

\section{Discussion and conclusions}
\label{sec:conclusions}

We have explored how long-range deformations in a black-hole geometry
affect the oscillation, trapping, and decay of a massive field.
We chose an exact logarithmic deformation of the
Reissner-Nordstr\"om spacetime as a concrete example for these
questions. Its known RN limit allows us to follow the effects
of the deformation, while a neutral, minimally coupled test
scalar isolates the role of the geometry. We studied both QNMs,
which probe scattering and relaxation, and QBSs, which describe
waves trapped outside the horizon and gradually absorbed by it. Our neutral-scalar QNM and QBS results also apply to the
corresponding positive-density Maxwell-PFDM geometry, under
the parameter identification in
Sec.~\ref{sec:log_ned_geometry} and with the same exterior
geometry and boundary conditions.

Our analytical calculation explains why a small scalar mass can
reduce QNM damping. For nonnegative NED coupling and a simple
outer horizon, the potential barrier moves outward, becomes
higher, and flattens. Within first-order WKB, these changes
weaken damping at order $\mu^2$ for every fixed finite multipole
and overtone. This local result applies near the massless limit.
The behavior of the exact spectrum must be checked using global
methods~\cite{OhashiSakagami2004,KonoplyaZhidenko2005,Konoplya2002RN}
and with due attention to WKB applicability~\cite{Konoplya2019}.

We have compared analytical estimates with global matching,
a generalized Leaver method, and time-domain Prony analysis.
For the sampled fundamental $\ell=2$ mode, increasing the scalar
mass raises the oscillation frequency and lengthens the decay
time. Global matching and Leaver agree closely, sixth-order
WKB improves the leading estimate~\cite{Konoplya:2003ii}, and
the accepted Prony frequencies agree within their numerical
sensitivities. In the global calculations, we retained the
logarithmic corrections in the outer boundary conditions as
well as in the potential. The waveforms also illustrate why
a persistent signal alone cannot establish an undamped mode. 
Massive-field tails can contribute to the late-time
response~\cite{Qian:2022kaq}.

For QBSs, we have found that binding and lifetime must be assessed
separately in the analytic approaches. Global matching and Leaver independently confirm
the long-lived reference state in Eq.~\eqref{eq:qbs-prd-reference}.
For this state, the selected local Pad\'e approximation~\cite{Konoplya2026} gives a binding-energy error
of about $0.04\%$, whereas leading Gamow tunnelling
underestimates damping by about $25\%$, even when supplied
with the global real frequency. A better lifetime estimate
therefore requires a more accurate treatment of horizon
absorption. Increasing the approximation order also does not
always improve the result, as our Pad\'e comparison illustrates.
Across the sampled fundamental states, increasing the scalar
mass or NED coupling strengthens binding but shortens the
lifetime. The wider parameter scans are collected in
Appendix~\ref{app:systematic-scans}.
Using the regularized
bilinear framework of
Refs.~\cite{Cannizzaro2024,Cannizzaro2025}, we find numerical
support for orthogonality of an independently computed
QNM-QBS pair. 
These results motivate further study of how long-range geometry
affects scattering, trapping, and horizon absorption in other
spacetimes. Near-extremal modes require further
boundary analysis. Extensions include charged
fields~\cite{Hod2017QBS}, radiating charged
systems~\cite{Verma2026RadiatingCharges}, coupled gravitational
and electromagnetic perturbations, and modified-gravity
backgrounds~\cite{LuiChung2026QNM,Paul2024EsGB,LuiChungTorresOrjuela2026EMRI}. The seven completions in Sec.~\ref{subsec:finite_mass_completion}
give us interesting ways to change the far field. All seven have finite
ADM mass, retain the charged central singularity, and satisfy the
linear mode-stability test outside a nonextremal horizon. C3 is the
one case in which the tangential pressure exceeds the density beyond
the crossover radius. The frequencies  reported here all
refer to the original logarithmic geometry. In future work, we will
vary the crossover radius in the completed solutions and compare the
shifts in the radiative modes with those of the more extended trapped
states.

\section{Acknowledgments}
We acknowledge the support of the Department of Physics, IIT Kharagpur. We  also acknowledge the assistance of AI tools in preparing the Latex tabulation of the detailed parameter scans in Appendix D generated from Mathematica notebooks. These tables have been carefully checked by the authors.

\section{Data Availability}
Data supporting the findings of this study are available from the authors upon reasonable request.

\appendix
\section{Contour independence of the globally matched spectrum}
\label{app_contour}

\begin{figure*}[t]
  \centering
  \includegraphics[width=\textwidth]{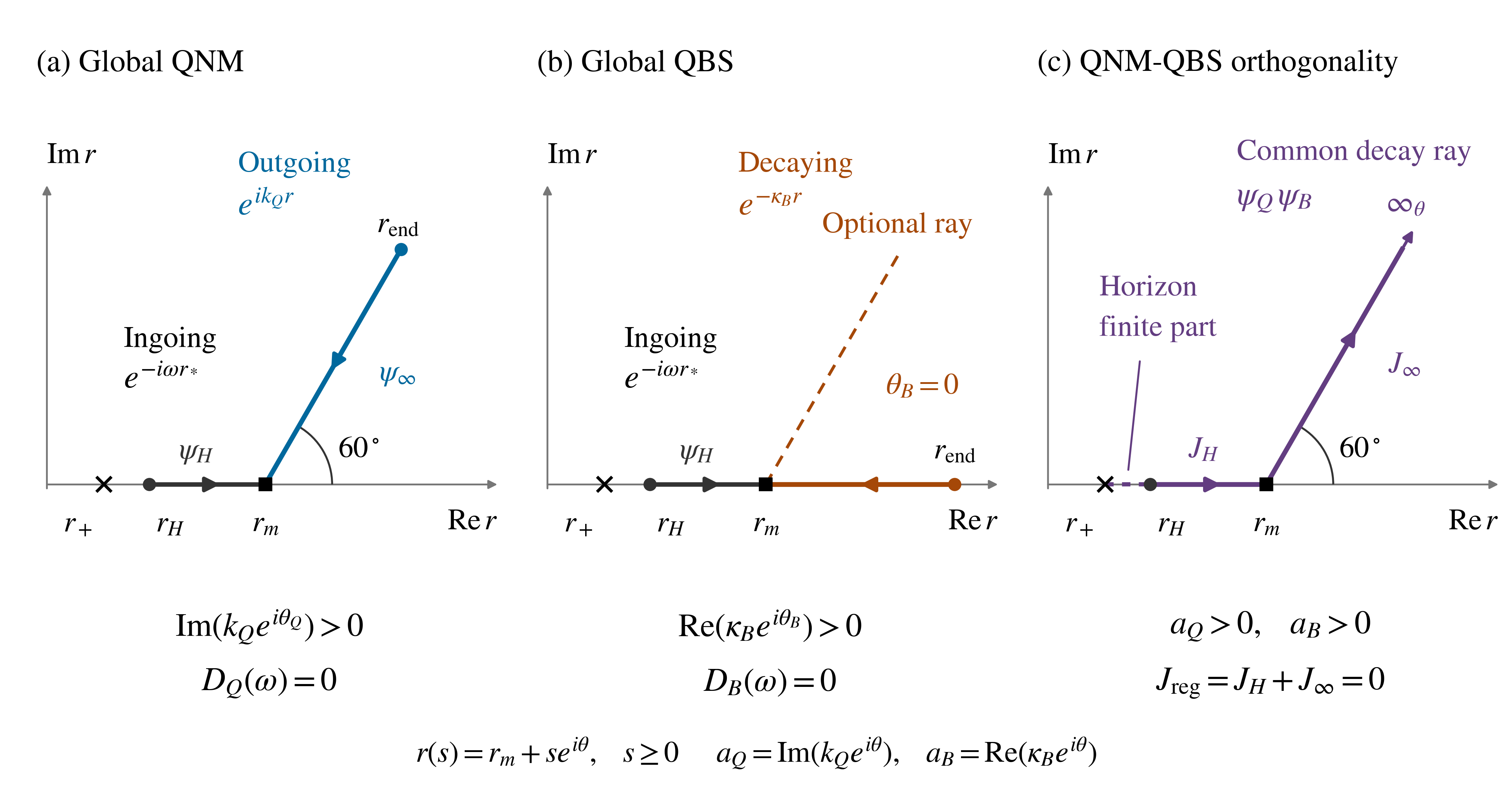}
  \caption{Schematic contours for (a) QNM matching, (b) QBS matching, and
  (c) the regularized QNM-QBS product. The horizon cutoff is
  $r_H=r_++\epsilon$, and squares mark $r_m$. In (a,b), arrows
  point toward the matching point; the dashed outer ray in (b)
  is optional. In (c), arrows follow the overlap integral along
  a ray where both modes decay; the dashed horizon segment
  denotes the finite-part contribution. Figures not to scale.}
  \label{fig:global-contours}
\end{figure*}

Here we explain why changing the outer integration contour leaves
the exact frequencies obtained in Secs.~\ref{subsec:logned-global}
and \ref{sec:qbs} unchanged. The matching point $r_m$ and the
selected QNM or QBS boundary condition are kept fixed.
Figure~\ref{fig:global-contours} shows the contours used for
matching and for the orthogonality calculation.

For a straight outer ray \footnote{ Derivatives are taken with respect to $r$; on a ray,
$\partial_r=e^{-i\theta}\partial_s$},
\begin{equation}
 r_\theta(s)=r_m+s e^{i\theta},\qquad s\geq0,
 \label{eq:contour-ray}
\end{equation}
the leading QNM and QBS factors, $e^{ikr}$ and $e^{-\kappa r}$,
respectively, decay when
\begin{equation}
 \begin{aligned}
 a_Q&\equiv\Im(k e^{i\theta})>0 &&\text{(QNM)},\\
 a_B&\equiv\Re(\kappa e^{i\theta})>0 &&\text{(QBS)}.
 \end{aligned}
 \label{eq:contour-decay-sector}
\end{equation}
The values of $k$ and $\kappa$ are those specified in the
matching calculation. The logarithmic corrections in
Eqs.~\eqref{eq:logned-far} and \eqref{eq:qbs-prd-tail} remain
part of the boundary data, but do not change these strict decay
conditions. Note that for a QBS, $\theta=0$ is already allowed, so matching
can be performed along the real axis.

We compare contours that can be continuously deformed into one
another without crossing $r=0$, a zero of $f$, or a branch cut.
The branches of $\ln(r/r_0)$, $k$, and $\kappa$ remain fixed,
and the outer directions stay within the same connected sector
where the selected solution decays. We exclude the threshold
$\omega^2=\mu^2$, where the outer asymptotic expansion must be
reconsidered.

The proof follows directly from the radial
equation~\eqref{eq:schrodinger}. For two solutions $u$ and $v$
at the same frequency,
\begin{equation}
 \mathcal W[u,v]=f(vu'-uv'),\qquad
 \frac{d\mathcal W[u,v]}{dr}=0.
 \label{eq:contour-equal-frequency-W}
\end{equation}
If both are decaying outer solutions in the chosen sector,
$\mathcal W\to0$ at infinity. Thus $\mathcal W=0$ everywhere,
and the solutions differ only by a constant factor (independent of r) which cancels out in the logarithmic drivative  $f\psi_\infty'/\psi_\infty$, while the horizon
solution is unchanged. Hence the mismatch defined in
Eqs.~\eqref{eq:logned-match} and \eqref{eq:qbs-prd-match} satisfies
\begin{equation}
 D_{\theta_1}(\omega)=D_{\theta_2}(\omega)
 \quad\Longrightarrow\quad
 \omega_n^{(\theta_1)}=\omega_n^{(\theta_2)}
 \label{eq:contour-isospectral}
\end{equation}
where $\theta_1$ and $\theta_2$ denote the angles of two different
allowed outer rays (contours). Thus, the matching condition selects the same frequencies
for both rays. This holds separately for the QNM and QBS
boundary conditions. 

Computed frequencies can still vary slightly with the ray angle
because of finite endpoints, truncated boundary expansions, and
integration errors. Convergence also becomes more difficult near
a boundary of the decay sector, where the relevant $a_Q$ or $a_B$
approaches zero. Varying the angle, endpoint, series order, and
integration tolerance checks numerical stability.

For the mixed bilinear in Sec.~\ref{sec:ortho}, both decay
conditions must hold on one common ray. The choice
$\theta=60^\circ$ satisfies them for the reference pair. 

\section{Horizon and far-zone expansions for global matching}
\label{app_series}

We now give the boundary expansions used in the global matching
calculations of Secs.~\ref{subsec:logned-global} and \ref{sec:qbs}.
We first recall the radial equation and boundary conditions, then
construct the ingoing solution near the horizon and the outgoing
solution at large radius. The main point is that the logarithmic
correction must be included in the far-zone boundary data.

\subsection{Radial equation and boundary conditions}

We consider a neutral, minimally coupled test scalar of mass
$\mu\geq0$ on the geometry
\begin{align}
 ds^2&=-f(r)dt^2+\frac{dr^2}{f(r)}+r^2d\Omega_2^2,
 \label{eq:uqnm_metric}\\
 f(r)&=1-\frac{2M+A\ln(r/r_0)}{r}+\frac{q^2}{r^2},
 &A&=\frac{4q^{3/2}\zeta}{3}.
 \label{eq:uqnm_f}
\end{align}
Here $q\geq0$ and $r_0>0$. When varying $A$, we keep $M$, $q$,
and $r_0$ fixed unless stated otherwise. We assume a nonextremal
outer horizon, with $f'_+>0$ and $f>0$ outside it. The horizon
expansion below does not apply at extremality.

We use $G=c=\hbar=1$: $M$, $q$, $A$, and $r_0$ have dimensions
of length, while $\mu$ and $\omega$ have dimensions of inverse
length. We also recall that the spherical mass function is
\begin{equation}
 m(r)=\frac r2(1-f)
 =M+\frac A2\ln\frac r{r_0}-\frac{q^2}{2r}.
 \label{eq:uqnm_mass_function}
\end{equation}
For $A>0$, this grows logarithmically even though $f\to1$.
We therefore treat $M$ as a parameter at fixed $r_0$, rather
than a finite standard ADM mass.

To keep the expressions compact, we write
\begin{equation}
 \begin{aligned}
 \Lambda&=\ell(\ell+1),&L&=\ell+\frac12,\\
 \mathcal L(r)&=\ln\frac r{r_0},&b(r)&=2M+A\mathcal L(r).
 \end{aligned}
 \label{eq:uqnm_notation}
\end{equation}
Primes denote derivatives with respect to $r$. The metric
derivatives we need are
\begin{align}
 f'&=\frac{b-A}{r^2}-\frac{2q^2}{r^3},
 \nonumber\\
 f''&=\frac{-2b+3A}{r^3}+\frac{6q^2}{r^4},
 \nonumber\\
 f'''&=\frac{6b-11A}{r^4}-\frac{24q^2}{r^5}.
 \label{eq:uqnm_fderivatives}
\end{align}

We separate the scalar field as
\begin{equation}
 \Phi=e^{-i\omega t}Y_{\ell m}(\theta,\phi)\frac{\Psi(r)}r,
 \qquad \omega=\omega_R+i\omega_I,
 \label{eq:uqnm_separation}
\end{equation}
which gives the radial wave equation
\begin{equation}
 \frac{d^2\Psi}{dr_*^2}+[\omega^2-V(r)]\Psi=0,
 \qquad \frac{dr_*}{dr}=\frac1f,
 \label{eq:uqnm_radial}
\end{equation}
with potential
\begin{align}
 V(r)&=f(r)\left[\mu^2+\frac{\Lambda}{r^2}+\frac{f'}r\right]
 \label{eq:uqnm_V}\\
 &=\frac{r^2-2Mr-Ar\mathcal L+q^2}{r^6}
 \nonumber\\
 &\quad\times\left[\mu^2r^4+\Lambda r^2+2Mr
 +Ar(\mathcal L-1)-2q^2\right].
 \label{eq:uqnm_Vfactorized}
\end{align}
For the boundary expansions, it is convenient to use $r$ directly:
\begin{equation}
 f^2\Psi''+ff'\Psi'+(\omega^2-V)\Psi=0.
 \label{eq:uqnm_radial_r}
\end{equation}

Both QNMs and QBSs are ingoing at the horizon,
\begin{equation}
 \Psi_{\rm H}\sim e^{-i\omega r_*}.
 \label{eq:uqnm_horizon_BC}
\end{equation}
Their outer boundary conditions differ. For QNMs, we select the
outgoing solution. Since $V\to\mu^2$ at infinity, its radial
momentum satisfies
\begin{equation}
 k_\infty^2=\omega^2-\mu^2.
 \label{eq:uqnm_kinfty}
\end{equation}
We fix the square-root by starting with real $\omega>\mu$
and $k_\infty>0$, then continuing analytically to the QNM
frequencies. In the positive-frequency radiative sector with
$\omega_I<0$, this gives $\Re k_\infty>0$ and
$\Im k_\infty<0$. The outgoing radial function then grows along
the real $r$ axis. This spatial growth is compatible with decay
in time, which is set by $e^{-i\omega t}$.

For QBSs, we instead require decay at large radius:
\begin{equation}
 \kappa=\sqrt{\mu^2-\omega^2},\qquad \Re\kappa>0,
 \qquad \Psi_{\rm dec}\sim e^{-\kappa r}
 \label{eq:uqnm_QBS_BC}
\end{equation}
at leading exponential order. This factor specifies the decaying
mode; the full outer solution also includes long-range
corrections, as in Sec.~\ref{sec:qbs}.

\subsection{Ingoing expansion near the horizon}

We begin just outside the horizon and write $x=r-r_+$. Expanding
the metric gives
\begin{equation}
 f=c_1x+c_2x^2+O(x^3),\qquad
 c_1=f'_+,\qquad c_2=\frac12f''_+,
\end{equation}
so $r_*\sim\ln x/c_1$. The ingoing condition therefore fixes
the leading power of the radial solution:
\begin{equation}
 \Psi_{\rm H}=x^\nu[1+a_1x+O(x^2)],
 \qquad \nu=-\frac{i\omega}{c_1},
 \label{eq:uqnm_Frobenius}
\end{equation}
Substituting this series into the radial equation gives the
first correction,
\begin{equation}
 a_1=\frac{\mu^2+\Lambda/r_+^2+c_1/r_+}
 {c_1(1+2\nu)}-\frac{c_2\nu}{c_1}.
 \label{eq:uqnm_a1}
\end{equation}
Using $f(r_+)=0$, we can write the metric coefficients as
\begin{equation}
 c_1=\frac{r_+^2-Ar_+-q^2}{r_+^3},\qquad
 c_2=\frac{-2r_+^2+3Ar_++4q^2}{2r_+^4}.
 \label{eq:uqnm_horizon_coefficients}
\end{equation}
Higher coefficients follow by continuing the same substitution.
We evaluate the truncated series a small distance outside $r_+$
to start the outward integration.

\subsection{Outgoing expansion at large radius}

We now construct the outer boundary data for the QNM calculation.
For a massive field, the factor $e^{ik_\infty r_*}$ does not
contain the full long-range phase. To find the missing terms,
we work with the logarithmic derivative,
\begin{equation}
 Y=\frac{\Psi'}\Psi,
 \qquad
 Y'+Y^2+\frac{f'}fY+\frac{\omega^2-V}{f^2}=0.
 \label{eq:uqnm_Riccati}
\end{equation}
This removes the arbitrary overall amplitude. For fixed
$k_\infty\ne0$, we expand the outgoing solution as
\begin{equation}
 Y_{\rm out}=ik_\infty+\frac{p_1(\mathcal L)}r
 +\frac{p_2(\mathcal L)}{r^2}
 +O\!\left(\frac{1+|\mathcal L|^3}{r^3}\right),
 \label{eq:uqnm_Yout}
\end{equation}
where the first two coefficients are
\begin{align}
 p_1&=\frac{i(2k_\infty^2+\mu^2)b}{2k_\infty},
 \label{eq:uqnm_p1}\\
 p_2&=\frac{\mu^2(b-A)}{4k_\infty^2}
 \nonumber\\
 &\quad+\frac{i}{2k_\infty}
 \bigg[
 \left(2k_\infty^2+\mu^2-\frac{\mu^4}{4k_\infty^2}\right)b^2
 \nonumber\\
 &\hspace{33mm}-(2k_\infty^2+\mu^2)q^2-\Lambda
 \bigg].
 \label{eq:uqnm_p2}
\end{align}
The coefficients depend on $r$ through
$\mathcal L=\ln(r/r_0)$. We therefore use
$d\mathcal L/dr=1/r$ when substituting the series into
Eq.~\eqref{eq:uqnm_Riccati}. Matching each power of $1/r$ gives
the coefficients in turn. For example, the $r^{-2}$ terms give
\begin{align}
 2ik_\infty p_2={}&-\big[\partial_{\mathcal L}p_1-p_1+p_1^2\big]
 -ik_\infty(b-A)
 \nonumber\\
 &-(3k_\infty^2+2\mu^2)b^2
 +(2k_\infty^2+\mu^2)q^2+\Lambda.
 \label{eq:uqnm_p2_recursion}
\end{align}
The expressions above satisfy the equation through order
$r^{-2}$, including all logarithmic terms at these orders.
If we write the same coefficients as polynomials,
$p_1=a_{11}\mathcal L+a_{10}$ and
$p_2=a_{22}\mathcal L^2+a_{21}\mathcal L+a_{20}$, the coefficient
of $\mathcal L$ in $p_2$ is
\begin{equation}
 \begin{aligned}
 a_{21}=\frac{i}{2k_\infty}\big[&
 (2a_{10}-1)a_{11}+iAk_\infty\\
 &+12AMk_\infty^2+8AM\mu^2\big].
 \end{aligned}
 \label{eq:uqnm_a21}
\end{equation}

To see what these corrections mean for the wave itself, we
integrate the first two terms of Eq.~\eqref{eq:uqnm_Yout}:
\begin{align}
 \ln\Psi_{\rm out}={}&ik_\infty r
 +i\frac{2\omega^2-\mu^2}{2k_\infty}
 \left[2M\mathcal L+\frac A2\mathcal L^2\right]
 \nonumber\\
 &+\mathrm{const}
 +O\!\left(\frac{1+|\mathcal L|^2}{r}\right).
 \label{eq:uqnm_outgoing_phase}
\end{align}
The logarithmic deformation produces the term proportional to
$\mathcal L^2$. We can compare this with the large-radius
tortoise coordinate,
\begin{equation}
 r_*=r+2M\mathcal L+\frac A2\mathcal L^2
 +\mathrm{const}
 +O\!\left(\frac{1+|\mathcal L|^2}{r}\right),
\end{equation}
to separate the usual $k_\infty r_*$ phase from the additional
massive-field contribution, proportional to
$\mu^2/(2k_\infty)$. This is why a plane-wave factor alone is
insufficient. We retain these long-range terms in the outer
boundary data used for global matching. The expansion assumes
fixed nonzero $k_\infty$ and the limit $k_\infty=0$
requires separate treatment.

\section{Scalar-mass corrections to the barrier-WKB frequency}
\label{app:uqnm_damping_sign}

We now explain the small-mass result of
Sec.~\ref{subsec:uqnm_mass_response}. We follow the barrier peak
as the scalar mass increases and use its height and curvature
to find the WKB frequency correction. The key step is to show
that the curvature correction is positive. This leads to the
longer decay time found in the main text.

We write $s=\mu^2$ and keep the background, multipole $\ell$,
and overtone $n$ fixed. The subscript $b$ denotes the massless
peak at $r_b$, while $p$ denotes the peak at $r_p(s)$.
As in the main text, $v_b$ is the massless peak height and
$k_b<0$ is its curvature in the tortoise coordinate.

\subsection{How scalar mass changes the peak}
\label{app:uqnm_peak_expansion}

We begin with a nondegenerate maximum, so its position changes
smoothly with $s$ by the implicit-function theorem. We establish
below that such a maximum exists and is unique under our horizon
assumptions. Differentiating the peak condition
$\mathcal V'(r_p)+s f'(r_p)=0$ gives
\begin{align}
 \frac{dr_p}{ds}
 &=-\frac{f'_p}{\mathcal V''(r_p)+s f''_p},
 \label{eq:uqnm_small_mass_peak}\\
 \frac{dV_p}{ds}&=f_p,\qquad
 K_p=f_p^2[\mathcal V''(r_p)+s f''_p].
 \label{eq:uqnm_peak_height_mass_derivative}
\end{align}
In the height derivative,
the contribution from this motion vanishes because the radial
derivative of the potential is zero at the peak. The quantity
$K_p$ is the curvature there, measured in the tortoise coordinate.

At $s=0$, these relations give the expansions in
Eq.~\eqref{eq:uqnm_mass_peak_summary}. Differentiating $K_p$,
including the motion of $r_p$, gives its leading correction,
$k_s=(dK_p/ds)_{s=0}$:
\begin{equation}
 k_s=f_b^2f''_b-2f_b(f'_b)^2
       -\frac{f_b^2f'_b\mathcal V'''_b}{\mathcal V''_b}.
 \label{eq:uqnm_small_mass_K}
\end{equation}
We then differentiate the squared WKB frequency in
Eq.~\eqref{eq:uqnm_mass_wkb_definition} to obtain
\begin{equation}
 2\omega^{\rm WKB}\frac{d\omega^{\rm WKB}}{ds}
 =f_p+\frac{i\alpha_n}{\sqrt{-2K_p}}\frac{dK_p}{ds}.
 \label{eq:massapp_squared_frequency_derivative}
\end{equation}
Evaluating this expression at $s=0$ gives the complex frequency
correction in Eq.~\eqref{eq:uqnm_small_mass_complex}.

\subsection{How the WKB frequency changes}
\label{app:uqnm_full_frequency_response}
\label{app:uqnm_real_mass_sign}
\label{app:mass_physical_criterion}

We next separate the change in oscillation frequency from the
change in damping. For the massless mode, the real and imaginary
parts of the WKB frequency satisfy
\begin{equation}
 \omega_{R,b}^2-\omega_{I,b}^2=v_b,\qquad
 -2\omega_{R,b}\omega_{I,b}=\alpha_n\sqrt{-2k_b}.
 \label{eq:massapp_square_root_parts}
\end{equation}
Using these relations in Eq.~\eqref{eq:uqnm_small_mass_complex},
we find the coefficients of the $\mu^2$ corrections:
\begin{align}
 C_{R,b}&=\frac{f_b\omega_{R,b}}{2|\omega_b^{\rm WKB}|^2}
 \left(1-\frac{k_s\omega_{I,b}^2}{(-k_b)f_b}\right),
 \label{eq:mass_physical_real_factor_appendix}\\
 C_{I,b}&=\frac{-f_b\omega_{I,b}}{2|\omega_b^{\rm WKB}|^2}
 \left(1+\frac{k_s\omega_{R,b}^2}{(-k_b)f_b}\right)>0.
 \label{eq:massapp_response_parts_appendix}
\end{align}
We prove $k_s>0$ below. Since $f_b>0$, $k_b<0$, and
$\omega_{I,b}<0$, both the prefactor and the bracket in the
second line are positive. Thus $C_{I,b}>0$: a small scalar mass
makes the imaginary frequency less negative, reduces the damping
rate, and increases the amplitude-decay time
$\tau=1/|\omega_I|$. This conclusion uses the full complex square
root in first-order WKB and does not require a weak-damping
approximation. The real-frequency correction has a separate sign
condition, set by the bracket in the first line.

\subsection{Why the curvature correction is positive}
\label{app:uqnm_exterior_barrier}

We now prove $k_s>0$. We assume $A\geq0$ and a simple outer
horizon, with $f(r_+)=0$ and $f'_+>0$, as in the main text.
The proof has two parts: we use the horizon conditions to bound
the quantities at the peak, then show that every factor in
$k_s$ has the required sign.

We start by dividing Eq.~\eqref{eq:uqnm_small_mass_K} by $f_bv_b$:
\begin{equation}
 \frac{k_s}{f_bv_b}
 =\frac{f_bf''_b}{v_b}-\frac{2(f'_b)^2}{v_b}
 -\frac{f_bf'_b}{v_b}\frac{\mathcal V'''_b}{\mathcal V''_b}.
 \label{eq:massapp_normalized_start}
\end{equation}
To make the signs easier to track, we introduce three
dimensionless ratios, with $\Lambda=\ell(\ell+1)$:
\begin{equation}
 j=\frac{\Lambda}{rf'},\qquad
 \chi=\frac{A/r+2q^2/r^2}{rf'},\qquad
 h=\frac{q^2}{r^3f'},
 \label{eq:massapp_chi}
\end{equation}
These ratios vary with $r$; we keep their radial dependence when
taking derivatives. In expressions for the peak, we evaluate
them at $r_b$. Using $v_b=f_bf'_b(j+1)/r_b$, we can write
\begin{equation}
 \frac{k_s}{f_bv_b}
 =\frac{1}{j+1}\left[
 \frac{r_bf''_b}{f'_b}-\frac{2r_bf'_b}{f_b}
 -\frac{r_b\mathcal V'''_b}{\mathcal V''_b}\right].
 \label{eq:massapp_dimensionless_response}
\end{equation}
We take the derivatives at fixed $\ell$ and only then impose
$\mathcal V'_b=0$. The result is
\begin{equation}
 \frac{k_s}{f_bv_b}=\frac{P}{(j+1)^2E},
 \label{eq:massapp_peak_factorization}
\end{equation}
where
\begin{align}
 E={}&(6-2\chi)(j+1)^2
 +[(\chi-2)^2+2h](j+1)\nonumber\\
 &+2(\chi-1)^2,
 \label{eq:massapp_E_explicit}\\
 P={}&6\chi^3-4\chi^2-20\chi+24+14h(\chi-1)\nonumber\\
 &+\bigl[3\chi^3+7\chi^2-56\chi+64
       +(18\chi-22)h\bigr]j\nonumber\\
 &+\bigl[\chi^3-3\chi^2-28\chi+48
       +4h(\chi-3)\bigr]j^2\nonumber\\
 &+(12-4\chi-2\chi^2-4h)j^3.
 \label{eq:massapp_polynomials}
\end{align}
This form isolates the sign question: once we show that
$f_b$, $v_b$, $E$, and $P$ are positive, $k_s>0$ follows
immediately.

\paragraph{What the horizon conditions imply.}
At the horizon, the metric gives
\begin{equation}
 r_+f'_+=1-\frac{A}{r_+}-\frac{q^2}{r_+^2}>0.
 \label{eq:massapp_horizon_metric}
\end{equation}
Outside it, we have
\begin{equation}
 r^2f'(r)=r_+^2f'_++A\ln\frac{r}{r_+}
 +2q^2\left(\frac{1}{r_+}-\frac{1}{r}\right)>0,
 \label{eq:massapp_exterior_fp}
\end{equation}
where the right-hand side is positive because $A\geq0$ and
$f'_+>0$. Thus $f'>0$, and $f$ increases from zero at the
horizon to one at infinity. Hence $0<f<1$ throughout the exterior.
In particular,
\begin{equation}
 f_b>0,\qquad
 v_b=f_b\left[\frac{\ell(\ell+1)}{r_b^2}
                  +\frac{f'_b}{r_b}\right]>0,
 \label{eq:massapp_positive_height}
\end{equation}
including $\ell=0$. The definitions in
Eq.~\eqref{eq:massapp_chi} give $j\geq0$ and
$\chi-2h=A/(r^2f')\geq0$, so $0\leq h\leq\chi/2$.
At any exterior stationary point, the condition
$\mathcal V'=0$ also gives
\begin{equation}
 \frac{f}{rf'}=\frac{j+1}{2j+3-\chi},
 \qquad 2j+3-\chi>0.
 \label{eq:massapp_stationary_ratio}
\end{equation}

\paragraph{Bounding $\chi$ at a stationary point.}
We still need an upper bound on $\chi$. To obtain it, we set
$\mathcal U=r^2f'$ and $y=\ln(r/r_+)$. The radial derivatives
satisfy
\[
 (r\partial_r)\mathcal U=\chi\mathcal U,
 \qquad (r\partial_r)^2\mathcal U=-\frac{2q^2}{r}\leq0,
\]
so $\mathcal U$ is concave as a function of logarithmic radius:
its graph lies below its tangent. Applying this fact between
the horizon and $r$ gives
\begin{equation}
 \frac{\mathcal U(re^{-u})}{\mathcal U(r)}
 \leq1-\chi u,\qquad 0\leq u\leq y.
 \label{eq:massapp_tangent_bound}
\end{equation}
Here $\chi$ is evaluated at the fixed radius $r$. At the horizon,
$u=y$ and $\mathcal U(r_+)>0$, so $\chi y<1$.

Suppose now that $\chi\geq4/3$ at a stationary point. Then
$y<3/4$. Integrating $f'$ from the horizon, with
$\rho=re^{-u}$, gives
\begin{align}
 \frac{f}{rf'}
 &=\int_0^y e^u\frac{\mathcal U(re^{-u})}{\mathcal U(r)}\,du
 \nonumber\\
 &\leq\int_0^y e^u(1-4u/3)\,du\nonumber\\
 &<\int_0^{3/4}e^u(1-4u/3)\,du
 =\frac{4e^{3/4}-7}{3}<\frac12.
 \label{eq:massapp_concavity_bound}
\end{align}
The final inequality uses $e^{3/4}<17/8$. But the stationary-point
condition, Eq.~\eqref{eq:massapp_stationary_ratio}, gives
\[
 \frac{f}{rf'}-\frac12
 =\frac{\chi-1}{2(2j+3-\chi)}>0,
\]
which requires the same ratio to be above $1/2$. This
contradiction rules out $\chi\geq4/3$. Thus every
exterior stationary point lies in the range
\begin{equation}
 j\geq0,\qquad 0\leq\chi<\frac43,\qquad
 0\leq h\leq\frac{\chi}{2}.
 \label{eq:massapp_peak_bound}
\end{equation}

\paragraph{Checking the signs of $E$ and $P$.}
Within this range, $6-2\chi>10/3$ and $j+1>0$. The first term
in $E$ is therefore positive, and its other terms are
nonnegative. Hence $E>0$.

This also tells us the shape of the massless barrier. Direct
differentiation at fixed $\ell$ gives, at a stationary point,
\begin{equation}
 \mathcal V''_b=-\frac{(f'_b)^2E}{r_b^2(2j+3-\chi)}<0.
 \label{eq:massapp_stationary_derivatives}
\end{equation}
We obtain this identity without dividing by $\mathcal V''_b$,
so it establishes that the curvature is strictly negative.
Every exterior stationary point is therefore a nondegenerate
maximum. Since $\mathcal V$ is positive outside the horizon
and tends to zero at both ends, at least one maximum exists.
Two maxima would require a stationary point between them that
is not a strict maximum. The exterior massless peak is therefore
unique, and the smooth-peak expansion used above is justified.

We check $P$ by treating it as a polynomial in $j$. Each
coefficient is linear in $h$, so its minimum at fixed $\chi$
occurs at $h=0$ or $h=\chi/2$. At both endpoints, the
coefficients of $j$, $j^2$, and $j^3$ decrease as $\chi$
runs from zero to $4/3$. Evaluating them at $\chi=4/3$ gives
the lower bounds $80/9$, $88/27$, and $4/9$, respectively.
For the constant coefficient, we find
\begin{align}
 P\big|_{j=h=0}
 &=2(\chi+2)\left[3\left(\frac43-\chi\right)^2+\frac23\right]
 \geq\frac83,\nonumber\\
 P\big|_{j=0,\,h=\chi/2}
 &=3\bigl[(\chi-1)^2(2\chi+5)+3-\chi\bigr]\geq5.
 \label{eq:massapp_constant_coefficient_bounds}
\end{align}
Since $j\geq0$, these bounds give
\begin{equation}
 P\geq\frac83+\frac{80}{9}j+\frac{88}{27}j^2+\frac49j^3>0.
 \label{eq:massapp_coefficient_bounds}
\end{equation}
All the required factors are now positive, so
Eq.~\eqref{eq:massapp_peak_factorization} yields
\begin{equation}
 k_s=\frac{f_bv_bP}{(j+1)^2E}>0.
 \label{eq:massapp_ks_positive}
\end{equation}

We have therefore shown that $k_s>0$ for every finite
$\ell\geq0$ in the nonextremal $A\geq0$ family, including
the Reissner-Nordstr\"om and Schwarzschild limits. Because
$k_b<0$, this positive correction makes the peak curvature
less negative: the barrier becomes less sharply curved in the
tortoise coordinate. Through
Eq.~\eqref{eq:massapp_response_parts_appendix}, this gives
$C_{I,b}>0$ and a longer amplitude-decay time at order $\mu^2$
for each fixed finite overtone. Finally, note that the conclusion only holds for
small-mass expansion of the first-order barrier-WKB frequency, while arbitrary mass results follow from the full numerics described in the main text.

\section{Systematic parameter scans of quasinormal and quasibound modes}
\label{app:systematic-scans}

We now present broader QNM and QBS scans that extend the selected
comparisons in the main text. We set $M=r_0=1$, use the time
dependence $e^{-i\omega t}$, and report the frequencies as
$M\omega$. We write the positive amplitude damping rate as
$M\Gamma=-M\operatorname{Im}\omega$. For simplicity, the table headings use $q$, $\zeta$, and $\mu$
for the dimensionless quantities $q/M$, $\zeta\sqrt M$, and
$M\mu$, respectively. Changing $q$ at fixed $\zeta$ also changes
the logarithmic correction through $A=4q^{3/2}\zeta/3$.
Setting $q=0$ gives Schwarzschild, while $\zeta=0$ gives RN.

We calculate the QBSs separately, imposing ingoing waves at the
horizon and spatial decay at infinity. We vary the scalar mass,
NED coupling, charge, multipole, and radial level, including
the reference states discussed in Sec.~\ref{sec:qbs}.
The tables compare global matching and Leaver with Bohr-Sommerfeld
(BS), two Pad\'e approximations, and the first-order RN expansion
(RN1). We include Leaver results for all 67 distinct QBS states.

\subsection{Quasinormal-mode tables}

We include all 61 points from the fundamental-mode mass scan
in the main text. We also consider the first two overtones,
multipoles up to $\ell=5$, charges up to $q/M=0.9$, and
couplings up to $\zeta\sqrt M=0.2$.

The last column gives
$\Delta_{LG}=|M\omega_L-M\omega_g|$, the difference between
the Leaver and global-matching frequencies before rounding.
The printed frequencies may therefore agree even when
$\Delta_{LG}$ is nonzero.

A dagger ($\dagger$) marks a WKB estimate outside our adopted
applicability range or a Prony estimate that fails our
reliability checks. A dash ($\text{---}^{\dagger}$) means that
we could not reliably resolve the Prony frequency (please see main text for our acceptance conditions). Unmarked
Prony entries pass these checks.

\begin{table*}[!htbp]
\centering\scriptsize
\caption{Fundamental QNM mass scan: $q=0.5$, $\zeta=0.05$, $\ell=2$, $n=0$ (part 1 of 3). $\Delta_{LG}=|M\omega_L-M\omega_g|$; WKB daggers mark  applicability limits. Prony daggers mark failed reliability under our acceptance criterion ; $\text{---}^{\dagger}$ means unresolved (see the text).}
\label{tab:scan-qnm-mass_fundamental-1}
\setlength{\tabcolsep}{1.8pt}
\renewcommand{\arraystretch}{1.12}
\begin{tabular}{rrrrrr}
\toprule
$\mu$ & $M\omega_6$ & $M\omega_g$ & $M\omega_L$ & $M\omega_P$ & $\Delta_{LG}$\\
\midrule
0.000 & $0.4991602-0.0960485i$ & $0.4991618-0.0960418i$ & $0.4991618-0.0960418i$ & $0.499165-0.096040i$ & $4.7\times10^{-14}$\\
0.010 & $0.4991898-0.0960387i$ & $0.4991913-0.0960320i$ & $0.4991913-0.0960320i$ & $0.499195-0.096030i$ & $4.8\times10^{-14}$\\
0.020 & $0.4992783-0.0960093i$ & $0.4992799-0.0960027i$ & $0.4992799-0.0960027i$ & $0.499284-0.096000i$ & $4.7\times10^{-14}$\\
0.030 & $0.4994260-0.0959604i$ & $0.4994276-0.0959537i$ & $0.4994276-0.0959537i$ & $0.499431-0.095951i$ & $4.7\times10^{-14}$\\
0.040 & $0.4996328-0.0958918i$ & $0.4996343-0.0958851i$ & $0.4996343-0.0958851i$ & $0.499638-0.095883i$ & $4.8\times10^{-14}$\\
0.050 & $0.4998987-0.0958036i$ & $0.4999002-0.0957969i$ & $0.4999002-0.0957969i$ & $0.499904-0.095795i$ & $4.7\times10^{-14}$\\
0.060 & $0.5002237-0.0956957i$ & $0.5002253-0.0956891i$ & $0.5002253-0.0956891i$ & $0.500229-0.095687i$ & $4.7\times10^{-14}$\\
0.070 & $0.5006080-0.0955682i$ & $0.5006095-0.0955615i$ & $0.5006095-0.0955615i$ & $0.500613-0.095559i$ & $4.8\times10^{-14}$\\
0.080 & $0.5010515-0.0954209i$ & $0.5010530-0.0954142i$ & $0.5010530-0.0954142i$ & $0.501057-0.095412i$ & $4.7\times10^{-14}$\\
0.090 & $0.5015543-0.0952538i$ & $0.5015559-0.0952472i$ & $0.5015559-0.0952472i$ & $0.501560-0.095245i$ & $4.8\times10^{-14}$\\
0.100 & $0.5021165-0.0950669i$ & $0.5021181-0.0950603i$ & $0.5021181-0.0950603i$ & $0.502122-0.095058i$ & $4.6\times10^{-14}$\\
0.110 & $0.5027382-0.0948602i$ & $0.5027397-0.0948535i$ & $0.5027397-0.0948535i$ & $0.502744-0.094851i$ & $4.7\times10^{-14}$\\
0.120 & $0.5034194-0.0946334i$ & $0.5034209-0.0946268i$ & $0.5034209-0.0946268i$ & $0.503425-0.094624i$ & $4.6\times10^{-14}$\\
0.130 & $0.5041602-0.0943866i$ & $0.5041618-0.0943800i$ & $0.5041618-0.0943800i$ & $0.504166-0.094378i$ & $4.6\times10^{-14}$\\
0.140 & $0.5049607-0.0941197i$ & $0.5049623-0.0941131i$ & $0.5049623-0.0941131i$ & $0.504966-0.094111i$ & $3.8\times10^{-14}$\\
0.150 & $0.5058211-0.0938326i$ & $0.5058227-0.0938259i$ & $0.5058227-0.0938259i$ & $0.505827-0.093824i$ & $3.9\times10^{-14}$\\
0.160 & $0.5067415-0.0935251i$ & $0.5067431-0.0935185i$ & $0.5067431-0.0935185i$ & $0.506747-0.093516i$ & $4.1\times10^{-14}$\\
0.170 & $0.5077219-0.0931973i$ & $0.5077235-0.0931907i$ & $0.5077235-0.0931907i$ & $0.507727-0.093188i$ & $4.2\times10^{-14}$\\
0.180 & $0.5087626-0.0928489i$ & $0.5087642-0.0928423i$ & $0.5087642-0.0928423i$ & $0.508768-0.092840i$ & $4.2\times10^{-14}$\\
0.190 & $0.5098636-0.0924799i$ & $0.5098652-0.0924733i$ & $0.5098652-0.0924733i$ & $0.509869-0.092471i$ & $4.3\times10^{-14}$\\
0.200 & $0.5110252-0.0920901i$ & $0.5110268-0.0920835i$ & $0.5110268-0.0920835i$ & $0.511031-0.092081i$ & $4.3\times10^{-14}$\\
\bottomrule
\end{tabular}
\end{table*}
\begin{table*}[!htbp]
\centering\scriptsize
\caption{Fundamental QNM mass scan: $q=0.5$, $\zeta=0.05$, $\ell=2$, $n=0$ (part 2 of 3). $\Delta_{LG}=|M\omega_L-M\omega_g|$; WKB daggers mark  applicability limits. Prony daggers mark failed reliability under our acceptance criterion; $\text{---}^{\dagger}$ means unresolved (see the text).}
\label{tab:scan-qnm-mass_fundamental-2}
\setlength{\tabcolsep}{1.8pt}
\renewcommand{\arraystretch}{1.12}
\begin{tabular}{rrrrrr}
\toprule
$\mu$ & $M\omega_6$ & $M\omega_g$ & $M\omega_L$ & $M\omega_P$ & $\Delta_{LG}$\\
\midrule
0.210 & $0.5122474-0.0916794i$ & $0.5122490-0.0916728i$ & $0.5122490-0.0916728i$ & $0.512253-0.091670i$ & $4.3\times10^{-14}$\\
0.220 & $0.5135305-0.0912477i$ & $0.5135321-0.0912411i$ & $0.5135321-0.0912411i$ & $0.513536-0.091238i$ & $4.3\times10^{-14}$\\
0.230 & $0.5148746-0.0907946i$ & $0.5148762-0.0907881i$ & $0.5148762-0.0907881i$ & $0.514880-0.090785i$ & $4.1\times10^{-14}$\\
0.240 & $0.5162799-0.0903202i$ & $0.5162816-0.0903137i$ & $0.5162816-0.0903137i$ & $0.516286-0.090310i$ & $3.8\times10^{-14}$\\
0.250 & $0.5177467-0.0898243i$ & $0.5177483-0.0898177i$ & $0.5177483-0.0898177i$ & $0.517753-0.089814i$ & $3.8\times10^{-14}$\\
0.260 & $0.5192751-0.0893065i$ & $0.5192767-0.0892999i$ & $0.5192767-0.0892999i$ & $0.519281-0.089297i$ & $3.8\times10^{-14}$\\
0.270 & $0.5208653-0.0887667i$ & $0.5208670-0.0887602i$ & $0.5208670-0.0887602i$ & $0.520871-0.088757i$ & $3.8\times10^{-14}$\\
0.280 & $0.5225176-0.0882048i$ & $0.5225193-0.0881983i$ & $0.5225193-0.0881983i$ & $0.522524-0.088195i$ & $3.9\times10^{-14}$\\
0.290 & $0.5242323-0.0876204i$ & $0.5242340-0.0876139i$ & $0.5242340-0.0876139i$ & $0.524239-0.087611i$ & $3.9\times10^{-14}$\\
0.300 & $0.5260095-0.0870134i$ & $0.5260112-0.0870069i$ & $0.5260112-0.0870069i$ & $0.526017-0.087003i$ & $3.8\times10^{-14}$\\
0.310 & $0.5278496-0.0863834i$ & $0.5278513-0.0863769i$ & $0.5278513-0.0863769i$ & $0.527857-0.086371i$ & $3.6\times10^{-14}$\\
0.320 & $0.5297527-0.0857302i$ & $0.5297544-0.0857237i$ & $0.5297544-0.0857237i$ & $0.529761-0.085720i$ & $3.6\times10^{-14}$\\
0.330 & $0.5317193-0.0850535i$ & $0.5317210-0.0850470i$ & $0.5317210-0.0850470i$ & $0.531726-0.085044i$ & $3.6\times10^{-14}$\\
0.340 & $0.5337496-0.0843529i$ & $0.5337513-0.0843465i$ & $0.5337513-0.0843465i$ & $0.533758-0.084344i$ & $3.4\times10^{-14}$\\
0.350 & $0.5358438-0.0836282i$ & $0.5358455-0.0836218i$ & $0.5358455-0.0836218i$ & $0.535852-0.083620i$ & $3.5\times10^{-14}$\\
0.360 & $0.5380024-0.0828790i$ & $0.5380041-0.0828727i$ & $0.5380041-0.0828727i$ & $0.538010-0.082871i$ & $3.3\times10^{-14}$\\
0.370 & $0.5402256-0.0821049i$ & $0.5402273-0.0820986i$ & $0.5402273-0.0820986i$ & $0.540233-0.082097i$ & $3.3\times10^{-14}$\\
0.380 & $0.5425137-0.0813056i$ & $0.5425155-0.0812993i$ & $0.5425155-0.0812993i$ & $0.542522-0.081298i$ & $3.2\times10^{-14}$\\
0.390 & $0.5448672-0.0804805i$ & $0.5448690-0.0804742i$ & $0.5448690-0.0804742i$ & $0.544874-0.080472i$ & $3.2\times10^{-14}$\\
0.400 & $0.5472864-0.0796293i$ & $0.5472881-0.0796230i$ & $0.5472881-0.0796230i$ & $0.547292-0.079621i$ & $3.1\times10^{-14}$\\
\bottomrule
\end{tabular}
\end{table*}
\begin{table*}[!htbp]
\centering\scriptsize
\caption{Fundamental QNM mass scan: $q=0.5$, $\zeta=0.05$, $\ell=2$, $n=0$ (part 3 of 3). $\Delta_{LG}=|M\omega_L-M\omega_g|$; WKB daggers mark  applicability limits. Prony daggers mark failed reliability under our acceptance criterion; $\text{---}^{\dagger}$ means unresolved (see the text).}
\label{tab:scan-qnm-mass_fundamental-3}
\setlength{\tabcolsep}{1.8pt}
\renewcommand{\arraystretch}{1.12}
\begin{tabular}{rrrrrr}
\toprule
$\mu$ & $M\omega_6$ & $M\omega_g$ & $M\omega_L$ & $M\omega_P$ & $\Delta_{LG}$\\
\midrule
0.410 & $0.5497716-0.0787514i$ & $0.5497734-0.0787452i$ & $0.5497734-0.0787452i$ & $0.549775-0.078743i$ & $2.9\times10^{-14}$\\
0.420 & $0.5523233-0.0778464i$ & $0.5523250-0.0778402i$ & $0.5523250-0.0778402i$ & $0.552326-0.077837i$ & $3.0\times10^{-14}$\\
0.430 & $0.5549417-0.0769138i$ & $0.5549435-0.0769076i$ & $0.5549435-0.0769076i$ & $0.554949-0.076904i$ & $3.0\times10^{-14}$\\
0.440 & $0.5576274-0.0759528i$ & $0.5576292-0.0759467i$ & $0.5576292-0.0759467i$ & $(0.557628-0.075944i)^{\dagger}$ & $2.9\times10^{-14}$\\
0.450 & $0.5603807-0.0749630i$ & $0.5603825-0.0749569i$ & $0.5603825-0.0749569i$ & $(0.560364-0.074951i)^{\dagger}$ & $2.9\times10^{-14}$\\
0.460 & $0.5632020-0.0739437i$ & $0.5632038-0.0739376i$ & $0.5632038-0.0739376i$ & $(0.563199-0.073933i)^{\dagger}$ & $2.9\times10^{-14}$\\
0.470 & $0.5660918-0.0728941i$ & $0.5660936-0.0728881i$ & $0.5660936-0.0728881i$ & $(0.566089-0.072868i)^{\dagger}$ & $2.8\times10^{-14}$\\
0.480 & $0.5690504-0.0718136i$ & $0.5690522-0.0718076i$ & $0.5690522-0.0718076i$ & $(0.569060-0.071798i)^{\dagger}$ & $2.7\times10^{-14}$\\
0.490 & $0.5720782-0.0707012i$ & $0.5720801-0.0706953i$ & $0.5720801-0.0706953i$ & $(0.572100-0.070692i)^{\dagger}$ & $3.0\times10^{-14}$\\
0.500 & $0.5751758-0.0695562i$ & $0.5751776-0.0695504i$ & $0.5751776-0.0695504i$ & $(0.575224-0.069529i)^{\dagger}$ & $2.8\times10^{-14}$\\
0.510 & $0.5783433-0.0683777i$ & $0.5783452-0.0683721i$ & $0.5783452-0.0683721i$ & $(0.578429-0.068388i)^{\dagger}$ & $3.2\times10^{-14}$\\
0.520 & $0.5815813-0.0671647i$ & $0.5815833-0.0671592i$ & $0.5815833-0.0671592i$ & $(0.581847-0.067114i)^{\dagger}$ & $3.1\times10^{-14}$\\
0.530 & $0.5848901-0.0659161i$ & $0.5848922-0.0659108i$ & $0.5848922-0.0659108i$ & $(0.584998-0.065953i)^{\dagger}$ & $3.1\times10^{-14}$\\
0.540 & $0.5882701-0.0646307i$ & $0.5882723-0.0646258i$ & $0.5882723-0.0646258i$ & $(0.588304-0.064585i)^{\dagger}$ & $3.3\times10^{-14}$\\
0.550 & $0.5917214-0.0633074i$ & $0.5917239-0.0633031i$ & $0.5917239-0.0633031i$ & $(0.591695-0.063238i)^{\dagger}$ & $3.6\times10^{-14}$\\
0.560 & $0.5952442-0.0619448i$ & $0.5952474-0.0619414i$ & $0.5952474-0.0619414i$ & $(0.595193-0.061689i)^{\dagger}$ & $3.8\times10^{-14}$\\
0.570 & $0.5988387-0.0605413i$ & $0.5988429-0.0605395i$ & $0.5988429-0.0605395i$ & $(0.599644-0.058446i)^{\dagger}$ & $3.9\times10^{-14}$\\
0.580 & $0.6025047-0.0590952i$ & $0.6025106-0.0590958i$ & $0.6025106-0.0590958i$ & $(0.606061-0.057590i)^{\dagger}$ & $4.0\times10^{-14}$\\
0.590 & $0.6062415-0.0576044i$ & $0.6062507-0.0576091i$ & $0.6062507-0.0576091i$ & $(0.610972-0.056005i)^{\dagger}$ & $4.0\times10^{-14}$\\
0.600 & $0.6100479-0.0560659i$ & $0.6100633-0.0560778i$ & $0.6100633-0.0560778i$ & $(0.612698-0.049417i)^{\dagger}$ & $4.0\times10^{-14}$\\
\bottomrule
\end{tabular}
\end{table*}
\clearpage
\begin{table*}[!htbp]
\centering\scriptsize
\caption{QNM overtone mass scans: $q=0.5$, $\zeta=0.05$, $\ell=2$ (part 1 of 2). $\Delta_{LG}=|M\omega_L-M\omega_g|$; WKB daggers mark  applicability limits. Prony daggers mark failed reliability under our acceptance criterion; $\text{---}^{\dagger}$ means unresolved (see the text).}
\label{tab:scan-qnm-mass_overtones-1}
\setlength{\tabcolsep}{1.8pt}
\renewcommand{\arraystretch}{1.12}
\begin{tabular}{rrrrrrr}
\toprule
$n$ & $\mu$ & $M\omega_6$ & $M\omega_g$ & $M\omega_L$ & $M\omega_P$ & $\Delta_{LG}$\\
\midrule
1 & 0.000 & $0.4810428-0.2929730i$ & $0.4810434-0.2929507i$ & $0.4810434-0.2929507i$ & $(0.480967-0.293967i)^{\dagger}$ & $9.4\times10^{-13}$\\
1 & 0.050 & $0.4814395-0.2924357i$ & $0.4814400-0.2924134i$ & $0.4814400-0.2924134i$ & $(0.481445-0.292421i)^{\dagger}$ & $9.5\times10^{-13}$\\
1 & 0.100 & $0.4826272-0.2908203i$ & $0.4826276-0.2907977i$ & $0.4826276-0.2907977i$ & $(0.480637-0.291226i)^{\dagger}$ & $9.3\times10^{-13}$\\
1 & 0.150 & $0.4845993-0.2881155i$ & $0.4845995-0.2880928i$ & $0.4845995-0.2880928i$ & $(0.483491-0.287105i)^{\dagger}$ & $9.1\times10^{-13}$\\
1 & 0.200 & $0.4873449-0.2843031i$ & $0.4873436-0.2842803i$ & $0.4873436-0.2842803i$ & $(0.491308-0.283143i)^{\dagger}$ & $8.7\times10^{-13}$\\
1 & 0.250 & $0.4908494-0.2793590i$ & $0.4908411-0.2793347i$ & $0.4908411-0.2793347i$ & $(0.504538-0.283539i)^{\dagger}$ & $8.1\times10^{-13}$\\
1 & 0.300 & $0.4950956-0.2732551i$ & $0.4950644-0.2732234i$ & $0.4950644-0.2732234i$ & $(0.495500-0.275395i)^{\dagger}$ & $7.5\times10^{-13}$\\
1 & 0.350 & $0.5000640-0.2659631i$ & $0.4999742-0.2659079i$ & $0.4999742-0.2659079i$ & $(0.498687-0.268035i)^{\dagger}$ & $6.6\times10^{-13}$\\
1 & 0.400 & $0.5057331-0.2574604i$ & $0.5055156-0.2573461i$ & $0.5055156-0.2573461i$ & $\text{---}^{\dagger}$ & $6.1\times10^{-13}$\\
1 & 0.450 & $0.5120654-0.2477356i^{\dagger}$ & $0.5116138-0.2474965i$ & $0.5116138-0.2474965i$ & $\text{---}^{\dagger}$ & $4.9\times10^{-13}$\\
1 & 0.500 & $0.5188908-0.2367722i^{\dagger}$ & $0.5181704-0.2363268i$ & $0.5181704-0.2363268i$ & $\text{---}^{\dagger}$ & $3.4\times10^{-13}$\\
1 & 0.550 & $0.5249459-0.2243835i^{\dagger}$ & $0.5250625-0.2238264i$ & $0.5250625-0.2238264i$ & $\text{---}^{\dagger}$ & $2.7\times10^{-13}$\\
1 & 0.600 & $0.5178256-0.2093544i^{\dagger}$ & $0.5321505-0.2100228i$ & $0.5321505-0.2100228i$ & $\text{---}^{\dagger}$ & $3.0\times10^{-13}$\\
\bottomrule
\end{tabular}
\end{table*}
\begin{table*}[!htbp]
\centering\scriptsize
\caption{QNM overtone mass scans: $q=0.5$, $\zeta=0.05$, $\ell=2$ (part 2 of 2). $\Delta_{LG}=|M\omega_L-M\omega_g|$; WKB daggers mark  applicability limits. Prony daggers mark failed reliability under our acceptance criterion; $\text{---}^{\dagger}$ means unresolved (see the text).}
\label{tab:scan-qnm-mass_overtones-2}
\setlength{\tabcolsep}{1.8pt}
\renewcommand{\arraystretch}{1.12}
\begin{tabular}{rrrrrrr}
\toprule
$n$ & $\mu$ & $M\omega_6$ & $M\omega_g$ & $M\omega_L$ & $M\omega_P$ & $\Delta_{LG}$\\
\midrule
2 & 0.000 & $0.4503738-0.5027238i^{\dagger}$ & $0.4504442-0.5025957i$ & $0.4504442-0.5025957i$ & $\text{---}^{\dagger}$ & $1.2\times10^{-11}$\\
2 & 0.050 & $0.4504330-0.5022276i^{\dagger}$ & $0.4504956-0.5020978i$ & $0.4504956-0.5020978i$ & $\text{---}^{\dagger}$ & $1.2\times10^{-11}$\\
2 & 0.100 & $0.4506018-0.5007381i^{\dagger}$ & $0.4506457-0.5006054i$ & $0.4506457-0.5006054i$ & $\text{---}^{\dagger}$ & $1.2\times10^{-11}$\\
2 & 0.150 & $0.4508639-0.4982498i^{\dagger}$ & $0.4508824-0.4981226i$ & $0.4508824-0.4981226i$ & $\text{---}^{\dagger}$ & $1.2\times10^{-11}$\\
2 & 0.200 & $0.4512273-0.4947461i^{\dagger}$ & $0.4511857-0.4946571i$ & $0.4511857-0.4946571i$ & $\text{---}^{\dagger}$ & $1.5\times10^{-11}$\\
2 & 0.250 & $0.4517829-0.4901869i^{\dagger}$ & $0.4515283-0.4902213i$ & $0.4515283-0.4902213i$ & $\text{---}^{\dagger}$ & $1.9\times10^{-11}$\\
2 & 0.300 & $0.4528016-0.4844901i^{\dagger}$ & $0.4518762-0.4848345i$ & $0.4518762-0.4848345i$ & $\text{---}^{\dagger}$ & $3.1\times10^{-11}$\\
2 & 0.350 & $0.4548796-0.4775049i^{\dagger}$ & $0.4521902-0.4785245i$ & $0.4521902-0.4785245i$ & $\text{---}^{\dagger}$ & $4.1\times10^{-11}$\\
2 & 0.400 & $0.4591109-0.4689851i^{\dagger}$ & $0.4524291-0.4713302i$ & $0.4524291-0.4713302i$ & $\text{---}^{\dagger}$ & $6.1\times10^{-11}$\\
2 & 0.450 & $0.4670363-0.4586300i^{\dagger}$ & $0.4525527-0.4633027i$ & $0.4525527-0.4633027i$ & $\text{---}^{\dagger}$ & $7.3\times10^{-11}$\\
2 & 0.500 & $0.4786063-0.4467111i^{\dagger}$ & $0.4525272-0.4545059i$ & $0.4525272-0.4545059i$ & $\text{---}^{\dagger}$ & $8.9\times10^{-11}$\\
2 & 0.550 & $0.4749965-0.4399824i^{\dagger}$ & $0.4523306-0.4450145i$ & $0.4523306-0.4450145i$ & $\text{---}^{\dagger}$ & $1.6\times10^{-10}$\\
2 & 0.600 & $0.2729325-0.5709762i^{\dagger}$ & $0.4519575-0.4349099i$ & $0.4519575-0.4349099i$ & $\text{---}^{\dagger}$ & $2.2\times10^{-10}$\\
\bottomrule
\end{tabular}
\end{table*}
\clearpage
\begin{table*}[!htbp]
\centering\scriptsize
\caption{QNM multipole and overtone scans: $q=0.5$, $\zeta=0.05$ (part 1 of 2). $\Delta_{LG}=|M\omega_L-M\omega_g|$; WKB daggers mark  applicability limits. Prony daggers mark failed reliability under our acceptance criterion; $\text{---}^{\dagger}$ means unresolved (see the text).}
\label{tab:scan-qnm-multipoles-1}
\setlength{\tabcolsep}{1.8pt}
\renewcommand{\arraystretch}{1.12}
\begin{tabular}{rrrrrrrr}
\toprule
$\ell$ & $n$ & $\mu$ & $M\omega_6$ & $M\omega_g$ & $M\omega_L$ & $M\omega_P$ & $\Delta_{LG}$\\
\midrule
0 & 0 & 0.000 & $0.1140580-0.0999186i^{\dagger}$ & $0.1140106-0.1035928i$ & $0.1140106-0.1035928i$ & $0.114017-0.103587i$ & $3.1\times10^{-13}$\\
0 & 0 & 0.050 & $0.1146065-0.0979717i^{\dagger}$ & $0.1145806-0.1016565i$ & $0.1145806-0.1016565i$ & $(0.114785-0.101561i)^{\dagger}$ & $2.8\times10^{-13}$\\
0 & 0 & 0.100 & $0.1159613-0.0923264i^{\dagger}$ & $0.1160518-0.0959614i$ & $0.1160518-0.0959614i$ & $(0.116915-0.091831i)^{\dagger}$ & $1.8\times10^{-12}$\\
1 & 0 & 0.000 & $0.3023879-0.0969372i$ & $0.3023982-0.0968659i$ & $0.3023982-0.0968659i$ & $0.302399-0.096864i$ & $2.3\times10^{-14}$\\
1 & 0 & 0.100 & $0.3066128-0.0944785i$ & $0.3066238-0.0944074i$ & $0.3066238-0.0944074i$ & $0.306625-0.094406i$ & $6.1\times10^{-14}$\\
1 & 0 & 0.200 & $0.3193761-0.0868644i$ & $0.3193891-0.0867946i$ & $0.3193891-0.0867946i$ & $0.319388-0.086794i$ & $8.5\times10^{-14}$\\
1 & 1 & 0.000 & $0.2762913-0.3027697i^{\dagger}$ & $0.2762402-0.3026025i$ & $0.2762402-0.3026025i$ & $(0.268749-0.299548i)^{\dagger}$ & $5.7\times10^{-13}$\\
1 & 1 & 0.100 & $0.2766542-0.2995264i^{\dagger}$ & $0.2766052-0.2993649i$ & $0.2766052-0.2993649i$ & $\text{---}^{\dagger}$ & $5.7\times10^{-13}$\\
1 & 1 & 0.200 & $0.2777731-0.2896136i^{\dagger}$ & $0.2774257-0.2897709i$ & $0.2774257-0.2897709i$ & $\text{---}^{\dagger}$ & $5.7\times10^{-13}$\\
1 & 2 & 0.000 & $0.2453171-0.5328345i^{\dagger}$ & $0.2436575-0.5314269i$ & $0.2436575-0.5314269i$ & $\text{---}^{\dagger}$ & $4.9\times10^{-11}$\\
1 & 2 & 0.100 & $0.2446540-0.5309452i^{\dagger}$ & $0.2428523-0.5298853i$ & $0.2428523-0.5298853i$ & $\text{---}^{\dagger}$ & $4.8\times10^{-11}$\\
1 & 2 & 0.200 & $0.2417569-0.5180299i^{\dagger}$ & $0.2404936-0.5254076i$ & $0.2404936-0.5254076i$ & $\text{---}^{\dagger}$ & $6.5\times10^{-11}$\\
2 & 0 & 0.000 & $0.4991602-0.0960485i$ & $0.4991618-0.0960418i$ & $0.4991618-0.0960418i$ & $0.499165-0.096040i$ & $4.7\times10^{-14}$\\
2 & 0 & 0.100 & $0.5021165-0.0950669i$ & $0.5021181-0.0950603i$ & $0.5021181-0.0950603i$ & $0.502122-0.095058i$ & $4.6\times10^{-14}$\\
2 & 0 & 0.200 & $0.5110252-0.0920901i$ & $0.5110268-0.0920835i$ & $0.5110268-0.0920835i$ & $0.511031-0.092081i$ & $4.3\times10^{-14}$\\
2 & 1 & 0.000 & $0.4810428-0.2929730i$ & $0.4810434-0.2929507i$ & $0.4810434-0.2929507i$ & $(0.480967-0.293967i)^{\dagger}$ & $9.4\times10^{-13}$\\
2 & 1 & 0.100 & $0.4826272-0.2908203i$ & $0.4826276-0.2907977i$ & $0.4826276-0.2907977i$ & $(0.480637-0.291226i)^{\dagger}$ & $9.3\times10^{-13}$\\
2 & 1 & 0.200 & $0.4873449-0.2843031i$ & $0.4873436-0.2842803i$ & $0.4873436-0.2842803i$ & $(0.491308-0.283143i)^{\dagger}$ & $8.7\times10^{-13}$\\
2 & 2 & 0.000 & $0.4503738-0.5027238i^{\dagger}$ & $0.4504442-0.5025957i$ & $0.4504442-0.5025957i$ & $\text{---}^{\dagger}$ & $1.2\times10^{-11}$\\
2 & 2 & 0.100 & $0.4506018-0.5007381i^{\dagger}$ & $0.4506457-0.5006054i$ & $0.4506457-0.5006054i$ & $\text{---}^{\dagger}$ & $1.2\times10^{-11}$\\
2 & 2 & 0.200 & $0.4512273-0.4947461i^{\dagger}$ & $0.4511857-0.4946571i$ & $0.4511857-0.4946571i$ & $\text{---}^{\dagger}$ & $1.5\times10^{-11}$\\
3 & 0 & 0.000 & $0.6969910-0.0958067i$ & $0.6969911-0.0958057i$ & $0.6969911-0.0958057i$ & $0.697000-0.095802i$ & $1.5\times10^{-14}$\\
3 & 0 & 0.100 & $0.6991946-0.0952910i$ & $0.6991947-0.0952901i$ & $0.6991947-0.0952901i$ & $0.699204-0.095286i$ & $1.7\times10^{-14}$\\
3 & 0 & 0.200 & $0.7058215-0.0937358i$ & $0.7058217-0.0937349i$ & $0.7058217-0.0937349i$ & $0.705831-0.093731i$ & $8.2\times10^{-15}$\\
\bottomrule
\end{tabular}
\end{table*}
\begin{table*}[!htbp]
\centering\scriptsize
\caption{QNM multipole and overtone scans: $q=0.5$, $\zeta=0.05$ (part 2 of 2). $\Delta_{LG}=|M\omega_L-M\omega_g|$; WKB daggers mark  applicability limits. Prony daggers mark failed reliability under our acceptance criterion; $\text{---}^{\dagger}$ means unresolved (see the text).}
\label{tab:scan-qnm-multipoles-2}
\setlength{\tabcolsep}{1.8pt}
\renewcommand{\arraystretch}{1.12}
\begin{tabular}{rrrrrrrr}
\toprule
$\ell$ & $n$ & $\mu$ & $M\omega_6$ & $M\omega_g$ & $M\omega_L$ & $M\omega_P$ & $\Delta_{LG}$\\
\midrule
3 & 1 & 0.000 & $0.6835439-0.2899474i$ & $0.6835441-0.2899440i$ & $0.6835441-0.2899440i$ & $(0.683620-0.289936i)^{\dagger}$ & $1.3\times10^{-12}$\\
3 & 1 & 0.100 & $0.6851697-0.2886314i$ & $0.6851698-0.2886280i$ & $0.6851698-0.2886280i$ & $(0.685232-0.288554i)^{\dagger}$ & $1.3\times10^{-12}$\\
3 & 1 & 0.200 & $0.6900489-0.2846608i$ & $0.6900490-0.2846574i$ & $0.6900490-0.2846574i$ & $(0.690451-0.284837i)^{\dagger}$ & $1.1\times10^{-12}$\\
3 & 2 & 0.000 & $0.6587563-0.4912644i$ & $0.6587753-0.4912558i$ & $0.6587753-0.4912558i$ & $\text{---}^{\dagger}$ & $8.7\times10^{-11}$\\
3 & 2 & 0.100 & $0.6595678-0.4896672i$ & $0.6595848-0.4896573i$ & $0.6595848-0.4896573i$ & $\text{---}^{\dagger}$ & $7.0\times10^{-11}$\\
3 & 2 & 0.200 & $0.6619807-0.4848583i$ & $0.6619918-0.4848462i$ & $0.6619918-0.4848462i$ & $\text{---}^{\dagger}$ & $6.8\times10^{-11}$\\
4 & 0 & 0.000 & $0.8951657-0.0957077i$ & $0.8951658-0.0957075i$ & $0.8951658-0.0957075i$ & $0.895185-0.095701i$ & $4.9\times10^{-14}$\\
4 & 0 & 0.100 & $0.8969101-0.0953920i$ & $0.8969101-0.0953917i$ & $0.8969101-0.0953917i$ & $0.896930-0.095385i$ & $5.1\times10^{-14}$\\
4 & 0 & 0.200 & $0.9021510-0.0944417i$ & $0.9021510-0.0944415i$ & $0.9021510-0.0944415i$ & $0.902171-0.094435i$ & $4.3\times10^{-14}$\\
4 & 1 & 0.000 & $0.8845440-0.2886672i$ & $0.8845441-0.2886665i$ & $0.8845441-0.2886665i$ & $0.884489-0.288554i$ & $1.6\times10^{-12}$\\
4 & 1 & 0.100 & $0.8859995-0.2878087i$ & $0.8859996-0.2878079i$ & $0.8859996-0.2878079i$ & $(0.885907-0.287651i)^{\dagger}$ & $1.7\times10^{-12}$\\
4 & 1 & 0.200 & $0.8903702-0.2852240i$ & $0.8903702-0.2852233i$ & $0.8903702-0.2852233i$ & $(0.890419-0.285054i)^{\dagger}$ & $1.5\times10^{-12}$\\
4 & 2 & 0.000 & $0.8642993-0.4861192i$ & $0.8643039-0.4861191i$ & $0.8643039-0.4861191i$ & $\text{---}^{\dagger}$ & $9.4\times10^{-11}$\\
4 & 2 & 0.100 & $0.8652865-0.4849414i$ & $0.8652909-0.4849410i$ & $0.8652909-0.4849410i$ & $\text{---}^{\dagger}$ & $9.4\times10^{-11}$\\
4 & 2 & 0.200 & $0.8682439-0.4813967i$ & $0.8682476-0.4813956i$ & $0.8682476-0.4813956i$ & $\text{---}^{\dagger}$ & $8.5\times10^{-11}$\\
5 & 0 & 0.000 & $1.0934956-0.0956576i$ & $1.0934956-0.0956575i$ & $1.0934956-0.0956575i$ & $1.093530-0.095648i$ & $2.2\times10^{-14}$\\
5 & 0 & 0.100 & $1.0949355-0.0954449i$ & $1.0949356-0.0954449i$ & $1.0949356-0.0954449i$ & $1.094970-0.095436i$ & $1.6\times10^{-14}$\\
5 & 0 & 0.200 & $1.0992598-0.0948056i$ & $1.0992599-0.0948055i$ & $1.0992599-0.0948055i$ & $1.099295-0.094796i$ & $2.3\times10^{-14}$\\
5 & 1 & 0.000 & $1.0847375-0.2880115i$ & $1.0847375-0.2880113i$ & $1.0847375-0.2880113i$ & $1.084733-0.287983i$ & $2.3\times10^{-12}$\\
5 & 1 & 0.100 & $1.0860143-0.2874142i$ & $1.0860143-0.2874140i$ & $1.0860143-0.2874140i$ & $1.086034-0.287382i$ & $2.3\times10^{-12}$\\
5 & 1 & 0.200 & $1.0898479-0.2856183i$ & $1.0898479-0.2856181i$ & $1.0898479-0.2856181i$ & $(1.089834-0.285575i)^{\dagger}$ & $2.2\times10^{-12}$\\
5 & 2 & 0.000 & $1.0677689-0.4834215i$ & $1.0677703-0.4834219i$ & $1.0677703-0.4834219i$ & $(1.066770-0.482781i)^{\dagger}$ & $1.1\times10^{-10}$\\
5 & 2 & 0.100 & $1.0687621-0.4825484i$ & $1.0687634-0.4825486i$ & $1.0687634-0.4825486i$ & $(1.067524-0.480883i)^{\dagger}$ & $1.0\times10^{-10}$\\
5 & 2 & 0.200 & $1.0717415-0.4799228i$ & $1.0717426-0.4799229i$ & $1.0717426-0.4799229i$ & $\text{---}^{\dagger}$ & $1.0\times10^{-10}$\\
\bottomrule
\end{tabular}
\end{table*}
\clearpage
\begin{table*}[!htbp]
\centering\scriptsize
\caption{QNM charge scans: $\zeta=0.05$, $\ell=2$ (part 1 of 2). $\Delta_{LG}=|M\omega_L-M\omega_g|$; WKB daggers mark  applicability limits. Prony daggers mark failed reliability under our acceptance criterion; $\text{---}^{\dagger}$ means unresolved (see the text).}
\label{tab:scan-qnm-charge-1}
\setlength{\tabcolsep}{1.8pt}
\renewcommand{\arraystretch}{1.12}
\begin{tabular}{rrrrrrrr}
\toprule
$q$ & $n$ & $\mu$ & $M\omega_6$ & $M\omega_g$ & $M\omega_L$ & $M\omega_P$ & $\Delta_{LG}$\\
\midrule
0.000 & 0 & 0.000 & $0.4836419-0.0967661i$ & $0.4836439-0.0967588i$ & $0.4836439-0.0967588i$ & $0.483647-0.096757i$ & $3.8\times10^{-14}$\\
0.000 & 0 & 0.200 & $0.4963246-0.0923964i$ & $0.4963266-0.0923892i$ & $0.4963266-0.0923892i$ & $0.496330-0.092387i$ & $3.6\times10^{-14}$\\
0.000 & 1 & 0.000 & $0.4638467-0.2956269i$ & $0.4638506-0.2956039i$ & $0.4638506-0.2956039i$ & $(0.463700-0.295763i)^{\dagger}$ & $5.0\times10^{-13}$\\
0.000 & 1 & 0.200 & $0.4702865-0.2862014i$ & $0.4702863-0.2861769i$ & $0.4702863-0.2861769i$ & $(0.481913-0.287632i)^{\dagger}$ & $4.5\times10^{-13}$\\
0.000 & 2 & 0.000 & $0.4303858-0.5086999i^{\dagger}$ & $0.4305441-0.5085584i$ & $0.4305441-0.5085584i$ & $\text{---}^{\dagger}$ & $2.7\times10^{-12}$\\
0.000 & 2 & 0.200 & $0.4309889-0.5002975i^{\dagger}$ & $0.4309570-0.5002095i$ & $0.4309570-0.5002095i$ & $\text{---}^{\dagger}$ & $7.1\times10^{-12}$\\
0.200 & 0 & 0.000 & $0.4853194-0.0965078i$ & $0.4853213-0.0965006i$ & $0.4853213-0.0965006i$ & $0.485325-0.096498i$ & $2.7\times10^{-14}$\\
0.200 & 0 & 0.200 & $0.4978857-0.0921993i$ & $0.4978877-0.0921923i$ & $0.4978877-0.0921923i$ & $0.497891-0.092190i$ & $2.8\times10^{-14}$\\
0.200 & 1 & 0.000 & $0.4658219-0.2947625i$ & $0.4658252-0.2947400i$ & $0.4658252-0.2947400i$ & $(0.465693-0.294887i)^{\dagger}$ & $1.5\times10^{-13}$\\
0.200 & 1 & 0.200 & $0.4722483-0.2854464i$ & $0.4722479-0.2854225i$ & $0.4722479-0.2854225i$ & $(0.483690-0.286283i)^{\dagger}$ & $7.7\times10^{-14}$\\
0.200 & 2 & 0.000 & $0.4328549-0.5069827i^{\dagger}$ & $0.4329976-0.5068446i$ & $0.4329976-0.5068446i$ & $\text{---}^{\dagger}$ & $2.2\times10^{-11}$\\
0.200 & 2 & 0.200 & $0.4334972-0.4986359i^{\dagger}$ & $0.4334634-0.4985489i$ & $0.4334634-0.4985489i$ & $\text{---}^{\dagger}$ & $2.4\times10^{-11}$\\
0.400 & 0 & 0.000 & $0.4926857-0.0962190i$ & $0.4926874-0.0962121i$ & $0.4926874-0.0962121i$ & $0.492691-0.096210i$ & $2.1\times10^{-14}$\\
0.400 & 0 & 0.200 & $0.5048641-0.0921049i$ & $0.5048659-0.0920982i$ & $0.5048659-0.0920982i$ & $0.504870-0.092096i$ & $2.7\times10^{-14}$\\
0.400 & 1 & 0.000 & $0.4739641-0.2936649i$ & $0.4739659-0.2936428i$ & $0.4739659-0.2936428i$ & $(0.473892-0.294336i)^{\dagger}$ & $6.2\times10^{-14}$\\
0.400 & 1 & 0.200 & $0.4803240-0.2847058i$ & $0.4803231-0.2846828i$ & $0.4803231-0.2846828i$ & $(0.484775-0.284331i)^{\dagger}$ & $5.7\times10^{-14}$\\
0.400 & 2 & 0.000 & $0.4422905-0.5044354i^{\dagger}$ & $0.4423913-0.5043039i$ & $0.4423913-0.5043039i$ & $\text{---}^{\dagger}$ & $2.2\times10^{-11}$\\
0.400 & 2 & 0.200 & $0.4430482-0.4962895i^{\dagger}$ & $0.4430103-0.4962019i$ & $0.4430103-0.4962019i$ & $\text{---}^{\dagger}$ & $2.6\times10^{-11}$\\
0.500 & 0 & 0.000 & $0.4991602-0.0960485i$ & $0.4991618-0.0960418i$ & $0.4991618-0.0960418i$ & $0.499165-0.096040i$ & $4.7\times10^{-14}$\\
0.500 & 0 & 0.200 & $0.5110252-0.0920901i$ & $0.5110268-0.0920835i$ & $0.5110268-0.0920835i$ & $0.511031-0.092081i$ & $4.3\times10^{-14}$\\
0.500 & 1 & 0.000 & $0.4810428-0.2929730i$ & $0.4810434-0.2929507i$ & $0.4810434-0.2929507i$ & $(0.480967-0.293967i)^{\dagger}$ & $9.4\times10^{-13}$\\
0.500 & 1 & 0.200 & $0.4873449-0.2843031i$ & $0.4873436-0.2842803i$ & $0.4873436-0.2842803i$ & $(0.491308-0.283143i)^{\dagger}$ & $8.7\times10^{-13}$\\
0.500 & 2 & 0.000 & $0.4503738-0.5027238i^{\dagger}$ & $0.4504442-0.5025957i$ & $0.4504442-0.5025957i$ & $\text{---}^{\dagger}$ & $1.2\times10^{-11}$\\
0.500 & 2 & 0.200 & $0.4512273-0.4947461i^{\dagger}$ & $0.4511857-0.4946571i$ & $0.4511857-0.4946571i$ & $\text{---}^{\dagger}$ & $1.5\times10^{-11}$\\
\bottomrule
\end{tabular}
\end{table*}
\begin{table*}[!htbp]
\centering\scriptsize
\caption{QNM charge scans: $\zeta=0.05$, $\ell=2$ (part 2 of 2). $\Delta_{LG}=|M\omega_L-M\omega_g|$; WKB daggers mark  applicability limits. Prony daggers mark failed reliability under our acceptance criterion; $\text{---}^{\dagger}$ means unresolved (see the text).}
\label{tab:scan-qnm-charge-2}
\setlength{\tabcolsep}{1.8pt}
\renewcommand{\arraystretch}{1.12}
\begin{tabular}{rrrrrrrr}
\toprule
$q$ & $n$ & $\mu$ & $M\omega_6$ & $M\omega_g$ & $M\omega_L$ & $M\omega_P$ & $\Delta_{LG}$\\
\midrule
0.600 & 0 & 0.000 & $0.5080761-0.0957909i$ & $0.5080774-0.0957844i$ & $0.5080774-0.0957844i$ & $0.508081-0.095782i$ & $2.6\times10^{-14}$\\
0.600 & 0 & 0.200 & $0.5195308-0.0920366i$ & $0.5195322-0.0920301i$ & $0.5195322-0.0920301i$ & $0.519536-0.092028i$ & $2.5\times10^{-14}$\\
0.600 & 1 & 0.000 & $0.4907559-0.2919577i$ & $0.4907549-0.2919349i$ & $0.4907549-0.2919349i$ & $(0.490787-0.292367i)^{\dagger}$ & $2.2\times10^{-13}$\\
0.600 & 1 & 0.200 & $0.4969813-0.2836696i$ & $0.4969792-0.2836468i$ & $0.4969792-0.2836468i$ & $(0.495418-0.285047i)^{\dagger}$ & $1.3\times10^{-13}$\\
0.600 & 2 & 0.000 & $0.4614036-0.5002860i^{\dagger}$ & $0.4614381-0.5001612i$ & $0.4614381-0.5001612i$ & $\text{---}^{\dagger}$ & $4.5\times10^{-11}$\\
0.600 & 2 & 0.200 & $0.4623901-0.4925346i^{\dagger}$ & $0.4623434-0.4924438i$ & $0.4623434-0.4924438i$ & $\text{---}^{\dagger}$ & $4.9\times10^{-11}$\\
0.700 & 0 & 0.000 & $0.5201993-0.0953110i$ & $0.5202004-0.0953046i$ & $0.5202004-0.0953046i$ & $0.520204-0.095302i$ & $2.9\times10^{-14}$\\
0.700 & 0 & 0.200 & $0.5311245-0.0918223i$ & $0.5311256-0.0918161i$ & $0.5311256-0.0918161i$ & $0.531130-0.091814i$ & $3.0\times10^{-14}$\\
0.700 & 1 & 0.000 & $0.5039278-0.2901822i$ & $0.5039248-0.2901586i$ & $0.5039248-0.2901586i$ & $(0.504004-0.290308i)^{\dagger}$ & $1.0\times10^{-13}$\\
0.700 & 1 & 0.200 & $0.5100560-0.2823954i$ & $0.5100527-0.2823723i$ & $0.5100527-0.2823723i$ & $(0.508558-0.283336i)^{\dagger}$ & $1.2\times10^{-13}$\\
0.700 & 2 & 0.000 & $0.4762799-0.4963039i^{\dagger}$ & $0.4762745-0.4961835i$ & $0.4762745-0.4961835i$ & $\text{---}^{\dagger}$ & $3.9\times10^{-11}$\\
0.700 & 2 & 0.200 & $0.4774554-0.4888551i^{\dagger}$ & $0.4774022-0.4887640i$ & $0.4774022-0.4887640i$ & $\text{---}^{\dagger}$ & $4.3\times10^{-11}$\\
0.800 & 0 & 0.000 & $0.5368610-0.0942847i$ & $0.5368615-0.0942787i$ & $0.5368615-0.0942787i$ & $0.536866-0.094276i$ & $2.2\times10^{-14}$\\
0.800 & 0 & 0.200 & $0.5470975-0.0911491i$ & $0.5470981-0.0911432i$ & $0.5470981-0.0911432i$ & $0.547103-0.091141i$ & $2.6\times10^{-14}$\\
0.800 & 1 & 0.000 & $0.5219442-0.2866052i$ & $0.5219384-0.2865825i$ & $0.5219384-0.2865825i$ & $(0.522100-0.286771i)^{\dagger}$ & $1.0\times10^{-12}$\\
0.800 & 1 & 0.200 & $0.5279560-0.2794911i$ & $0.5279507-0.2794695i$ & $0.5279507-0.2794695i$ & $(0.526449-0.280506i)^{\dagger}$ & $8.6\times10^{-13}$\\
0.800 & 2 & 0.000 & $0.4964021-0.4888441i^{\dagger}$ & $0.4963585-0.4887378i$ & $0.4963585-0.4887378i$ & $\text{---}^{\dagger}$ & $1.1\times10^{-10}$\\
0.800 & 2 & 0.200 & $0.4978633-0.4818095i^{\dagger}$ & $0.4978068-0.4817288i$ & $0.4978068-0.4817288i$ & $\text{---}^{\dagger}$ & $1.2\times10^{-10}$\\
0.900 & 0 & 0.000 & $0.5606686-0.0917698i$ & $0.5606683-0.0917653i$ & $0.5606683-0.0917653i$ & $0.560673-0.091763i$ & $5.5\times10^{-14}$\\
0.900 & 0 & 0.200 & $0.5699774-0.0891373i$ & $0.5699770-0.0891330i$ & $0.5699770-0.0891330i$ & $0.569982-0.089130i$ & $3.9\times10^{-14}$\\
0.900 & 1 & 0.000 & $0.5471923-0.2782254i$ & $0.5471847-0.2782128i$ & $0.5471847-0.2782128i$ & $(0.547214-0.278361i)^{\dagger}$ & $4.2\times10^{-12}$\\
0.900 & 1 & 0.200 & $0.5530937-0.2720790i$ & $0.5530871-0.2720680i$ & $0.5530871-0.2720680i$ & $(0.554436-0.272708i)^{\dagger}$ & $3.5\times10^{-12}$\\
0.900 & 2 & 0.000 & $0.5234039-0.4724235i^{\dagger}$ & $0.5233499-0.4723675i$ & $0.5233499-0.4723675i$ & $\text{---}^{\dagger}$ & $2.8\times10^{-10}$\\
0.900 & 2 & 0.200 & $0.5253741-0.4659906i^{\dagger}$ & $0.5253401-0.4659581i$ & $0.5253401-0.4659581i$ & $\text{---}^{\dagger}$ & $2.4\times10^{-10}$\\
\bottomrule
\end{tabular}
\end{table*}
\clearpage
\begin{table*}[!htbp]
\centering\scriptsize
\caption{QNM coupling scans: $q=0.5$, $\ell=2$ (part 1 of 2). $\Delta_{LG}=|M\omega_L-M\omega_g|$; WKB daggers mark  applicability limits. Prony daggers mark failed reliability under our acceptance criterion; $\text{---}^{\dagger}$ means unresolved (see the text).}
\label{tab:scan-qnm-coupling-1}
\setlength{\tabcolsep}{1.8pt}
\renewcommand{\arraystretch}{1.12}
\begin{tabular}{rrrrrrrr}
\toprule
$\zeta$ & $n$ & $\mu$ & $M\omega_6$ & $M\omega_g$ & $M\omega_L$ & $M\omega_P$ & $\Delta_{LG}$\\
\midrule
0.000 & 0 & 0.000 & $0.5059663-0.0979492i$ & $0.5059680-0.0979421i$ & $0.5059680-0.0979421i$ & $0.505972-0.097940i$ & $3.2\times10^{-14}$\\
0.000 & 0 & 0.200 & $0.5177546-0.0939989i$ & $0.5177564-0.0939918i$ & $0.5177564-0.0939918i$ & $0.517760-0.093989i$ & $3.0\times10^{-14}$\\
0.000 & 1 & 0.000 & $0.4873058-0.2988412i$ & $0.4873066-0.2988174i$ & $0.4873066-0.2988174i$ & $(0.487281-0.299424i)^{\dagger}$ & $2.1\times10^{-13}$\\
0.000 & 1 & 0.200 & $0.4935296-0.2902101i$ & $0.4935284-0.2901857i$ & $0.4935284-0.2901857i$ & $(0.497866-0.288768i)^{\dagger}$ & $2.2\times10^{-13}$\\
0.000 & 2 & 0.000 & $0.4557563-0.5129963i^{\dagger}$ & $0.4558350-0.5128607i$ & $0.4558350-0.5128607i$ & $\text{---}^{\dagger}$ & $2.2\times10^{-11}$\\
0.000 & 2 & 0.200 & $0.4565655-0.5050924i^{\dagger}$ & $0.4565270-0.5049938i$ & $0.4565270-0.5049938i$ & $\text{---}^{\dagger}$ & $3.4\times10^{-11}$\\
0.010 & 0 & 0.000 & $0.5046018-0.0975665i$ & $0.5046034-0.0975594i$ & $0.5046034-0.0975594i$ & $0.504607-0.097557i$ & $4.3\times10^{-14}$\\
0.010 & 0 & 0.200 & $0.5164052-0.0936146i$ & $0.5164069-0.0936076i$ & $0.5164069-0.0936076i$ & $0.516411-0.093605i$ & $3.9\times10^{-14}$\\
0.010 & 1 & 0.000 & $0.4860511-0.2976592i$ & $0.4860519-0.2976358i$ & $0.4860519-0.2976358i$ & $(0.486039-0.298283i)^{\dagger}$ & $6.7\times10^{-13}$\\
0.010 & 1 & 0.200 & $0.4922905-0.2890205i$ & $0.4922892-0.2889965i$ & $0.4922892-0.2889965i$ & $(0.496552-0.287637i)^{\dagger}$ & $6.4\times10^{-13}$\\
0.010 & 2 & 0.000 & $0.4546795-0.5109265i^{\dagger}$ & $0.4547566-0.5107925i$ & $0.4547566-0.5107925i$ & $\text{---}^{\dagger}$ & $8.7\times10^{-12}$\\
0.010 & 2 & 0.200 & $0.4554976-0.5030080i^{\dagger}$ & $0.4554584-0.5029114i$ & $0.4554584-0.5029114i$ & $\text{---}^{\dagger}$ & $1.4\times10^{-11}$\\
0.025 & 0 & 0.000 & $0.5025581-0.0969948i$ & $0.5025597-0.0969879i$ & $0.5025597-0.0969879i$ & $0.502563-0.096986i$ & $4.3\times10^{-14}$\\
0.025 & 0 & 0.200 & $0.5143844-0.0930405i$ & $0.5143860-0.0930337i$ & $0.5143860-0.0930337i$ & $0.514390-0.093031i$ & $4.1\times10^{-14}$\\
0.025 & 1 & 0.000 & $0.4841711-0.2958941i$ & $0.4841717-0.2958711i$ & $0.4841717-0.2958711i$ & $(0.484156-0.296426i)^{\dagger}$ & $8.0\times10^{-13}$\\
0.025 & 1 & 0.200 & $0.4904338-0.2872438i$ & $0.4904326-0.2872203i$ & $0.4904326-0.2872203i$ & $(0.494637-0.285928i)^{\dagger}$ & $7.2\times10^{-13}$\\
0.025 & 2 & 0.000 & $0.4530647-0.5078362i^{\dagger}$ & $0.4531392-0.5077045i$ & $0.4531392-0.5077045i$ & $\text{---}^{\dagger}$ & $6.0\times10^{-12}$\\
0.025 & 2 & 0.200 & $0.4538959-0.4998956i^{\dagger}$ & $0.4538559-0.4998019i$ & $0.4538559-0.4998019i$ & $\text{---}^{\dagger}$ & $1.6\times10^{-11}$\\
0.050 & 0 & 0.000 & $0.4991602-0.0960485i$ & $0.4991618-0.0960418i$ & $0.4991618-0.0960418i$ & $0.499165-0.096040i$ & $4.7\times10^{-14}$\\
0.050 & 0 & 0.200 & $0.5110252-0.0920901i$ & $0.5110268-0.0920835i$ & $0.5110268-0.0920835i$ & $0.511031-0.092081i$ & $4.3\times10^{-14}$\\
0.050 & 1 & 0.000 & $0.4810428-0.2929730i$ & $0.4810434-0.2929507i$ & $0.4810434-0.2929507i$ & $(0.480967-0.293967i)^{\dagger}$ & $9.4\times10^{-13}$\\
0.050 & 1 & 0.200 & $0.4873449-0.2843031i$ & $0.4873436-0.2842803i$ & $0.4873436-0.2842803i$ & $(0.491308-0.283143i)^{\dagger}$ & $8.7\times10^{-13}$\\
0.050 & 2 & 0.000 & $0.4503738-0.5027238i^{\dagger}$ & $0.4504442-0.5025957i$ & $0.4504442-0.5025957i$ & $\text{---}^{\dagger}$ & $1.2\times10^{-11}$\\
0.050 & 2 & 0.200 & $0.4512273-0.4947461i^{\dagger}$ & $0.4511857-0.4946571i$ & $0.4511857-0.4946571i$ & $\text{---}^{\dagger}$ & $1.5\times10^{-11}$\\
\bottomrule
\end{tabular}
\end{table*}
\begin{table*}[!htbp]
\centering\scriptsize
\caption{QNM coupling scans: $q=0.5$, $\ell=2$ (part 2 of 2). $\Delta_{LG}=|M\omega_L-M\omega_g|$; WKB daggers mark  applicability limits. Prony daggers mark failed reliability under our acceptance criterion; $\text{---}^{\dagger}$ means unresolved (see the text).}
\label{tab:scan-qnm-coupling-2}
\setlength{\tabcolsep}{1.8pt}
\renewcommand{\arraystretch}{1.12}
\begin{tabular}{rrrrrrrr}
\toprule
$\zeta$ & $n$ & $\mu$ & $M\omega_6$ & $M\omega_g$ & $M\omega_L$ & $M\omega_P$ & $\Delta_{LG}$\\
\midrule
0.075 & 0 & 0.000 & $0.4957729-0.0951104i$ & $0.4957744-0.0951040i$ & $0.4957744-0.0951040i$ & $0.495778-0.095102i$ & $2.6\times10^{-14}$\\
0.075 & 0 & 0.200 & $0.5076773-0.0911477i$ & $0.5076788-0.0911413i$ & $0.5076788-0.0911413i$ & $0.507683-0.091139i$ & $2.7\times10^{-14}$\\
0.075 & 1 & 0.000 & $0.4779213-0.2900778i$ & $0.4779218-0.2900563i$ & $0.4779218-0.2900563i$ & $(0.477860-0.291164i)^{\dagger}$ & $2.5\times10^{-13}$\\
0.075 & 1 & 0.200 & $0.4842630-0.2813879i$ & $0.4842616-0.2813659i$ & $0.4842616-0.2813659i$ & $(0.482411-0.282859i)^{\dagger}$ & $1.5\times10^{-13}$\\
0.075 & 2 & 0.000 & $0.4476840-0.4976589i^{\dagger}$ & $0.4477505-0.4975344i$ & $0.4477505-0.4975344i$ & $\text{---}^{\dagger}$ & $3.3\times10^{-11}$\\
0.075 & 2 & 0.200 & $0.4485598-0.4896437i^{\dagger}$ & $0.4485169-0.4895593i$ & $0.4485169-0.4895593i$ & $\text{---}^{\dagger}$ & $3.2\times10^{-11}$\\
0.100 & 0 & 0.000 & $0.4923965-0.0941805i$ & $0.4923979-0.0941743i$ & $0.4923979-0.0941743i$ & $0.492401-0.094172i$ & $2.0\times10^{-14}$\\
0.100 & 0 & 0.200 & $0.5043409-0.0902131i$ & $0.5043424-0.0902070i$ & $0.5043424-0.0902070i$ & $0.504346-0.090205i$ & $2.5\times10^{-14}$\\
0.100 & 1 & 0.000 & $0.4748067-0.2872087i$ & $0.4748072-0.2871879i$ & $0.4748072-0.2871879i$ & $(0.474731-0.288319i)^{\dagger}$ & $5.7\times10^{-13}$\\
0.100 & 1 & 0.200 & $0.4811884-0.2784982i$ & $0.4811870-0.2784769i$ & $0.4811870-0.2784769i$ & $(0.479114-0.279886i)^{\dagger}$ & $4.3\times10^{-13}$\\
0.100 & 2 & 0.000 & $0.4449954-0.4926415i^{\dagger}$ & $0.4450582-0.4925204i$ & $0.4450582-0.4925204i$ & $\text{---}^{\dagger}$ & $4.3\times10^{-11}$\\
0.100 & 2 & 0.200 & $0.4458938-0.4845885i^{\dagger}$ & $0.4458496-0.4845086i$ & $0.4458496-0.4845086i$ & $\text{---}^{\dagger}$ & $4.2\times10^{-11}$\\
0.150 & 0 & 0.000 & $0.4856767-0.0923453i$ & $0.4856780-0.0923394i$ & $0.4856780-0.0923394i$ & $0.485681-0.092337i$ & $2.7\times10^{-14}$\\
0.150 & 0 & 0.200 & $0.4977035-0.0883679i$ & $0.4977049-0.0883621i$ & $0.4977049-0.0883621i$ & $0.497708-0.088360i$ & $2.9\times10^{-14}$\\
0.150 & 1 & 0.000 & $0.4685994-0.2815484i$ & $0.4685997-0.2815288i$ & $0.4685997-0.2815288i$ & $(0.468553-0.282476i)^{\dagger}$ & $3.2\times10^{-13}$\\
0.150 & 1 & 0.200 & $0.4750621-0.2727953i$ & $0.4750606-0.2727755i$ & $0.4750606-0.2727755i$ & $(0.472718-0.274064i)^{\dagger}$ & $3.0\times10^{-13}$\\
0.150 & 2 & 0.000 & $0.4396231-0.4827490i^{\dagger}$ & $0.4396788-0.4826344i$ & $0.4396788-0.4826344i$ & $\text{---}^{\dagger}$ & $7.0\times10^{-12}$\\
0.150 & 2 & 0.200 & $0.4405669-0.4746191i^{\dagger}$ & $0.4405206-0.4745479i$ & $0.4405206-0.4745479i$ & $\text{---}^{\dagger}$ & $1.3\times10^{-11}$\\
0.200 & 0 & 0.000 & $0.4790027-0.0905428i$ & $0.4790039-0.0905373i$ & $0.4790039-0.0905373i$ & $0.479007-0.090535i$ & $2.8\times10^{-14}$\\
0.200 & 0 & 0.200 & $0.4911148-0.0865543i$ & $0.4911161-0.0865489i$ & $0.4911161-0.0865489i$ & $0.491119-0.086547i$ & $2.9\times10^{-14}$\\
0.200 & 1 & 0.000 & $0.4624228-0.2759919i$ & $0.4624230-0.2759736i$ & $0.4624230-0.2759736i$ & $(0.460342-0.277738i)^{\dagger}$ & $5.5\times10^{-14}$\\
0.200 & 1 & 0.200 & $0.4689679-0.2671943i$ & $0.4689663-0.2671758i$ & $0.4689663-0.2671758i$ & $(0.466467-0.268192i)^{\dagger}$ & $6.6\times10^{-14}$\\
0.200 & 2 & 0.000 & $0.4342589-0.4730457i^{\dagger}$ & $0.4343081-0.4729372i$ & $0.4343081-0.4729372i$ & $\text{---}^{\dagger}$ & $1.6\times10^{-11}$\\
0.200 & 2 & 0.200 & $0.4352486-0.4648374i^{\dagger}$ & $0.4352007-0.4647745i$ & $0.4352007-0.4647745i$ & $\text{---}^{\dagger}$ & $1.6\times10^{-11}$\\
\bottomrule
\end{tabular}
\end{table*}
\clearpage
\begin{table*}[!htbp]
\centering\scriptsize
\caption{High mass fundamental QNM coupling scan: $q=0.5$, $\ell=2$, $n=0$, $\mu=0.6$. $\Delta_{LG}=|M\omega_L-M\omega_g|$; WKB daggers mark  applicability limits. Prony daggers mark failed reliability under our acceptance criterion; $\text{---}^{\dagger}$ means unresolved (see the text).}
\label{tab:scan-qnm-coupling_massive-1}
\setlength{\tabcolsep}{1.8pt}
\renewcommand{\arraystretch}{1.12}
\begin{tabular}{rrrrrr}
\toprule
$\zeta$ & $M\omega_6$ & $M\omega_g$ & $M\omega_L$ & $M\omega_P$ & $\Delta_{LG}$\\
\midrule
0.000 & $0.6160361-0.0582619i$ & $0.6160462-0.0582671i$ & $0.6160462-0.0582671i$ & $(0.621157-0.056282i)^{\dagger}$ & $3.7\times10^{-14}$\\
0.010 & $0.6148330-0.0578222i$ & $0.6148439-0.0578284i$ & $0.6148439-0.0578284i$ & $(0.618146-0.053628i)^{\dagger}$ & $3.7\times10^{-14}$\\
0.025 & $0.6130334-0.0571631i$ & $0.6130458-0.0571712i$ & $0.6130458-0.0571712i$ & $(0.616632-0.052720i)^{\dagger}$ & $3.9\times10^{-14}$\\
0.050 & $0.6100479-0.0560659i$ & $0.6100633-0.0560778i$ & $0.6100633-0.0560778i$ & $(0.612698-0.049417i)^{\dagger}$ & $4.0\times10^{-14}$\\
0.075 & $0.6070795-0.0549701i$ & $0.6070990-0.0549867i$ & $0.6070990-0.0549867i$ & $(0.611737-0.048469i)^{\dagger}$ & $3.5\times10^{-14}$\\
0.100 & $0.6041281-0.0538753i$ & $0.6041532-0.0538978i$ & $0.6041532-0.0538978i$ & $(0.609826-0.047280i)^{\dagger}$ & $3.4\times10^{-14}$\\
0.150 & $0.5982755-0.0516858i^{\dagger}$ & $0.5983184-0.0517260i$ & $0.5983184-0.0517260i$ & $(0.607422-0.048672i)^{\dagger}$ & $3.6\times10^{-14}$\\
0.200 & $0.5924846-0.0494910i^{\dagger}$ & $0.5925609-0.0495609i$ & $0.5925609-0.0495609i$ & $(0.600862-0.043291i)^{\dagger}$ & $3.6\times10^{-14}$\\
\bottomrule
\end{tabular}
\end{table*}
\clearpage

\subsection{Quasibound-state tables}
We first vary the scalar mass over
$\mu=0.300,0.325,\ldots,0.650$ for the radial levels
$n_r=0,1,2$. We keep $q=0.5$, $\zeta=0.01$, and $\ell=2$ fixed. Next, we study how the NED coupling affects the fundamental state. We use $\zeta=0,0.0125,\ldots,0.1000$ and also include
$\zeta=0.01$, keeping $q=0.5$, $\ell=2$, and $\mu=0.6$ fixed. We then vary the charge over $q=0,0.1,\ldots,0.9$, with $\zeta=0.01$, $\ell=2$, and $\mu=0.6$. Finally, we compare multipoles $\ell=1,\ldots,5$ for the
fundamental radial level, $n_r=0$. We keep $q=0.5$ and
$\zeta=0.01$ fixed and adjust the scalar mass so that
$\mu/(\ell+1)=0.15$.
We list the real frequency and positive damping rate separately, so
that small differences in damping remain visible. All errors are computed wrt the
unrounded global method results. For Leaver, these errors measure numerical
agreement between two solvers. An asterisk ($\ast$) identifies  six particular
entries that required higher bit arithmetic; .
These comprise $n_r=1$ at $\mu=0.3$, $n_r=2$ at
$\mu=0.3,0.325,0.35,0.375$, and the $\ell=5$ entry.
The other 61 entries use the usual calculation. All 67 damping
rates pass the adopted $10^{-3}$ relative convergence and global-agreement
checks. The largest Leaver/global damping difference in this catalogue is
$0.06747\%$; 
BS and Pad\'e damping rates use leading Gamow tunnelling, while RN1
expands around the complex RN pole. 

\begin{table*}[!htbp]
\centering\footnotesize
\caption{Mass scan with $n_r=0$ at $q=0.5$, $\zeta=0.01$, $\ell=2$. Leaver and global matching solve the complex boundary-value problem. BS and both Pad\'e columns use leading Gamow decay; RN1 expands the complex RN pole. $\Gamma=-\operatorname{Im}\omega$. A dagger denotes an inadmissible method entry according to our criterion.}
\label{tab:scan-qbs-mass-0}
\setlength{\tabcolsep}{3pt}
\renewcommand{\arraystretch}{0.96}
\textit{Real frequency $M\omega_R$}\par\smallskip
\begin{tabular}{rrrrrrr}
\toprule
$M\mu$ & BS & $[2/2]$ & $[8/8]$ & RN1 & Leaver & Global\\
\midrule
0.3000 & 0.298365889 & 0.298415108 & 0.298415228 & 0.298415355 & 0.298415228 & 0.298415228\\
0.3250 & 0.322912160 & 0.322973813 & 0.322973960 & 0.322974113 & 0.322973960 & 0.322973960\\
0.3500 & 0.347377864 & 0.347453548 & 0.347453723 & 0.347453904 & 0.347453723 & 0.347453723\\
0.3750 & 0.371754837 & 0.371846063 & 0.371846255 & 0.371846474 & 0.371846260 & 0.371846260\\
0.4000 & 0.396034306 & 0.396142408 & 0.396142692 & 0.396142866 & 0.396142615 & 0.396142615\\
0.4250 & 0.420206766 & 0.420332778 & 0.420333039 & 0.420333253 & 0.420332959 & 0.420332959\\
0.4500 & 0.444261817 & 0.444406286 & 0.444393316 & 0.444406708 & 0.444406366 & 0.444406366\\
0.4750 & 0.468187944 & 0.468350661 & 0.468353671 & 0.468350879 & 0.468350481 & 0.468350481\\
0.5000 & 0.491972235 & 0.492151800 & 0.492152062 & 0.492151505 & 0.492151041 & 0.492151041\\
0.5250 & 0.515599969 & 0.515793074 & 0.515791891 & 0.515791703 & 0.515791159 & 0.515791159\\
0.5500 & 0.539054036 & 0.539254171 & 0.539247863 & 0.539250891 & 0.539250247 & 0.539250247\\
0.5750 & 0.562314034 & 0.562508938 & 0.562148014 & 0.562503269 & 0.562502498 & 0.562502498\\
0.6000 & 0.585354821 & 0.585520634 & 0.585534685 & 0.585516049 & 0.585515116 & 0.585515116\\
0.6250 & 0.608143921 & 0.608228462 & 0.608205952 & 0.608248582 & 0.608247456 & 0.608247456\\
0.6500 & 0.630636300 & 0.630486960 & 0.630176202 & 0.630655500 & 0.630654178 & 0.630654178\\
\bottomrule
\end{tabular}
\par\vspace{8pt}
\textit{Decay rate $M\Gamma$}\par\smallskip
\begin{tabular}{rrrrrrr}
\toprule
$M\mu$ & BS & $[2/2]$ & $[8/8]$ & RN1 & Leaver & Global\\
\midrule
0.3000 & $2.2596\times10^{-11}$ & $2.2358\times10^{-11}$ & $2.2357\times10^{-11}$ & $2.5177\times10^{-11}$ & $2.5227\times10^{-11}$ & $2.5227\times10^{-11}$\\
0.3250 & $9.6018\times10^{-11}$ & $9.5103\times10^{-11}$ & $9.5101\times10^{-11}$ & $1.0930\times10^{-10}$ & $1.0952\times10^{-10}$ & $1.0951\times10^{-10}$\\
0.3500 & $3.7918\times10^{-10}$ & $3.7598\times10^{-10}$ & $3.7597\times10^{-10}$ & $4.4045\times10^{-10}$ & $4.4134\times10^{-10}$ & $4.4134\times10^{-10}$\\
0.3750 & $1.4054\times10^{-9}$ & $1.3953\times10^{-9}$ & $1.3952\times10^{-9}$ & $1.6645\times10^{-9}$ & $1.6679\times10^{-9}$ & $1.6679\times10^{-9}$\\
0.4000 & $4.9282\times10^{-9}$ & $4.8989\times10^{-9}$ & $4.8988\times10^{-9}$ & $5.9467\times10^{-9}$ & $5.9591\times10^{-9}$ & $5.9591\times10^{-9}$\\
0.4250 & $1.6453\times10^{-8}$ & $1.6378\times10^{-8}$ & $1.6378\times10^{-8}$ & $2.0215\times10^{-8}$ & $2.0258\times10^{-8}$ & $2.0258\times10^{-8}$\\
0.4500 & $5.2569\times10^{-8}$ & $5.2404\times10^{-8}$ & $5.2423\times10^{-8}$ & $6.5728\times10^{-8}$ & $6.5870\times10^{-8}$ & $6.5870\times10^{-8}$\\
0.4750 & $1.6142\times10^{-7}$ & $1.6116\times10^{-7}$ & $1.6115\times10^{-7}$ & $2.0526\times10^{-7}$ & $2.0571\times10^{-7}$ & $2.0571\times10^{-7}$\\
0.5000 & $4.7799\times10^{-7}$ & $4.7793\times10^{-7}$ & $4.7793\times10^{-7}$ & $6.1767\times10^{-7}$ & $6.1905\times10^{-7}$ & $6.1905\times10^{-7}$\\
0.5250 & $1.3687\times10^{-6}$ & $1.3705\times10^{-6}$ & $1.3705\times10^{-6}$ & $1.7951\times10^{-6}$ & $1.7992\times10^{-6}$ & $1.7992\times10^{-6}$\\
0.5500 & $3.7979\times10^{-6}$ & $3.8080\times10^{-6}$ & $3.8077\times10^{-6}$ & $5.0445\times10^{-6}$ & $5.0559\times10^{-6}$ & $5.0559\times10^{-6}$\\
0.5750 & $1.0228\times10^{-5}$ & $1.0265\times10^{-5}$ & $1.0191\times10^{-5}$ & $1.3697\times10^{-5}$ & $1.3728\times10^{-5}$ & $1.3728\times10^{-5}$\\
0.6000 & $2.6758\times10^{-5}$ & $2.6862\times10^{-5}$ & $2.6870\times10^{-5}$ & $3.5820\times10^{-5}$ & $3.5899\times10^{-5}$ & $3.5899\times10^{-5}$\\
0.6250 & $6.8007\times10^{-5}$ & $6.8166\times10^{-5}$ & $6.8124\times10^{-5}$ & $8.9544\times10^{-5}$ & $8.9728\times10^{-5}$ & $8.9728\times10^{-5}$\\
0.6500 & $1.6767\times10^{-4}$ & $1.6688\times10^{-4}$ & $1.6514\times10^{-4}$ & $2.1129\times10^{-4}$ & $2.1167\times10^{-4}$ & $2.1167\times10^{-4}$\\
\bottomrule
\end{tabular}
\end{table*}
\begin{table*}[!htbp]
\centering\footnotesize
\caption{Relative errors for the mass scan with $n_r=0$ at $q=0.5$, $\zeta=0.01$, $\ell=2$. Errors are percentages relative to the global method's binding energy and decay rate, evaluated before rounding. The implicit zero global-error column is omitted.}
\label{tab:scan-qbs-errors-mass-0}
\setlength{\tabcolsep}{3pt}
\renewcommand{\arraystretch}{0.96}
\textit{Binding error $100\epsilon_B$}\par\smallskip
\begin{tabular}{rrrrrr}
\toprule
$M\mu$ & BS & $[2/2]$ & $[8/8]$ & RN1 & Leaver\\
\midrule
0.3000 & 3.113 & 0.007539 & $3.024\times10^{-6}$ & 0.008002 & $1.751\times10^{-11}$\\
0.3250 & 3.05 & 0.007251 & $3.415\times10^{-6}$ & 0.007532 & $9.042\times10^{-11}$\\
0.3500 & 2.979 & 0.006851 & $8.505\times10^{-6}$ & 0.007134 & $8.960\times10^{-10}$\\
0.3750 & 2.899 & 0.006265 & $1.510\times10^{-4}$ & 0.006798 & $9.505\times10^{-11}$\\
0.4000 & 2.808 & 0.005358 & 0.002015 & 0.006518 & $3.749\times10^{-9}$\\
0.4250 & 2.704 & 0.003891 & 0.001713 & 0.006289 & $8.183\times10^{-10}$\\
0.4500 & 2.584 & 0.001438 & 0.2333 & 0.006111 & $1.538\times10^{-10}$\\
0.4750 & 2.444 & 0.002717 & 0.04797 & 0.005984 & $1.085\times10^{-11}$\\
0.5000 & 2.278 & 0.009678 & 0.01301 & 0.005915 & $1.132\times10^{-11}$\\
0.5250 & 2.076 & 0.0208 & 0.007957 & 0.005914 & $9.645\times10^{-12}$\\
0.5500 & 1.825 & 0.0365 & 0.02218 & 0.005994 & $5.164\times10^{-12}$\\
0.5750 & 1.508 & 0.05153 & 2.836 & 0.006171 & $6.218\times10^{-12}$\\
0.6000 & 1.107 & 0.03809 & 0.1351 & 0.006437 & $3.832\times10^{-12}$\\
0.6250 & 0.618 & 0.1134 & 0.2477 & 0.00672 & $8.615\times10^{-12}$\\
0.6500 & 0.09242 & 0.8644 & 2.471 & 0.006833 & $4.017\times10^{-12}$\\
\bottomrule
\end{tabular}
\par\vspace{8pt}
\textit{Damping error $100\epsilon_\Gamma$}\par\smallskip
\begin{tabular}{rrrrrr}
\toprule
$M\mu$ & BS & $[2/2]$ & $[8/8]$ & RN1 & Leaver\\
\midrule
0.3000 & 10.43 & 11.38 & 11.38 & 0.1984 & $4.475\times10^{-4}$\\
0.3250 & 12.32 & 13.16 & 13.16 & 0.1998 & $7.224\times10^{-4}$\\
0.3500 & 14.08 & 14.81 & 14.81 & 0.2019 & $2.098\times10^{-4}$\\
0.3750 & 15.74 & 16.35 & 16.35 & 0.2047 & $1.843\times10^{-5}$\\
0.4000 & 17.3 & 17.79 & 17.79 & 0.2079 & $3.057\times10^{-5}$\\
0.4250 & 18.78 & 19.15 & 19.15 & 0.2114 & $4.572\times10^{-7}$\\
0.4500 & 20.19 & 20.44 & 20.41 & 0.2151 & $1.139\times10^{-5}$\\
0.4750 & 21.53 & 21.66 & 21.66 & 0.2188 & $1.001\times10^{-7}$\\
0.5000 & 22.79 & 22.8 & 22.8 & 0.2222 & $2.582\times10^{-8}$\\
0.5250 & 23.93 & 23.83 & 23.83 & 0.2249 & $2.274\times10^{-8}$\\
0.5500 & 24.88 & 24.68 & 24.69 & 0.2263 & $6.896\times10^{-9}$\\
0.5750 & 25.5 & 25.23 & 25.77 & 0.225 & $2.275\times10^{-9}$\\
0.6000 & 25.46 & 25.17 & 25.15 & 0.2189 & $4.317\times10^{-9}$\\
0.6250 & 24.21 & 24.03 & 24.08 & 0.2049 & $5.947\times10^{-9}$\\
0.6500 & 20.79 & 21.16 & 21.99 & 0.1807 & $9.449\times10^{-11}$\\
\bottomrule
\end{tabular}
\end{table*}
\clearpage
\begin{table*}[!htbp]
\centering\footnotesize
\caption{Mass scan with $n_r=1$ at $q=0.5$, $\zeta=0.01$, $\ell=2$. Leaver and global matching solve the complex boundary-value problem. BS and both Pad\'e columns use leading Gamow decay; RN1 expands the complex RN pole. $\Gamma=-\operatorname{Im}\omega$. A dagger denotes an inadmissible method entry according to our criterion.}
\label{tab:scan-qbs-mass-1}
\setlength{\tabcolsep}{3pt}
\renewcommand{\arraystretch}{0.96}
\textit{Real frequency $M\omega_R$}\par\smallskip
\begin{tabular}{rrrrrrr}
\toprule
$M\mu$ & BS & $[2/2]$ & $[8/8]$ & RN1 & Leaver & Global\\
\midrule
0.3000 & 0.299086528 & 0.299107394 & 0.299107412 & 0.299107501 & $(0.299107412)^{\ast}$ & 0.299107412\\
0.3250 & 0.323832832 & 0.323858967 & 0.323859024 & 0.323859132 & 0.323859024 & 0.323859024\\
0.3500 & 0.348534047 & 0.348566114 & 0.348566240 & 0.348566368 & 0.348566240 & 0.348566240\\
0.3750 & 0.373185578 & 0.373224189 & 0.373224425 & 0.373224578 & 0.373224425 & 0.373224425\\
0.4000 & 0.397782477 & 0.397828143 & 0.397828550 & 0.397828728 & 0.397828548 & 0.397828548\\
0.4250 & 0.422319360 & 0.422372422 & 0.422373076 & 0.422373288 & 0.422373077 & 0.422373077\\
0.4500 & 0.446790308 & 0.446850836 & 0.446851771 & 0.446852080 & 0.446851833 & 0.446851833\\
0.4750 & 0.471188725 & 0.471256354 & 0.471258171 & 0.471258078 & 0.471257789 & 0.471257789\\
0.5000 & 0.495507139 & 0.495580811 & 0.495582953 & 0.495583089 & 0.495582751 & 0.495582751\\
0.5250 & 0.519736918 & 0.519814431 & 0.519816505 & 0.519817265 & 0.519816866 & 0.519816866\\
0.5500 & 0.543867825 & 0.543944993 & 0.543943046 & 0.543948355 & 0.543947881 & 0.543947881\\
0.5750 & 0.567887293 & 0.567956206 & 0.567905969 & 0.567960663 & 0.567960091 & 0.567960091\\
0.6000 & 0.591779098 & 0.591824037 & 0.592085000 & 0.591834049 & 0.591833355 & 0.591833355\\
0.6250 & 0.615520588 & 0.615506603 & 0.615635420 & 0.615544537 & 0.615543707 & 0.615543707\\
0.6500 & 0.639075330 & 0.638907557 & 0.639060927 & 0.639070068 & 0.639069136 & 0.639069136\\
\bottomrule
\end{tabular}
\par\vspace{8pt}
\textit{Decay rate $M\Gamma$}\par\smallskip
\begin{tabular}{rrrrrrr}
\toprule
$M\mu$ & BS & $[2/2]$ & $[8/8]$ & RN1 & Leaver & Global\\
\midrule
0.3000 & $1.4663\times10^{-11}$ & $1.4318\times10^{-11}$ & $1.4318\times10^{-11}$ & $1.5913\times10^{-11}$ & $(1.5948\times10^{-11})^{\ast}$ & $1.5948\times10^{-11}$\\
0.3250 & $6.3002\times10^{-11}$ & $6.1569\times10^{-11}$ & $6.1566\times10^{-11}$ & $6.9799\times10^{-11}$ & $7.0003\times10^{-11}$ & $6.9955\times10^{-11}$\\
0.3500 & $2.5195\times10^{-10}$ & $2.4643\times10^{-10}$ & $2.4641\times10^{-10}$ & $2.8464\times10^{-10}$ & $2.8529\times10^{-10}$ & $2.8528\times10^{-10}$\\
0.3750 & $9.4724\times10^{-10}$ & $9.2741\times10^{-10}$ & $9.2729\times10^{-10}$ & $1.0903\times10^{-9}$ & $1.0928\times10^{-9}$ & $1.0928\times10^{-9}$\\
0.4000 & $3.3753\times10^{-9}$ & $3.3082\times10^{-9}$ & $3.3076\times10^{-9}$ & $3.9551\times10^{-9}$ & $3.9644\times10^{-9}$ & $3.9644\times10^{-9}$\\
0.4250 & $1.1474\times10^{-8}$ & $1.1260\times10^{-8}$ & $1.1257\times10^{-8}$ & $1.3679\times10^{-8}$ & $1.3712\times10^{-8}$ & $1.3712\times10^{-8}$\\
0.4500 & $3.7410\times10^{-8}$ & $3.6762\times10^{-8}$ & $3.6752\times10^{-8}$ & $4.5349\times10^{-8}$ & $4.5461\times10^{-8}$ & $4.5461\times10^{-8}$\\
0.4750 & $1.1751\times10^{-7}$ & $1.1565\times10^{-7}$ & $1.1560\times10^{-7}$ & $1.4476\times10^{-7}$ & $1.4513\times10^{-7}$ & $1.4513\times10^{-7}$\\
0.5000 & $3.5695\times10^{-7}$ & $3.5192\times10^{-7}$ & $3.5177\times10^{-7}$ & $4.4648\times10^{-7}$ & $4.4765\times10^{-7}$ & $4.4765\times10^{-7}$\\
0.5250 & $1.0518\times10^{-6}$ & $1.0390\times10^{-6}$ & $1.0386\times10^{-6}$ & $1.3339\times10^{-6}$ & $1.3375\times10^{-6}$ & $1.3375\times10^{-6}$\\
0.5500 & $3.0140\times10^{-6}$ & $2.9842\times10^{-6}$ & $2.9850\times10^{-6}$ & $3.8645\times10^{-6}$ & $3.8749\times10^{-6}$ & $3.8749\times10^{-6}$\\
0.5750 & $8.4182\times10^{-6}$ & $8.3578\times10^{-6}$ & $8.4019\times10^{-6}$ & $1.0840\times10^{-5}$ & $1.0870\times10^{-5}$ & $1.0870\times10^{-5}$\\
0.6000 & $2.2955\times10^{-5}$ & $2.2869\times10^{-5}$ & $2.2351\times10^{-5}$ & $2.9269\times10^{-5}$ & $2.9346\times10^{-5}$ & $2.9346\times10^{-5}$\\
0.6250 & $6.1186\times10^{-5}$ & $6.1242\times10^{-5}$ & $6.0716\times10^{-5}$ & $7.5048\times10^{-5}$ & $7.5229\times10^{-5}$ & $7.5229\times10^{-5}$\\
0.6500 & $1.5943\times10^{-4}$ & $1.6077\times10^{-4}$ & $1.5955\times10^{-4}$ & $1.7871\times10^{-4}$ & $1.7907\times10^{-4}$ & $1.7907\times10^{-4}$\\
\bottomrule
\end{tabular}
\end{table*}
\begin{table*}[!htbp]
\centering\footnotesize
\caption{Relative errors for the mass scan with $n_r=1$ at $q=0.5$, $\zeta=0.01$, $\ell=2$. Errors are percentages relative to the global method's binding energy and decay rate, evaluated before rounding. The implicit zero global-error column is omitted.}
\label{tab:scan-qbs-errors-mass-1}
\setlength{\tabcolsep}{3pt}
\renewcommand{\arraystretch}{0.96}
\textit{Binding error $100\epsilon_B$}\par\smallskip
\begin{tabular}{rrrrrr}
\toprule
$M\mu$ & BS & $[2/2]$ & $[8/8]$ & RN1 & Leaver\\
\midrule
0.3000 & 2.34 & 0.00205 & $3.710\times10^{-7}$ & 0.009955 & $(1.866\times10^{-11})^{\ast}$\\
0.3250 & 2.296 & 0.005015 & $3.658\times10^{-7}$ & 0.009429 & $9.492\times10^{-9}$\\
0.3500 & 2.245 & 0.00873 & $3.187\times10^{-6}$ & 0.008982 & $4.746\times10^{-8}$\\
0.3750 & 2.188 & 0.01328 & $2.889\times10^{-5}$ & 0.008605 & $6.071\times10^{-9}$\\
0.4000 & 2.122 & 0.01869 & $7.641\times10^{-5}$ & 0.008292 & $1.161\times10^{-9}$\\
0.4250 & 2.045 & 0.0249 & $2.487\times10^{-5}$ & 0.008038 & $1.101\times10^{-8}$\\
0.4500 & 1.954 & 0.03165 & 0.001969 & 0.007844 & $3.183\times10^{-9}$\\
0.4750 & 1.846 & 0.03835 & 0.01019 & 0.007714 & $2.170\times10^{-9}$\\
0.5000 & 1.712 & 0.0439 & 0.004588 & 0.007659 & $1.675\times10^{-8}$\\
0.5250 & 1.542 & 0.04698 & 0.006972 & 0.007694 & $4.003\times10^{-9}$\\
0.5500 & 1.323 & 0.04771 & 0.07988 & 0.007844 & $1.282\times10^{-9}$\\
0.5750 & 1.034 & 0.05518 & 0.7688 & 0.008126 & $2.523\times10^{-10}$\\
0.6000 & 0.6644 & 0.1141 & 3.081 & 0.008501 & $1.835\times10^{-10}$\\
0.6250 & 0.2445 & 0.3924 & 0.9699 & 0.008769 & $2.008\times10^{-10}$\\
0.6500 & 0.05666 & 1.478 & 0.0751 & 0.008527 & $1.869\times10^{-10}$\\
\bottomrule
\end{tabular}
\par\vspace{8pt}
\textit{Damping error $100\epsilon_\Gamma$}\par\smallskip
\begin{tabular}{rrrrrr}
\toprule
$M\mu$ & BS & $[2/2]$ & $[8/8]$ & RN1 & Leaver\\
\midrule
0.3000 & 8.058 & 10.22 & 10.22 & 0.2203 & $(3.830\times10^{-7})^{\ast}$\\
0.3250 & 9.94 & 11.99 & 11.99 & 0.2227 & 0.06747\\
0.3500 & 11.69 & 13.62 & 13.63 & 0.2266 & 0.002851\\
0.3750 & 13.32 & 15.13 & 15.15 & 0.2311 & $1.493\times10^{-4}$\\
0.4000 & 14.86 & 16.55 & 16.57 & 0.2362 & 0.001189\\
0.4250 & 16.32 & 17.88 & 17.9 & 0.2418 & $1.982\times10^{-4}$\\
0.4500 & 17.71 & 19.14 & 19.16 & 0.2479 & $6.214\times10^{-6}$\\
0.4750 & 19.03 & 20.31 & 20.34 & 0.2543 & $5.571\times10^{-6}$\\
0.5000 & 20.26 & 21.39 & 21.42 & 0.2607 & $7.084\times10^{-7}$\\
0.5250 & 21.36 & 22.32 & 22.34 & 0.2666 & $2.048\times10^{-6}$\\
0.5500 & 22.22 & 22.99 & 22.97 & 0.2708 & $1.212\times10^{-7}$\\
0.5750 & 22.55 & 23.11 & 22.7 & 0.2711 & $1.808\times10^{-7}$\\
0.6000 & 21.78 & 22.07 & 23.84 & 0.2631 & $3.144\times10^{-8}$\\
0.6250 & 18.67 & 18.59 & 19.29 & 0.2406 & $9.124\times10^{-9}$\\
0.6500 & 10.96 & 10.22 & 10.9 & 0.2001 & $2.234\times10^{-9}$\\
\bottomrule
\end{tabular}
\end{table*}
\clearpage
\begin{table*}[!htbp]
\centering\footnotesize
\caption{Mass scan with $n_r=2$ at $q=0.5$, $\zeta=0.01$, $\ell=2$. Leaver and global matching solve the complex boundary-value problem. BS and both Pad\'e columns use leading Gamow decay; RN1 expands the complex RN pole. $\Gamma=-\operatorname{Im}\omega$. A dagger denotes an inadmissible method entry according to our criterion.}
\label{tab:scan-qbs-mass-2}
\setlength{\tabcolsep}{3pt}
\renewcommand{\arraystretch}{0.96}
\textit{Real frequency $M\omega_R$}\par\smallskip
\begin{tabular}{rrrrrrr}
\toprule
$M\mu$ & BS & $[2/2]$ & $[8/8]$ & RN1 & Leaver & Global\\
\midrule
0.3000 & 0.299418801 & 0.299429244 & 0.299429466 & 0.299429532 & $(0.299429466)^{\ast}$ & 0.299429466\\
0.3250 & 0.324257726 & 0.324270668 & 0.324271100 & 0.324271180 & $(0.324271100)^{\ast}$ & 0.324271100\\
0.3500 & 0.349068180 & 0.349083832 & 0.349084614 & 0.349084710 & $(0.349084614)^{\ast}$ & 0.349084614\\
0.3750 & 0.373847317 & 0.373865798 & 0.373867141 & 0.373867256 & $(0.373867142)^{\ast}$ & 0.373867142\\
0.4000 & 0.398592086 & 0.398613367 & 0.398615567 & 0.398615719 & 0.398615585 & 0.398615585\\
0.4250 & 0.423299179 & 0.423323008 & 0.423326567 & 0.423326712 & 0.423326555 & 0.423326555\\
0.4500 & 0.447964978 & 0.447990766 & 0.447996297 & 0.447996473 & 0.447996289 & 0.447996289\\
0.4750 & 0.472585464 & 0.472612117 & 0.472620488 & 0.472620744 & 0.472620529 & 0.472620529\\
0.5000 & 0.497156102 & 0.497181760 & 0.497193724 & 0.497194580 & 0.497194329 & 0.497194329\\
0.5250 & 0.521671666 & 0.521693275 & 0.521714037 & 0.521712057 & 0.521711762 & 0.521711762\\
0.5500 & 0.546125958 & 0.546138524 & 0.546166067 & 0.546165817 & 0.546165468 & 0.546165468\\
0.5750 & 0.570511338 & 0.570506479 & 0.570533923 & 0.570546453 & 0.570546035 & 0.570546035\\
0.6000 & 0.594817820 & 0.594780489 & 0.594861613 & 0.594842035 & 0.594841533 & 0.594841533\\
0.6250 & 0.619031020 & 0.618930172 & 0.619031814 & 0.619038998 & 0.619038408 & 0.619038408\\
0.6500 & 0.643125726 & 0.642878198 & 0.643046939 & 0.643126769 & 0.643126121 & 0.643126121\\
\bottomrule
\end{tabular}
\par\vspace{8pt}
\textit{Decay rate $M\Gamma$}\par\smallskip
\begin{tabular}{rrrrrrr}
\toprule
$M\mu$ & BS & $[2/2]$ & $[8/8]$ & RN1 & Leaver & Global\\
\midrule
0.3000 & $8.7737\times10^{-12}$ & $8.5802\times10^{-12}$ & $8.5761\times10^{-12}$ & $9.5280\times10^{-12}$ & $(9.5504\times10^{-12})^{\ast}$ & $9.5504\times10^{-12}$\\
0.3250 & $3.7857\times10^{-11}$ & $3.7052\times10^{-11}$ & $3.7026\times10^{-11}$ & $4.1958\times10^{-11}$ & $(4.2058\times10^{-11})^{\ast}$ & $4.2058\times10^{-11}$\\
0.3500 & $1.5212\times10^{-10}$ & $1.4903\times10^{-10}$ & $1.4888\times10^{-10}$ & $1.7187\times10^{-10}$ & $(1.7229\times10^{-10})^{\ast}$ & $1.7229\times10^{-10}$\\
0.3750 & $5.7508\times10^{-10}$ & $5.6402\times10^{-10}$ & $5.6321\times10^{-10}$ & $6.6173\times10^{-10}$ & $(6.6338\times10^{-10})^{\ast}$ & $6.6338\times10^{-10}$\\
0.4000 & $2.0620\times10^{-9}$ & $2.0249\times10^{-9}$ & $2.0211\times10^{-9}$ & $2.4146\times10^{-9}$ & $2.4198\times10^{-9}$ & $2.4207\times10^{-9}$\\
0.4250 & $7.0588\times10^{-9}$ & $6.9425\times10^{-9}$ & $6.9251\times10^{-9}$ & $8.4064\times10^{-9}$ & $8.4273\times10^{-9}$ & $8.4284\times10^{-9}$\\
0.4500 & $2.3198\times10^{-8}$ & $2.2857\times10^{-8}$ & $2.2783\times10^{-8}$ & $2.8080\times10^{-8}$ & $2.8161\times10^{-8}$ & $2.8156\times10^{-8}$\\
0.4750 & $7.3525\times10^{-8}$ & $7.2596\times10^{-8}$ & $7.2303\times10^{-8}$ & $9.0400\times10^{-8}$ & $9.0655\times10^{-8}$ & $9.0650\times10^{-8}$\\
0.5000 & $2.2560\times10^{-7}$ & $2.2331\times10^{-7}$ & $2.2224\times10^{-7}$ & $2.8151\times10^{-7}$ & $2.8231\times10^{-7}$ & $2.8231\times10^{-7}$\\
0.5250 & $6.7237\times10^{-7}$ & $6.6754\times10^{-7}$ & $6.6290\times10^{-7}$ & $8.5007\times10^{-7}$ & $8.5256\times10^{-7}$ & $8.5256\times10^{-7}$\\
0.5500 & $1.9519\times10^{-6}$ & $1.9450\times10^{-6}$ & $1.9300\times10^{-6}$ & $2.4915\times10^{-6}$ & $2.4989\times10^{-6}$ & $2.4989\times10^{-6}$\\
0.5750 & $5.5327\times10^{-6}$ & $5.5390\times10^{-6}$ & $5.5033\times10^{-6}$ & $7.0721\times10^{-6}$ & $7.0932\times10^{-6}$ & $7.0932\times10^{-6}$\\
0.6000 & $1.5346\times10^{-5}$ & $1.5460\times10^{-5}$ & $1.5213\times10^{-5}$ & $1.9289\times10^{-5}$ & $1.9344\times10^{-5}$ & $1.9344\times10^{-5}$\\
0.6250 & $4.1736\times10^{-5}$ & $4.2431\times10^{-5}$ & $4.1731\times10^{-5}$ & $4.9673\times10^{-5}$ & $4.9802\times10^{-5}$ & $4.9802\times10^{-5}$\\
0.6500 & $1.1146\times10^{-4}$ & $1.1522\times10^{-4}$ & $1.1267\times10^{-4}$ & $1.1745\times10^{-4}$ & $1.1770\times10^{-4}$ & $1.1770\times10^{-4}$\\
\bottomrule
\end{tabular}
\end{table*}
\begin{table*}[!htbp]
\centering\footnotesize
\caption{Relative errors for the mass scan with $n_r=2$ at $q=0.5$, $\zeta=0.01$, $\ell=2$. Errors are percentages relative to the global method's binding energy and decay rate, evaluated before rounding. The zero global-error column is omitted.}
\label{tab:scan-qbs-errors-mass-2}
\setlength{\tabcolsep}{3pt}
\renewcommand{\arraystretch}{0.96}
\textit{Binding error $100\epsilon_B$}\par\smallskip
\begin{tabular}{rrrrrr}
\toprule
$M\mu$ & BS & $[2/2]$ & $[8/8]$ & RN1 & Leaver\\
\midrule
0.3000 & 1.869 & 0.03887 & $1.142\times10^{-6}$ & 0.01157 & $(2.715\times10^{-9})^{\ast}$\\
0.3250 & 1.835 & 0.05924 & $3.071\times10^{-6}$ & 0.01099 & $(2.361\times10^{-10})^{\ast}$\\
0.3500 & 1.795 & 0.0854 & $6.762\times10^{-6}$ & 0.01049 & $(2.426\times10^{-11})^{\ast}$\\
0.3750 & 1.75 & 0.1186 & $6.157\times10^{-5}$ & 0.01007 & $(0)^{\ast}$\\
0.4000 & 1.697 & 0.1602 & 0.001323 & 0.00971 & $6.979\times10^{-8}$\\
0.4250 & 1.636 & 0.2119 & $7.562\times10^{-4}$ & 0.009417 & $1.868\times10^{-7}$\\
0.4500 & 1.563 & 0.2757 & $3.847\times10^{-4}$ & 0.009188 & $3.179\times10^{-7}$\\
0.4750 & 1.474 & 0.3535 & 0.001715 & 0.009027 & $3.960\times10^{-7}$\\
0.5000 & 1.362 & 0.448 & 0.02154 & 0.008946 & $1.176\times10^{-7}$\\
0.5250 & 1.219 & 0.5622 & 0.06918 & 0.008962 & $1.962\times10^{-8}$\\
0.5500 & 1.03 & 0.7027 & 0.01563 & 0.009102 & $3.291\times10^{-8}$\\
0.5750 & 0.779 & 0.8881 & 0.2719 & 0.009379 & $2.824\times10^{-8}$\\
0.6000 & 0.4597 & 1.183 & 0.3893 & 0.009731 & $2.367\times10^{-8}$\\
0.6250 & 0.1239 & 1.816 & 0.1106 & 0.009895 & $1.970\times10^{-9}$\\
0.6500 & 0.005757 & 3.607 & 1.152 & 0.009424 & $1.678\times10^{-8}$\\
\bottomrule
\end{tabular}
\par\vspace{8pt}
\textit{Damping error $100\epsilon_\Gamma$}\par\smallskip
\begin{tabular}{rrrrrr}
\toprule
$M\mu$ & BS & $[2/2]$ & $[8/8]$ & RN1 & Leaver\\
\midrule
0.3000 & 8.134 & 10.16 & 10.2 & 0.2354 & $(3.126\times10^{-4})^{\ast}$\\
0.3250 & 9.988 & 11.9 & 11.97 & 0.2387 & $(9.528\times10^{-6})^{\ast}$\\
0.3500 & 11.71 & 13.5 & 13.59 & 0.243 & $(2.838\times10^{-7})^{\ast}$\\
0.3750 & 13.31 & 14.98 & 15.1 & 0.2483 & $(9.179\times10^{-9})^{\ast}$\\
0.4000 & 14.82 & 16.35 & 16.51 & 0.2544 & 0.03711\\
0.4250 & 16.25 & 17.63 & 17.84 & 0.2612 & 0.01283\\
0.4500 & 17.61 & 18.82 & 19.08 & 0.2686 & 0.01791\\
0.4750 & 18.89 & 19.92 & 20.24 & 0.2765 & 0.005229\\
0.5000 & 20.09 & 20.9 & 21.28 & 0.2844 & $8.473\times10^{-4}$\\
0.5250 & 21.14 & 21.7 & 22.25 & 0.292 & $3.478\times10^{-4}$\\
0.5500 & 21.89 & 22.17 & 22.77 & 0.2977 & $1.456\times10^{-4}$\\
0.5750 & 22 & 21.91 & 22.42 & 0.2983 & $2.349\times10^{-6}$\\
0.6000 & 20.67 & 20.08 & 21.36 & 0.2879 & $1.208\times10^{-5}$\\
0.6250 & 16.2 & 14.8 & 16.21 & 0.2584 & $2.942\times10^{-6}$\\
0.6500 & 5.295 & 2.104 & 4.27 & 0.2071 & $7.157\times10^{-7}$\\
\bottomrule
\end{tabular}
\end{table*}
\clearpage
\begin{table*}[!htbp]
\centering\footnotesize
\caption{Fundamental coupling scan at $q=0.5$, $\ell=2$, $\mu=0.6$. Leaver and global matching solve the complex boundary-value problem. BS and both Pad\'e columns use leading Gamow decay; RN1 expands the complex RN pole. $\Gamma=-\operatorname{Im}\omega$. A dagger denotes an inadmissible method entry according to our criterion.}
\label{tab:scan-qbs-coupling}
\setlength{\tabcolsep}{3pt}
\renewcommand{\arraystretch}{0.96}
\textit{Real frequency $M\omega_R$}\par\smallskip
\begin{tabular}{rrrrrrr}
\toprule
$\zeta\sqrt M$ & BS & $[2/2]$ & $[8/8]$ & RN1 & Leaver & Global\\
\midrule
0.0000 & 0.585591124 & 0.585761038 & 0.585776728 & 0.585755146 & 0.585755146 & 0.585755146\\
0.0100 & 0.585354821 & 0.585520634 & 0.585534685 & 0.585516049 & 0.585515116 & 0.585515116\\
0.0125 & 0.585295485 & 0.585460226 & 0.585473878 & 0.585456274 & 0.585454816 & 0.585454816\\
0.0250 & 0.584997235 & 0.585156317 & 0.585168013 & 0.585157403 & 0.585151549 & 0.585151549\\
0.0375 & 0.584696347 & 0.584849252 & 0.584858995 & 0.584858532 & 0.584845311 & 0.584845311\\
0.0500 & 0.584392794 & 0.584538962 & 0.584546690 & 0.584559661 & 0.584536067 & 0.584536067\\
0.0625 & 0.584086548 & 0.584225377 & 0.584230963 & 0.584260789 & 0.584223783 & 0.584223783\\
0.0750 & 0.583777580 & 0.583908416 & 0.583911669 & 0.583961918 & 0.583908423 & 0.583908423\\
0.0875 & 0.583465857 & 0.583587992 & 0.583588655 & 0.583663047 & 0.583589949 & 0.583589949\\
0.1000 & 0.583151348 & 0.583264011 & 0.583261749 & 0.583364175 & 0.583268324 & 0.583268324\\
\bottomrule
\end{tabular}
\par\vspace{8pt}
\textit{Decay rate $M\Gamma$}\par\smallskip
\begin{tabular}{rrrrrrr}
\toprule
$\zeta\sqrt M$ & BS & $[2/2]$ & $[8/8]$ & RN1 & Leaver & Global\\
\midrule
0.0000 & $2.4974\times10^{-5}$ & $2.5069\times10^{-5}$ & $2.5077\times10^{-5}$ & $3.3510\times10^{-5}$ & $3.3510\times10^{-5}$ & $3.3510\times10^{-5}$\\
0.0100 & $2.6758\times10^{-5}$ & $2.6862\times10^{-5}$ & $2.6870\times10^{-5}$ & $3.5820\times10^{-5}$ & $3.5899\times10^{-5}$ & $3.5899\times10^{-5}$\\
0.0125 & $2.7224\times10^{-5}$ & $2.7330\times10^{-5}$ & $2.7338\times10^{-5}$ & $3.6398\times10^{-5}$ & $3.6522\times10^{-5}$ & $3.6522\times10^{-5}$\\
0.0250 & $2.9674\times10^{-5}$ & $2.9791\times10^{-5}$ & $2.9799\times10^{-5}$ & $3.9286\times10^{-5}$ & $3.9794\times10^{-5}$ & $3.9794\times10^{-5}$\\
0.0375 & $3.2342\times10^{-5}$ & $3.2469\times10^{-5}$ & $3.2477\times10^{-5}$ & $4.2175\times10^{-5}$ & $4.3348\times10^{-5}$ & $4.3348\times10^{-5}$\\
0.0500 & $3.5246\times10^{-5}$ & $3.5384\times10^{-5}$ & $3.5391\times10^{-5}$ & $4.5063\times10^{-5}$ & $4.7207\times10^{-5}$ & $4.7207\times10^{-5}$\\
0.0625 & $3.8407\times10^{-5}$ & $3.8556\times10^{-5}$ & $3.8561\times10^{-5}$ & $4.7952\times10^{-5}$ & $5.1396\times10^{-5}$ & $5.1396\times10^{-5}$\\
0.0750 & $4.1848\times10^{-5}$ & $4.2006\times10^{-5}$ & $4.2010\times10^{-5}$ & $5.0840\times10^{-5}$ & $5.5941\times10^{-5}$ & $5.5941\times10^{-5}$\\
0.0875 & $4.5594\times10^{-5}$ & $4.5761\times10^{-5}$ & $4.5762\times10^{-5}$ & $5.3728\times10^{-5}$ & $6.0872\times10^{-5}$ & $6.0872\times10^{-5}$\\
0.1000 & $4.9672\times10^{-5}$ & $4.9845\times10^{-5}$ & $4.9841\times10^{-5}$ & $5.6617\times10^{-5}$ & $6.6218\times10^{-5}$ & $6.6218\times10^{-5}$\\
\bottomrule
\end{tabular}
\end{table*}
\begin{table*}[!htbp]
\centering\footnotesize
\caption{Relative errors for the fundamental coupling scan at $q=0.5$, $\ell=2$, $\mu=0.6$. Errors are percentages relative to the global method's binding energy and decay rate, evaluated before rounding. The implicit zero global-error column is omitted.}
\label{tab:scan-qbs-errors-coupling}
\setlength{\tabcolsep}{3pt}
\renewcommand{\arraystretch}{0.96}
\textit{Binding error $100\epsilon_B$}\par\smallskip
\begin{tabular}{rrrrrr}
\toprule
$\zeta\sqrt M$ & BS & $[2/2]$ & $[8/8]$ & RN1 & Leaver\\
\midrule
0.0000 & 1.151 & 0.04136 & 0.1515 & $0$ & $3.897\times10^{-12}$\\
0.0100 & 1.107 & 0.03809 & 0.1351 & 0.006437 & $3.832\times10^{-12}$\\
0.0125 & 1.095 & 0.03719 & 0.1311 & 0.01002 & $1.527\times10^{-12}$\\
0.0250 & 1.039 & 0.03211 & 0.1109 & 0.03942 & $1.495\times10^{-12}$\\
0.0375 & 0.983 & 0.026 & 0.0903 & 0.08724 & $5.861\times10^{-12}$\\
0.0500 & 0.9265 & 0.01872 & 0.06869 & 0.1526 & $2.872\times10^{-12}$\\
0.0625 & 0.8699 & 0.0101 & 0.04551 & 0.2346 & $9.149\times10^{-12}$\\
0.0750 & 0.8131 & $4.273\times10^{-5}$ & 0.02018 & 0.3324 & $1.380\times10^{-11}$\\
0.0875 & 0.7562 & 0.01192 & 0.007882 & 0.4454 & $1.827\times10^{-11}$\\
0.1000 & 0.6991 & 0.02578 & 0.0393 & 0.5729 & $2.654\times10^{-12}$\\
\bottomrule
\end{tabular}
\par\vspace{8pt}
\textit{Damping error $100\epsilon_\Gamma$}\par\smallskip
\begin{tabular}{rrrrrr}
\toprule
$\zeta\sqrt M$ & BS & $[2/2]$ & $[8/8]$ & RN1 & Leaver\\
\midrule
0.0000 & 25.47 & 25.19 & 25.17 & $6.086\times10^{-10}$ & $2.228\times10^{-9}$\\
0.0100 & 25.46 & 25.17 & 25.15 & 0.2189 & $4.317\times10^{-9}$\\
0.0125 & 25.46 & 25.17 & 25.15 & 0.338 & $7.212\times10^{-10}$\\
0.0250 & 25.43 & 25.14 & 25.12 & 1.275 & $4.435\times10^{-10}$\\
0.0375 & 25.39 & 25.1 & 25.08 & 2.706 & $1.924\times10^{-10}$\\
0.0500 & 25.34 & 25.05 & 25.03 & 4.541 & $6.970\times10^{-10}$\\
0.0625 & 25.27 & 24.98 & 24.97 & 6.701 & $6.888\times10^{-10}$\\
0.0750 & 25.19 & 24.91 & 24.9 & 9.119 & $1.073\times10^{-8}$\\
0.0875 & 25.1 & 24.82 & 24.82 & 11.74 & $6.810\times10^{-9}$\\
0.1000 & 24.99 & 24.73 & 24.73 & 14.5 & $3.683\times10^{-9}$\\
\bottomrule
\end{tabular}
\end{table*}
\clearpage
\begin{table*}[!htbp]
\centering\footnotesize
\caption{Fundamental charge scan at $\zeta=0.01$, $\ell=2$, $\mu=0.6$. Leaver and global matching solve the complex boundary-value problem. BS and both Pad\'e columns use leading Gamow decay; RN1 expands around the complex RN pole. $\Gamma=-\operatorname{Im}\omega$.}
\label{tab:scan-qbs-charge}
\setlength{\tabcolsep}{3pt}
\renewcommand{\arraystretch}{0.96}
\textit{Real frequency $M\omega_R$}\par\smallskip
\begin{tabular}{rrrrrrr}
\toprule
$q/M$ & BS & $[2/2]$ & $[8/8]$ & RN1 & Leaver & Global\\
\midrule
0.0000 & 0.585290940 & 0.585404977 & 0.585392967 & 0.585415907 & 0.585415907 & 0.585415907\\
0.1000 & 0.585281817 & 0.585398063 & 0.585387487 & 0.585407991 & 0.585407983 & 0.585407983\\
0.2000 & 0.585279631 & 0.585403033 & 0.585396687 & 0.585409982 & 0.585409919 & 0.585409919\\
0.3000 & 0.585290246 & 0.585425106 & 0.585424759 & 0.585427892 & 0.585427682 & 0.585427682\\
0.4000 & 0.585315203 & 0.585464647 & 0.585471049 & 0.585463221 & 0.585462732 & 0.585462732\\
0.5000 & 0.585354821 & 0.585520634 & 0.585534685 & 0.585516049 & 0.585515116 & 0.585515116\\
0.6000 & 0.585408771 & 0.585591488 & 0.585619517 & 0.585585556 & 0.585583994 & 0.585583994\\
0.7000 & 0.585476323 & 0.585675466 & $0.585266421^{\dagger}$ & 0.585670251 & 0.585667862 & 0.585667862\\
0.8000 & 0.585556501 & 0.585770860 & 0.585758865 & 0.585768162 & 0.585764746 & 0.585764746\\
0.9000 & 0.585648181 & 0.585876083 & 0.585873349 & 0.585877059 & 0.585872419 & 0.585872419\\
\bottomrule
\end{tabular}
\par\vspace{8pt}
\textit{Decay rate $M\Gamma$}\par\smallskip
\begin{tabular}{rrrrrrr}
\toprule
$q/M$ & BS & $[2/2]$ & $[8/8]$ & RN1 & Leaver & Global\\
\midrule
0.0000 & $5.3213\times10^{-5}$ & $5.3354\times10^{-5}$ & $5.3340\times10^{-5}$ & $6.9974\times10^{-5}$ & $6.9974\times10^{-5}$ & $6.9974\times10^{-5}$\\
0.1000 & $5.2053\times10^{-5}$ & $5.2195\times10^{-5}$ & $5.2182\times10^{-5}$ & $6.8515\times10^{-5}$ & $6.8516\times10^{-5}$ & $6.8516\times10^{-5}$\\
0.2000 & $4.8336\times10^{-5}$ & $4.8476\times10^{-5}$ & $4.8469\times10^{-5}$ & $6.3800\times10^{-5}$ & $6.3808\times10^{-5}$ & $6.3808\times10^{-5}$\\
0.3000 & $4.2406\times10^{-5}$ & $4.2540\times10^{-5}$ & $4.2540\times10^{-5}$ & $5.6210\times10^{-5}$ & $5.6236\times10^{-5}$ & $5.6236\times10^{-5}$\\
0.4000 & $3.4932\times10^{-5}$ & $3.5054\times10^{-5}$ & $3.5059\times10^{-5}$ & $4.6535\times10^{-5}$ & $4.6586\times10^{-5}$ & $4.6586\times10^{-5}$\\
0.5000 & $2.6758\times10^{-5}$ & $2.6862\times10^{-5}$ & $2.6870\times10^{-5}$ & $3.5820\times10^{-5}$ & $3.5899\times10^{-5}$ & $3.5899\times10^{-5}$\\
0.6000 & $1.8768\times10^{-5}$ & $1.8848\times10^{-5}$ & $1.8860\times10^{-5}$ & $2.5214\times10^{-5}$ & $2.5312\times10^{-5}$ & $2.5312\times10^{-5}$\\
0.7000 & $1.1737\times10^{-5}$ & $1.1792\times10^{-5}$ & $(1.1672\times10^{-5})^{\dagger}$ & $1.5784\times10^{-5}$ & $1.5885\times10^{-5}$ & $1.5885\times10^{-5}$\\
0.8000 & $6.2223\times10^{-6}$ & $6.2543\times10^{-6}$ & $6.2526\times10^{-6}$ & $8.3405\times10^{-6}$ & $8.4242\times10^{-6}$ & $8.4242\times10^{-6}$\\
0.9000 & $2.4838\times10^{-6}$ & $2.4979\times10^{-6}$ & $2.4977\times10^{-6}$ & $3.2955\times10^{-6}$ & $3.3465\times10^{-6}$ & $3.3465\times10^{-6}$\\
\bottomrule
\end{tabular}
\end{table*}
\begin{table*}[!htbp]
\centering\footnotesize
\caption{Relative errors for the fundamental charge scan at $\zeta=0.01$, $\ell=2$, $\mu=0.6$. Errors are percentages relative to the global method's binding energy and decay rate, evaluated before rounding. The implicit zero global-error column is omitted. }
\label{tab:scan-qbs-errors-charge}
\setlength{\tabcolsep}{3pt}
\renewcommand{\arraystretch}{0.96}
\textit{Binding error $100\epsilon_B$}\par\smallskip
\begin{tabular}{rrrrrr}
\toprule
$q/M$ & BS & $[2/2]$ & $[8/8]$ & RN1 & Leaver\\
\midrule
0.0000 & 0.8569 & 0.07495 & 0.1573 & $0$ & $0$\\
0.1000 & 0.8646 & 0.06798 & 0.1405 & $5.429\times10^{-5}$ & $1.522\times10^{-12}$\\
0.2000 & 0.893 & 0.0472 & 0.09069 & $4.315\times10^{-4}$ & $7.609\times10^{-13}$\\
0.3000 & 0.9431 & 0.01768 & 0.02006 & 0.001441 & $6.857\times10^{-12}$\\
0.4000 & 1.015 & 0.01317 & 0.05722 & 0.003364 & $7.637\times10^{-12}$\\
0.5000 & 1.107 & 0.03809 & 0.1351 & 0.006437 & $3.832\times10^{-12}$\\
0.6000 & 1.215 & 0.05198 & 0.2464 & 0.01084 & $7.701\times10^{-13}$\\
0.7000 & 1.336 & 0.05305 & $2.801^{\dagger}$ & 0.01667 & $7.746\times10^{-13}$\\
0.8000 & 1.463 & 0.04295 & 0.04131 & 0.02399 & $1.560\times10^{-12}$\\
0.9000 & 1.587 & 0.02593 & 0.006586 & 0.03284 & $4.715\times10^{-12}$\\
\bottomrule
\end{tabular}
\par\vspace{8pt}
\textit{Damping error $100\epsilon_\Gamma$}\par\smallskip
\begin{tabular}{rrrrrr}
\toprule
$q/M$ & BS & $[2/2]$ & $[8/8]$ & RN1 & Leaver\\
\midrule
0.0000 & 23.95 & 23.75 & 23.77 & $0$ & $0$\\
0.1000 & 24.03 & 23.82 & 23.84 & 0.001656 & $1.753\times10^{-10}$\\
0.2000 & 24.25 & 24.03 & 24.04 & 0.01332 & $1.146\times10^{-9}$\\
0.3000 & 24.59 & 24.35 & 24.35 & 0.04548 & $2.712\times10^{-10}$\\
0.4000 & 25.02 & 24.75 & 24.74 & 0.1096 & $1.312\times10^{-9}$\\
0.5000 & 25.46 & 25.17 & 25.15 & 0.2189 & $4.317\times10^{-9}$\\
0.6000 & 25.86 & 25.54 & 25.49 & 0.3884 & $1.137\times10^{-9}$\\
0.7000 & 26.11 & 25.77 & $26.52^{\dagger}$ & 0.6372 & $1.935\times10^{-9}$\\
0.8000 & 26.14 & 25.76 & 25.78 & 0.9939 & $2.951\times10^{-9}$\\
0.9000 & 25.78 & 25.36 & 25.36 & 1.523 & $1.927\times10^{-8}$\\
\bottomrule
\end{tabular}
\end{table*}
\clearpage
\begin{table*}[!htbp]
\centering\footnotesize
\caption{Fundamental mode's $\ell$ scan at $q=0.5$, $\zeta=0.01$, $\mu/(\ell+1)=0.15$. Leaver and global matching solve the complex boundary-value problem. BS and both Pad\'e columns use leading Gamow decay; RN1 expands around the complex RN pole. $\Gamma=-\operatorname{Im}\omega$. A dagger denotes an inadmissible method entry according to our criterion}
\label{tab:scan-qbs-multipole}
\setlength{\tabcolsep}{3pt}
\renewcommand{\arraystretch}{0.96}
\textit{Real frequency $M\omega_R$}\par\smallskip
\begin{tabular}{rrrrrrrr}
\toprule
$\ell$ & $M\mu$ & BS & $[2/2]$ & $[8/8]$ & RN1 & Leaver & Global\\
\midrule
1 & 0.3000 & 0.295945715 & 0.296166322 & 0.296054011 & 0.296166796 & 0.296166541 & 0.296166541\\
2 & 0.4500 & 0.444261817 & 0.444406286 & 0.444393316 & 0.444406708 & 0.444406366 & 0.444406366\\
3 & 0.6000 & 0.592510982 & 0.592616089 & 0.592616145 & 0.592616584 & 0.592616145 & 0.592616145\\
4 & 0.7500 & 0.740736367 & 0.740818738 & 0.740818762 & 0.740819301 & 0.740818762 & 0.740818762\\
5 & 0.9000 & 0.888950847 & 0.889018524 & 0.889018536 & 0.889019175 & $(0.889018536)^{\ast}$ & 0.889018536\\
\bottomrule
\end{tabular}
\par\vspace{8pt}
\textit{Decay rate $M\Gamma$}\par\smallskip
\begin{tabular}{rrrrrrrr}
\toprule
$\ell$ & $M\mu$ & BS & $[2/2]$ & $[8/8]$ & RN1 & Leaver & Global\\
\midrule
1 & 0.3000 & $4.4623\times10^{-6}$ & $4.2490\times10^{-6}$ & $4.3594\times10^{-6}$ & $6.3039\times10^{-6}$ & $6.3107\times10^{-6}$ & $6.3107\times10^{-6}$\\
2 & 0.4500 & $5.2569\times10^{-8}$ & $5.2404\times10^{-8}$ & $5.2423\times10^{-8}$ & $6.5728\times10^{-8}$ & $6.5870\times10^{-8}$ & $6.5870\times10^{-8}$\\
3 & 0.6000 & $5.0034\times10^{-10}$ & $5.0611\times10^{-10}$ & $5.0611\times10^{-10}$ & $5.9566\times10^{-10}$ & $5.9779\times10^{-10}$ & $5.9779\times10^{-10}$\\
4 & 0.7500 & $4.3880\times10^{-12}$ & $4.4623\times10^{-12}$ & $4.4623\times10^{-12}$ & $5.09\times10^{-12}$ & $5.1139\times10^{-12}$ & $5.1138\times10^{-12}$\\
5 & 0.9000 & $3.69\times10^{-14}$ & $3.76\times10^{-14}$ & $3.76\times10^{-14}$ & $4.20\times10^{-14}$ & $(4.2317\times10^{-14})^{\ast}$ & $4.23\times10^{-14}$\\
\bottomrule
\end{tabular}
\end{table*}
\begin{table*}[!htbp]
\centering\footnotesize
\caption{Relative errors (wrt to global matching)  for the fundamental mode's $\ell$ scan, at $q=0.5$, $\zeta=0.01$, $\mu/(\ell+1)=0.15$. Errors are percentages relative to the global method's binding energy and decay rate, evaluated before rounding. The implicit zero global-error column is omitted.}
\label{tab:scan-qbs-errors-multipole}
\setlength{\tabcolsep}{3pt}
\renewcommand{\arraystretch}{0.96}
\textit{Binding error $100\epsilon_B$}\par\smallskip
\begin{tabular}{rrrrrrr}
\toprule
$\ell$ & $M\mu$ & BS & $[2/2]$ & $[8/8]$ & RN1 & Leaver\\
\midrule
1 & 0.3000 & 5.76 & 0.005708 & 2.935 & 0.006656 & $9.514\times10^{-10}$\\
2 & 0.4500 & 2.584 & 0.001438 & 0.2333 & 0.006111 & $1.538\times10^{-10}$\\
3 & 0.6000 & 1.424 & $7.551\times10^{-4}$ & $4.658\times10^{-6}$ & 0.005948 & $4.060\times10^{-11}$\\
4 & 0.7500 & 0.8974 & $2.593\times10^{-4}$ & $9.467\times10^{-8}$ & 0.005871 & $1.330\times10^{-11}$\\
5 & 0.9000 & 0.6164 & $1.022\times10^{-4}$ & $7.279\times10^{-11}$ & 0.005826 & $(2.022\times10^{-12})^{\ast}$\\
\bottomrule
\end{tabular}
\par\vspace{8pt}
\textit{Damping error $100\epsilon_\Gamma$}\par\smallskip
\begin{tabular}{rrrrrrr}
\toprule
$\ell$ & $M\mu$ & BS & $[2/2]$ & $[8/8]$ & RN1 & Leaver\\
\midrule
1 & 0.3000 & 29.29 & 32.67 & 30.92 & 0.1077 & $8.070\times10^{-9}$\\
2 & 0.4500 & 20.19 & 20.44 & 20.41 & 0.2151 & $1.139\times10^{-5}$\\
3 & 0.6000 & 16.3 & 15.34 & 15.34 & 0.3551 & $1.812\times10^{-4}$\\
4 & 0.7500 & 14.19 & 12.74 & 12.74 & 0.528 & 0.001945\\
5 & 0.9000 & 12.86 & 11.22 & 11.22 & 0.7343 & $(4.453\times10^{-5})^{\ast}$\\
\bottomrule
\end{tabular}
\end{table*}

\end{document}